\documentclass[aps,prb,times,twocolumn,amsmath,amssymb,superscriptaddress,floatfix,showpacs]{revtex4}

\usepackage{color}

\usepackage{graphicx}
\usepackage{subfigure}
\usepackage{bbold}
\usepackage{verbatim}
\usepackage{float}
\usepackage{enumerate}
\usepackage{amsfonts}
\usepackage{multirow}
\usepackage[colorlinks,bookmarks=false,citecolor=blue,linkcolor=red,urlcolor=blue]{hyperref}

\newcommand{\dg}{^\dagger}
\newcommand{\pdg}{^{\phantom\dagger}}

\newcommand{\bolk}{\mathbf{k}}
\newcommand{\bolK}{\mathbf{K}}
\newcommand{\bolp}{\mathbf{p}}
\newcommand{\bolq}{\mathbf{q}}

\newcommand{\bolQ}{\mathbf{Q}}

\newcommand{\bolr}{\mathbf{r}}

\newcommand{\ket}[1]{| #1 \rangle}  
\newcommand{\VEV}[1]{\langle #1 \rangle}  

\newsavebox{\dotdot}
\savebox{\dotdot}[3mm]{\shortstack{\circle*{0.8}\\ \\ \circle*{0.8}}}

\begin{document}


\title{
Ghost in the machine: Theory of inelastic neutron scattering in a field-induced spin-nematic state
}


\author{Andrew Smerald}

\affiliation{Institut de Th{\'e}orie des Ph{\'e}nom{\`e}nes Physiques, Ecole Polytechnique F{\'e}d{\'e}rale de Lausanne (EPFL), CH-1015 Lausanne, Switzerland}
\affiliation{Okinawa Institute for Science and Technology Graduate University, Onna-son, Okinawa 904-0495, Japan}

\author{Hiroaki T. Ueda}
\affiliation{Okinawa Institute for Science and Technology Graduate University, Onna-son, Okinawa 904-0495, Japan}

\author{Nic Shannon}
\affiliation{Okinawa Institute for Science and Technology Graduate University, Onna-son, Okinawa 904-0495, Japan}


\date{\today}


\begin{abstract}  
The spin-nematic state has proved elusive, due to the difficulty of
observing the order parameter in experiment.
In this article we develop a theory of spin excitations in a
field-induced spin-nematic state, and use it to show how a
spin-nematic order can be indentified using inelastic neutron
scattering.
We concentrate on 2-dimensional frustrated ferromagnets, for which a
two-sublattice, bond-centered spin-nematic state is predicted to exist
over a wide range of parameters.
First, to clarify the nature of spin-excitations, we introduce a
soluble spin-1 model, and use this to derive a continuum field theory,
applicable to any two-sublattice spin-nematic state.
We then parameterise this field theory, using diagrammatic
calculations for a realistic microscopic model of a spin-1/2
frustrated ferromagnet, and show how it can be used to make
predictions for inelastic neutron scattering.
As an example, we show quantitative predictions for inelastic
scattering of neutrons from BaCdVO(PO$_4$)$_2$, a promising candidate
to realise a spin-nematic state at an achievable $h\sim 4$T.
We show that in this material it is realistic to expect a ghostly
Goldstone mode, signalling spin-nematic order, to be visible in
experiment.
\end{abstract}


\maketitle


\section{Introduction}

\begin{figure}[ht]
\centering
\includegraphics[width=0.48\textwidth]{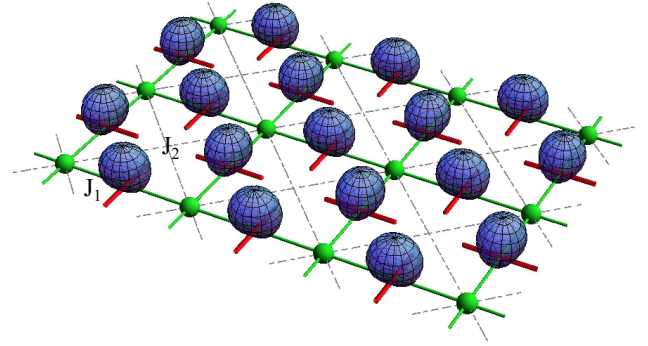}
\caption{\footnotesize{(Color online). 
Two-sublattice, bond-centered spin-nematic state, of the type found in spin-1/2 frustrated 
ferromagnets in applied magnetic field.
Spin fluctuations show quadrupolar character, visible in the probability
distribution for spin fluctuations on the bonds of the lattice, here represented 
by a blue surface.
Spin-nematic order is known to occur close to saturation 
in the \mbox{spin-1/2} $J_1$--$J_2$ model on a square-lattice,
$\mathcal{H}^{\sf S=1/2}_{\sf J_1-J_2}$~[Eq.~\ref{eq:HJ1J2}],
for ferromagnetic $J_1$ and antiferromagnetic $J_2$
\mbox{[\onlinecite{shannon06,ueda07,ueda-arXiv}]}. 
This model is believed to describe 
a number of quasi-two dimensional magnets, including 
BaCdVO(PO$_4$)$_2$ [\onlinecite{nath08}].
}}
\label{fig:J1J2h-nematic}
\end{figure}

The spin-nematic state is a ``hidden order'' of spin degrees of freedom, 
involving the ordering of spin-quadrupole moments, in the absence of conventional 
spin-dipole order [cf. Fig.~\ref{fig:J1J2h-nematic}].
The spin-nematic state was first proposed several decades ago\cite{blume69,chen71,andreev84},
and the theoretical possibility of spin-nematic order is now well-established, especially for spin-1/2 
frustrated magnets in applied magnetic field\cite{chubukov91-PRB,shannon06,momoi06,heidrich-meisner06,lu06,ueda07,vekua07,kecke07,kuzian07,hikihara08,sudan09,ueda09,zhitomirsky10,syromyatnikov12,sato13,ueda13,starykh14}.
Nevertheless, to date, the spin-nematic has generally lived up to its epithet, and remained 
well-hidden from experimental observation.


The reason why spin-nematic order is difficult to observe is that its order parameter, 
a quadrupole moment of spin, does not break time-reversal symmetry~\cite{andreev84}.
This means that spin-nematics are invisible to common probes of magnetism : they do 
not lead to magnetic Bragg peaks in elastic neutron scattering, asymmetry in muon spin 
resonance ($\mu$SR), or splitting of spectral lines in nuclear magnetic resonance (NMR) 
experiments.
In this article, we continue the program begun in \mbox{[\onlinecite{smerald13,smerald-arXiv,smerald-thesis}]} 
of exploring how the symmetries broken by spin-nematic order manifest themselves 
in its excitations, and how these excitations might be observed in experiment.
To this end we develop a general theory of inelastic neutron scattering from a spin-nematic 
state in applied magnetic field.


\begin{figure*}[ht]
\includegraphics[width=0.95\textwidth]{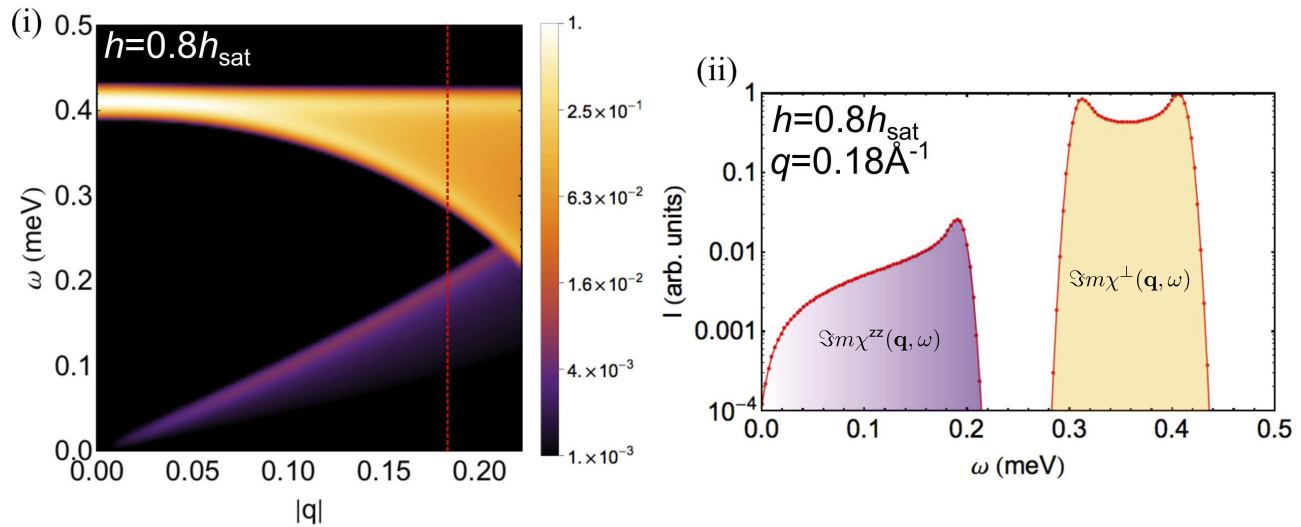}
\caption{\footnotesize{(Color online). 
Predictions for inelastic neutron scattering from a powder sample 
of a material
exhibiting bond-centered spin-nematic order in applied magnetic field, 
of the type shown in Fig.~\ref{fig:J1J2h-nematic}.
(i) Inelastic scattering at small $|{\bf q}|$, for magnetic field approaching 
the saturation value $h_{\sf sat}$.   
The existence of a spin-nematic state is heralded by a ghostly,
linearly-dispersing Goldstone-mode at low energy.
(ii) Inelastic scattering at fixed $|{\bf q}|=0.18$\AA$^{-1}$, showing 
the distribution of spectral weight as a function of frequency.
The Goldstone mode has at its maximum about 3\% of the intensity 
of the spin-wave mode at higher energy.
Predictions for the dynamical structure factor $\chi^{\alpha\beta}({\bf q}, \omega)$
were obtained using the methods described in Section~\ref{sec:neutrons} 
of this article, for parameters relevant to BaCdVO(PO$_4$)$_2$ [\onlinecite{nath08}], 
and powder-averaged.
Equivalent results for a single-crystal sample are shown in Fig.~\ref{fig:chiJ1J2}.
}}
\label{fig:chih0_8angint}
\end{figure*}

The scenario we explore, summarised in Fig.~\ref{fig:J1J2h-nematic} and Fig.~\ref{fig:chih0_8angint},  
is applicable to a wide range of materials. 
When a frustrated magnet is polarised by applied magnetic field, interactions between magnons 
can lead to the formation of a two-magnon bound state.
At lower values of magnetic field, this bound-state can condense, leading to spin-nematic order~\cite{chubukov91-PRB,shannon06,sindzingre09,ueda-arXiv} [Fig.~\ref{fig:J1J2h-nematic}].  
Since the spin-nematic order breaks spin-rotation invariance in the plane perpendicular to the
magnetic field, it must possess a Goldstone mode.
This has observable consequences --- the two-magnon bound state, invisible to neutrons in the polarised phase, transforms upon condensation
into a ghostly, linearly-dispersing Goldstone mode, which can be resolved in inelastic neutron scattering
[Fig.~\ref{fig:chih0_8angint}].  


%


For concreteness, in this article we concentrate on spin-1/2 frustrated ferromagnets
on a square lattice, taking our motivation from materials such as 
Pb$_2$VO(PO$_4$)$_2$\cite{kaul04,kaul05,tsirlin09,skoulatos09,nath09}
and SrZnVO(PO$_4$)$_2$\cite{tsirlin09,skoulatos09,bossoni11}.
We pay particular attention to the quasi-two dimensional frustrated magnet 
BaCdVO(PO$_4$)$_2$\cite{nath08,tsirlin09}, which has a saturation field of only 
\mbox{$h_{\sf sat} \approx 4\ \text{T}$}, and so is easily accessible to experiment. 
However, with small modifications, the same methods and conclusions can be generalised 
to other systems, such as coupled spin-half chains~\cite{sato13,starykh14,syromyatnikov12,zhitomirsky10}


The primary microscopic model we consider is the \mbox{spin-1/2} ``$J_1$--$J_2$'' 
Heisenberg model,
\begin{align}
\mathcal{H}^{\sf S=1/2}_{\sf J_1-J_2}=&
J_1 \sum_{\langle ij \rangle_1} {\bf S}_i.{\bf S}_j 
+ J_2 \sum_{\langle ij \rangle_2} {\bf S}_i.{\bf S}_j 
+ h \sum_i S_i^{\sf z},
\label{eq:HJ1J2}
\end{align}
where $\langle ij \rangle_1$ counts first-neighbour bonds, and $\langle ij \rangle_2$ second-neighbour bonds of a square lattice [see Fig.~\ref{fig:J1J2h-nematic}].
Magnetic field $h$ defines the $S^{\sf z}$ axis for spin.
This model 
is believed to describe several distinct families of quasi-2D materials,
including the vandates Pb$_2$VO(PO$_4$)$_2$\cite{kaul04,kaul05,tsirlin09,skoulatos09,nath09}, SrZnVO(PO$_4$)$_2$\cite{tsirlin09,skoulatos09,bossoni11} and 
BaCdVO(PO$_4$)$_2$\cite{nath08,tsirlin09}.


The ``$J_1$--$J_2$''  model 
$\mathcal{H}^{\sf S=1/2}_{\sf J_1-J_2}$ [Eq.~(\ref{eq:HJ1J2})] 
can be shown to support a spin-nematic ground state for {\it all} 
ferromagnetic $J_1 < 0$ and antiferromagnetic 
\mbox{$J_2 > 0.408 |J_1|$} \mbox{[\onlinecite{sindzingre09,ueda-arXiv}]}.
Spin-nematic order is formed through the condensation of bound pairs of magnons
out of the saturated state at $h = h_{\sf sat}$ [\onlinecite{shannon06}], and
is generally believed to be stable for a small range of fields approaching saturation, i.e. for $h \lesssim h_{\sf sat}$. 
However for a range of parameters \mbox{$ 0.408 < J_2/|J_1| \lesssim 0.7$},
the zero-field ($h = 0$) ground state of 
$\mathcal{H}^{\sf S=1/2}_{\sf J_1-J_2}$ [Eq.~(\ref{eq:HJ1J2})] 
is also a spin-nematic state \cite{shannon06,shindou09,ueda07,shindou11,shindou13}.


1d and quasi-1d frustrated ferromagnets have also been extensively studied in the search for the spin-nematic state.
Theoretically it has been shown that $J_1$-$J_2$, spin-1/2 chains in applied magnetic field, with \mbox{$J_1<0$} and \mbox{$J_2>0$}, demonstrate dominant quadrupolar correlations for a wide parameter range\cite{chubukov87,chubukov90,chubukov91-PRB,hikihara08,sudan09,kecke07,kuzian07,vekua07,heidrich-meisner06,chubukov91-JPCM,lu06,ueda09}.
In the presence of small interchain coupling this can lead to a long-range-ordered spin-nematic state at low temperature\cite{sato13,starykh14,syromyatnikov12,zhitomirsky10}.
The spin-nematic state is stabilised by magnetic field, and is most pronounced close to the saturation field. 
There have been a number of calculations of dynamical properties of such a spin-nematic state, with a view to providing experimental predictions\cite{sato09,sato11,sato13,starykh14,syromyatnikov12,zhitomirsky10}.
The material LiCuVO$_4$ is thought to approximately realise this model, and may show spin-nematic order close to saturation\cite{svistov10,zhitomirsky10,buttgen14}.
However, high-field NMR measurements have not yet detected evidence for such a state, and have shown that, if it does exist, it is limited to a very narrow field range\cite{buttgen14,smerald-arXiv}.


The spin-nematic state that appears in spin-1/2 models such as $\mathcal{H}^{\sf S=1/2}_{\sf J_1-J_2}$ [Eq.~\ref{eq:HJ1J2}]  is known as a bond-nematic.
While an individual spin-1/2 cannot have a quadrupole moment, a pair of neighbouring spin-1/2's can form a triplet, and thus develop a quadrupole moment living on the bond\cite{andreev84}.
If conventional dipole magnetism is suppressed, for example due to high frustration, quadrupolar or higher-order multipolar correlations can be revealed.
In $\mathcal{H}^{\sf S=1/2}_{\sf J_1-J_2}$ [Eq.~\ref{eq:HJ1J2}] the triplets organise themselves 
into a bond-centered antiferroquadrupolar (AFQ) order\cite{shannon06,shindou09,ueda07,shindou11,shindou13}.
The way in which this occurs is most easily understood at high values
of magnetic field, in the saturated state, where triplets are preformed.
Here magnons form bound states, which condense to give 
spin-nematic order as the magnetic field is lowered\cite{shannon06}.
In such a state the spin-dipole moment $\langle {\bf S} \rangle=0$, while the rank-2, symmetric, traceless tensor,
\begin{align}
Q_{ij}^{\alpha\beta} = S_i^\alpha S_j^\beta + S_i^\beta S_j^\alpha -\frac{2}{3}\delta^{\alpha\beta} {\bf S}_i \cdot {\bf S}_j,
\label{eq:OP}
\end{align}
with $\alpha,\beta = {\sf x,y,z}$, has entries with non-zero expectation value\cite{andreev84}.


\begin{figure}[t]
\centering
\includegraphics[width=0.45\textwidth]{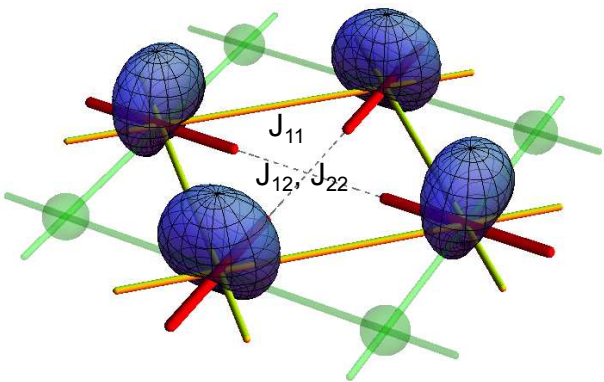}
\caption{\footnotesize{(Color online). 
Detail of a bond-centered, two-sublattice spin-nematic state in 
applied magnetic field.   
%
Green spheres represent spin-1/2 degrees of freedom at the vertices 
of a square lattice.
The bond-centred, nematic order parameter is represented by red cylinders, 
and the associated distribution of spin fluctuations by a blue surface.
A second, bond-centred lattice is introduced and shown in yellow.
In this article we consider both the site-centered spin-1/2 Heisenberg model 
$\mathcal{H}^{\sf S=1/2}_{\sf J_1-J_2}$ [Eq.~\ref{eq:HJ1J2}], and a bond-centered
spin-1 bilinear-biquadratic (bbq) model 
$\mathcal{H}^{\sf S=1}_{\sf bbq}$ [Eq.~(\ref{eq:H-BBQ})].  
Both exhibit the same form of spin-nematic order, in applied magnetic field.
}}
\label{fig:Spin1vsSpin12}
\end{figure}


At $h=0$, one possible approach to understanding the excitation spectrum of the spin-nematic state in
$\mathcal{H}_{\sf J1-J2}$ [Eq.~\ref{eq:HJ1J2}] is to work directly with the microscopic model, 
construct a lattice gauge theory, and then solve 
this within a large-N expansion scheme~\cite{shindou09,shindou11,shindou13}.
This approach has the advantage that, in principle, it can access all of the
different excitations of the spin-nematic state.
However, the solution of the lattice gauge theory is extremely involved,
which complicates the interpretation of experiments\cite{shindou13}.


A second possibility --- explained in detail in Refs.~[\onlinecite{smerald13,smerald-thesis}] --- is
to construct a continuum theory for the long-wavelength excitations of 
2-sublattice AFQ order.
This approach has the advantage of bringing the universal properties 
of the spin-nematic state to the fore, and of making clear predictions 
for inelastic neutron scattering.
However, being grounded in the symmetry of the order parameter, it cannot 
hope to describe the microscopic details of the underlying spin-1/2 model 
at high energies.
%


In this article we combine the continuum theory approach with a microscopic study of $\mathcal{H}_{\sf J1-J2}$ [Eq.~\ref{eq:HJ1J2}].
We use diagrammatic calculations to determine the magnetic dispersion spectrum of 1-magnon and 2-magnon excitations at and just below $h_{\sf sat}$.
This allows the continuum theory to be parametrised using $J_1$ and $J_2$, and therefore quantitative predictions to be made for inelastic neutron scattering experiments.


Inelastic neutron scattering measures the dynamical spin correlation function, which is defined as,
\begin{align}
&\Im m   \chi^{\alpha\beta}({\bf q}, \omega)  = \nonumber \\
&\qquad   \Im m \{ 
i \int_0^\infty dt 
e^{i \omega t} 
\langle \delta S_{\bf q}^\alpha(t) \delta S_{-{\bf q}}^\beta(0) \rangle
\},
\label{eq:im-chi}
\end{align}
where we have set $g\mu_{\sf B}=1$.
Thus the task of this article is to calculate this quantity, and we will do this for two complementary models of field-induced, spin-nematic order on the square lattice: $\mathcal{H}_{\sf J1-J2}$ [Eq.~\ref{eq:HJ1J2}] and a spin-1 model that is a generalisation of the bilinear-biquadratic (BBQ) Hamiltonian\cite{blume69,chen71}.
We now briefly review the route taken.


In Section~\ref{sec:flvwave} we introduce a spin-1, BBQ model with a partially polarised, 2-sublattice, spin-nematic ground state.
As shown in Fig.~\ref{fig:Spin1vsSpin12}, one way to view this is as an effective model describing spin-1 degrees freedom living at the bond centres of a spin-1/2 square lattice.
The lattice of bond centres also forms a square lattice, with a reduced lattice constant $b=a/\sqrt{2}$.
A major advantage of spin-1 models is that the excitation spectrum can be calculated within flavour-wave theory\cite{papanicolau84,papanicolau88,onufrieva85,lauchli06,tsunetsugu06}.
This allows predictions for $\Im m   \chi^{\alpha\beta}({\bf q}, \omega) $ [Eq.~\ref{eq:im-chi}] to be made.
While the primary motivation for considering this model is as a first step towards making predictions for spin-1/2 systems, spin-1 is also interesting in its own right\cite{lauchli06,tsunetsugu06}, and if real materials can be synthesised with large biquadratic interaction, the results presented here would be relevant.


In Section~\ref{sec:fieldtheory} we start from the spin-1 BBQ model studied in Section~\ref{sec:flvwave} and use it to derive a continuum field theory.
We first demonstrate that, at long wavelength, this exactly reproduces the flavour-wave results of Section~\ref{sec:flvwave}.
The power of the field theory is that it is a theory of the order parameter symmetry, and therefore describes the universal features of a partially polarised, 2-sublattice spin-nematic state.
Thus we recast the theory in terms of a minimal set of hydrodynamic parameters.
This renders the theory free of any particular microscopic model, and one can in principle parametrise it from any microscopic model with a partially polarised, 2-sublattice spin-nematic ground state or directly from experiment.


In Section~\ref{sec:spinmapping} we make a mapping between the effective spin-1 degrees of freedom on the bonds and spin-1/2 degrees of freedom on the sites.
This will allow predictions to be made for spin-1/2 frustrated ferromagnets, based on the theory developed in Section~\ref{sec:flvwave} and Section~\ref{sec:fieldtheory}.


In Section~\ref{sec:parametrisation} we consider $\mathcal{H}^{\sf S=1/2}_{\sf J_1-J_2}$ [Eq.~\ref{eq:HJ1J2}] from a microscopic perspective close to the saturation field.
We consider the condensation of magnons out of the fully-saturated state.
For a sizeable parameter range, condensation of bound-magnon pairs occurs at a higher magnetic field than the condensation of single magnons, and therefore a spin-nematic state is formed.
Diagrammatic calculations allow the critical field to be calculated, as well as the velocities and gaps of the excitation modes in the spin-nematic state.
These can then be used to parametrise the continuum theory.

  
In Section~\ref{sec:neutrons} we make predictions for inelastic neutron scattering experiments for materials described by $\mathcal{H}^{\sf S=1/2}_{\sf J_1-J_2}$ [Eq.~\ref{eq:HJ1J2}].
As a worked example we consider the material BaCdVO(PO$_4$)$_2$, which is expected to have a spin-nematic ground state close to saturation, and show quantitative experimental predictions.
%


Finally in Section~\ref{sec:conclusion} we conclude by showing that the detection of a ghostly Goldstone mode in BaCdVO(PO$_4$)$_2$ -- a characteristic signature of spin-nematic order -- is experimentally feasible using current instruments.


In Appendix~\ref{App:FT-h=0} we derive a non-linear-sigma-model field theory for the 2 sublattice AFQ state at $h=0$.
This is complementary to the continuum theory presented in Section~\ref{sec:fieldtheory}, which considerably simplifies at $h=0$ if a set of high-energy modes are eliminated by a Gaussian integral. 
This theory can then be compared to previous work considering the spin-nematic state of $\mathcal{H}^{\sf S=1/2}_{\sf J_1-J_2}$ [Eq.~\ref{eq:HJ1J2}] at $h=0$\cite{shindou09,shindou11,shindou13}.


\section{A spin-1 model for the 2-sublattice spin nematic in applied magnetic field}
\label{sec:flvwave}


Here we construct and solve a spin-1 bilinear-biquadratic model in applied magnetic field on the square lattice that supports the same type of spin-nematic state as is found in $\mathcal{H}^{\sf S=1/2}_{\sf J_1-J_2}$ [Eq.~\ref{eq:HJ1J2}].
While this may be relevant to spin-1 systems, the primary motivation is as an effective model of bond degrees of freedom in a spin-1/2 bond-nematic state.
We make the assumption that the 2-sublattice, spin-nematic state reduces to triplets on nearest-neighbour bonds and, at low energy, singlet degrees of freedom can be ignored. 
The exchange parameters in the spin-1 model are chosen such that a partially-polarised, 2-sublattice AFQ ground state is realised for the full magnetic field range $0<h<h_{\sf sat}$.


We use flavour-wave theory\cite{papanicolau84,papanicolau88,onufrieva85,lauchli06,tsunetsugu06} to determine the evolution of the magnetic dispersion and the imaginary part of the dynamic spin susceptibility as the magnetic field is varied. 
Here we present the results of linear flavour-wave theory (see Fig.~\ref{fig:bbq-chiqomh}).
At $h=0$ we have checked that the 2-sublattice AFQ ground state remains stable when interactions are included, but the details will be presented elsewhere\cite{smerald-unpub}.


\subsection{Definition of the model}


The model we introduce is a straightforward generalisation of the bilinear-biquadratic model introduced in Ref.~[\onlinecite{blume69,chen71}].
Effective spin-1 degrees of freedom are placed at the bond centres of a square lattice with lattice constant $a$.
The bond centres of the square lattice also form a square lattice, with lattice constant $b=a/\sqrt{2}$ (see Fig.~\ref{fig:Spin1vsSpin12}).
\begin{figure*}[ht]
\includegraphics[width=0.98\textwidth]{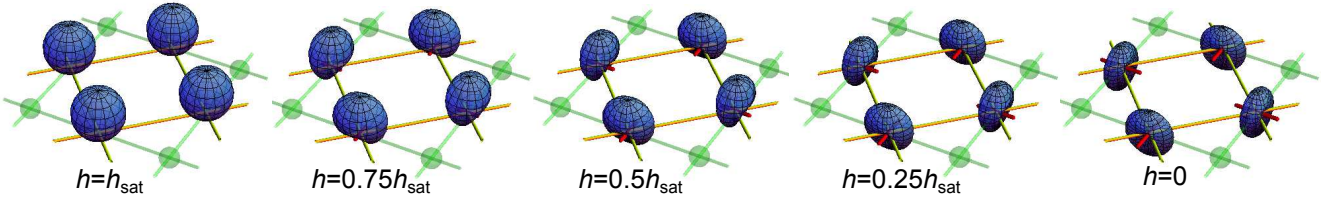}
\caption{\footnotesize{(Color online). 
Evoluiton of the 2-sublattice, AFQ spin-nematic state in applied magnetic field.
The distributions of spin fluctuations, shown by blue surfaces, are calculated within the mean-field ground state described by Eq.~\ref{eq:MFgs}.
Magnetic field is applied perpendicular to the 2-d plane, and polarises the magnetic moments.
Full polarisation is achieved at $h=h_{\sf sat}$.
The nematic order parameter is shown by red cylinders, and disappears at $h=h_{\sf sat}$.
In the case of a spin-1 system, the polarised moments and the quadrupolar order parameter exist at the vertices of the yellow lattice.
In the case of a spin-1/2 system, magnetic moments are associated with the vertices of the green lattice (shown by green spheres).
The nematic order parameter lives on the bonds, and one can define partially polarised spin-dipoles on the bonds by summing contributions from neighbouring sites (see Section~\ref{sec:spinmapping}).
Magnetic field values correspond to those in Fig.~\ref{fig:bbq-chiqomh} and Fig.~\ref{fig:FTspin1-chiqomh}.
}}
\label{fig:magnetisation}
\end{figure*}
%


The microscopic model is given by,
\begin{align}
&\mathcal{H}^{\sf S=1}_{\sf bbq}[{\bf S}]
= \sum_{\langle ij \rangle_1}  
J_{11} \left[ ({\bf S}_i.{\bf S}_j)^2 +{\bf S}_i.{\bf S}_j \right]  
-h\sum_iS^{\sf z}_i \nonumber \\
&-\sum_{\langle ij \rangle_2}  
\left[ 
J_{12} \left[ ({\bf S}_i.{\bf S}_j)^2 +{\bf S}_i.{\bf S}_j \right]
+ J_{22} ({\bf S}_i.{\bf S}_j)^2
\right],
\label{eq:H-BBQ}
\end{align}
where $\langle ij \rangle_1$ counts the first- and $\langle ij \rangle_2$ the 
second-neighbour bonds on the bond-centred lattice [see Fig.~\ref{fig:Spin1vsSpin12}] and ${\bf S}=(S^{\sf x},S^{\sf y},S^{\sf z})$ is the usual spin-1 operator.
We consider all interactions $J$ to be positive and $h$ is the applied magnetic field. 

It is useful to rewrite $\mathcal{H}^{\sf S=1}_{\sf bbq}[{\bf S}]$ [Eq.~\ref{eq:H-BBQ}] as,
\begin{align}
&\mathcal{H}^{\sf S=1}_{\sf bbq}[{\bf S},{\bf Q}] 
= \sum_{\langle ij \rangle_1}  
\frac{J_{11}}{2} \left( {\bf Q}_i.{\bf Q}_j +{\bf S}_i.{\bf S}_j \right)  
-h\sum_iS^{\sf z}_i \nonumber \\
&-\sum_{\langle ij \rangle_2}  
\left[ 
\frac{J_{12}}{2} \left( {\bf Q}_i.{\bf Q}_j +{\bf S}_i.{\bf S}_j \right) 
+ \frac{J_{22}}{2} \left( {\bf Q}_i.{\bf Q}_j -{\bf S}_i.{\bf S}_j \right)  
\right],
\label{eq:H-BBQ-SQ}
\end{align}
where,
\begin{align}
{\bf Q}&=
 \left(
\begin{array}{c}
Q^{\sf x^2-y^2} \\
Q^{\sf 3z^2-r^2} \\
Q^{\sf xy} \\
Q^{\sf yz} \\
Q^{\sf xz}
\end{array}
\right)=
 \left(
\begin{array}{c}
(S^{\sf x})^2 - (S^{\sf y})^2 \\
\frac{1}{\sqrt{3}}[2(S^{\sf z})^2-(S^{\sf x})^2 - (S^{\sf y})^2] \\
S^{\sf x}S^{\sf y}+S^{\sf y}S^{\sf x} \\
S^{\sf y}S^{\sf z}+S^{\sf z}S^{\sf y} \\
S^{\sf x}S^{\sf z}+S^{\sf z}S^{\sf x}
\end{array}
\right),
\label{eq:Q}
\end{align}
describes spin-quadrupole operators and a constant term has been dropped.
%


This model has a hidden {\sf SU(3)} symmetry for $J_{22}=0$ and $h=0$, a fact which is more easily understood if $\mathcal{H}^{\sf S=1}_{\sf bbq}[{\bf S}]$~[Eq.~\ref{eq:H-BBQ}] is expressed in terms of a director vector, ${\bf d}$.
This follows from noting that the wavefunction of a spin-1 on a site $j$ can be written as\cite{papanicolau88,ivanov03,ivanov07,ivanov08,smerald13},
\begin{align}
|{\bf d}_j \rangle= d_j^{\sf x} |x \rangle +d_j^{\sf y} |y \rangle+d_j^{\sf z} |z \rangle,
\label{eq:dvecs}
\end{align}
where,
\begin{align}
|x\rangle=i\frac{|1\rangle-|\bar{1}\rangle}{\sqrt{2}}, \ |y\rangle=\frac{|1\rangle+|\bar{1}\rangle}{\sqrt{2}}, \ |z\rangle=-i|0\rangle,
\end{align}
are linear superpositions of the usual spin-1 basis states. 
The director vector is defined by ${\bf d}_j=(d_j^{\sf x},d_j^{\sf y},d_j^{\sf z})$ and is normalised by requiring ${\bf d}_j.\bar{{\bf d}}_j=1$.
Assuming the total wavefunction can be site-factorised allows $\mathcal{H}^{\sf S=1}_{\sf bbq}[{\bf S}]$~[Eq.~\ref{eq:H-BBQ}] to be re-expressed as,
\begin{align}
 \mathcal{H}^{\sf S=1}_{\sf bbq}[{\bf d}] 
&= J_{11} \sum_{\langle ij\rangle_1} |{\bf d}_i.\bar{{\bf d}}_j|^2  
 - J_{12} \sum_{\langle ij\rangle_2} |{\bf d}_i.\bar{{\bf d}}_j|^2 \nonumber \\
&- J_{22} \sum_{\langle ij\rangle_2} |{\bf d}_i.{\bf d}_j|^2  
-ih\sum_i \left( d^{\sf x}_i \bar{d}^{\sf y}_i -d^{\sf y}_i \bar{d}^{\sf x}_i \right).
\label{eq:Hbbq-d}
\end{align} 
The first and second terms of $\mathcal{H}^{\sf S=1}_{\sf bbq}[{\bf d}]$ are ${\sf SU(3)}$ invariant,
and favour general two-sublattice states.
However, the interaction $J_{22}$ breaks the symmetry of the model down 
to ${\sf SU(2)}$, and, enforces two-sublattice AFQ order of the type shown 
in Fig.~\ref{fig:Spin1vsSpin12}.
The magnetic field further reduces the symmetry to {\sf U(1)} and, above a critical field $h_{\sf sat}$, favours a saturated paramagnetic state.

\begin{figure*}[ht]
\includegraphics[width=0.7\textwidth]{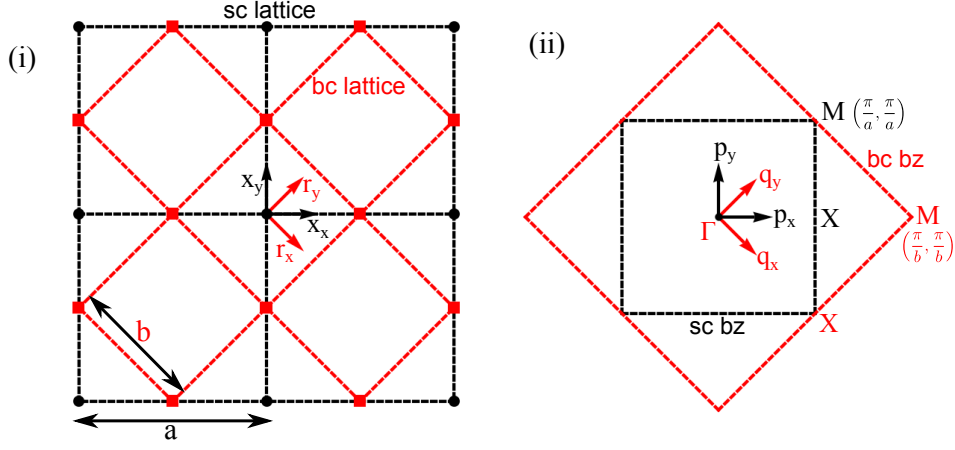}
\caption{\footnotesize{(Color online). 
Relationship between the site-centred lattice of $\mathcal{H}^{\sf S=1/2}_{\sf J_1-J_2}$ [Eq.~\ref{eq:HJ1J2}] and the bond-centred lattice of $\mathcal{H}^{\sf S=1}_{\sf bbq}[{\bf S}]$~[Eq.~\ref{eq:H-BBQ}] and associated Brillouin zones (bz).
(i) The site-centred lattice, with lattice constant $a$, is shown in black and the bond-centred lattice, with lattice constant $b=a/\sqrt{2}$, in red.
A coordinate system $(x_{\sf x},x_{\sf y})$ is associated with the sc lattice and $(r_{\sf x},r_{\sf y})$ with the bond-centred lattice.
(ii) The site-centred bz is shown in black and the bc bz in red.
The associated coordinates are $(p_{\sf x},p_{\sf y})$ and $(q_{\sf x},q_{\sf y})$.
High symmetry points in the site-centred bz are ${\bf p}_\Gamma=(0,0)$, ${\bf p}_{\sf X}=(\pi/a,0)$ and ${\bf p}_{\sf M}=(\pi/a,\pi/a)$ and in the bond-centred bz ${\bf q}_\Gamma=(0,0)$, ${\bf q}_{\sf X}=(\pi/b,0)$ and ${\bf q}_{\sf M}=(\pi/b,\pi/b)$. 
}}
\label{fig:bz}
\end{figure*}
%


An alternative rewriting of $\mathcal{H}^{\sf S=1}_{\sf bbq}[{\bf S}]$~[Eq.~\ref{eq:H-BBQ}] is in terms of a vector ${\bf e}$, which parametrises the usual spin-1 basis states.
The wavefunction on a site $j$ is written as,
\begin{align}
|{\bf e}_j \rangle= e_j^{\sf 1} |1 \rangle +e_j^{\sf 0} |0 \rangle+e_j^{\sf \bar{1}} |\bar{1} \rangle,
\label{eq:ebasis}
\end{align}
and ${\bf e}_j=(e_j^{\sf 1},e_j^{\sf 0},e_j^{\sf \bar{1}})$.
The wavefunction is normalised by requiring ${\bf e}_j.\bar{{\bf e}}_j=1$ .
Again assuming a site factorisation of the wavefunction, $\mathcal{H}^{\sf S=1}_{\sf bbq}[{\bf S}]$~[Eq.~\ref{eq:H-BBQ}] can be rewritten as,
\begin{align}
& \mathcal{H}^{\sf S=1}_{\sf bbq}[{\bf e}]
= J_{11} \sum_{\langle ij\rangle_1} |{\bf e}_i.\bar{{\bf e}}_j|^2  
 - J_{12} \sum_{\langle ij\rangle_2} |{\bf e}_i.\bar{{\bf e}}_j|^2 \nonumber \\
&\quad - J_{22} \sum_{\langle ij\rangle_2} |e_i^{\sf 1}e_j^{\sf \bar{1}}-e_i^{\sf 0}e_j^{\sf 0}+e_i^{\sf \bar{1}}e_j^{\sf 1}|^2  
-h\sum_i \left( |e_i^{\sf 1}|^2 - |e_i^{\sf \bar{1}}|^2 \right).
\label{eq:Hbbq-e}
\end{align}
At large values of $h$ is it clear that all sites will have ${\bf e}=(i,0,0)$, corresponding to a saturated paramagnet.

The relationship between the ${\bf e}$ vectors and the operators ${\bf S}$ and ${\bf Q}$ is given by,
\begin{align}
 &\left(
\begin{array}{c}
\langle S^{\sf x} \rangle \\
\langle S^{\sf y} \rangle \\
\langle S^{\sf z} \rangle \\
\langle Q^{\sf x^2-y^2} \rangle \\
\langle Q^{\sf 3z^2-r^2} \rangle \\
\langle Q^{\sf xy} \rangle \\
\langle Q^{\sf yz} \rangle \\
\langle Q^{\sf xz} \rangle 
\end{array}
\right)
=
 \left(
\begin{array}{c}
-\frac{1}{\sqrt{2}} (e^{1} \bar{e}^{0}+e^{0}\bar{e}^{1}+e^{0}\bar{e}^{\bar{1}}+e^{\bar{1}}\bar{e}^{0}) \\
\frac{i}{\sqrt{2}}  (e^{1} \bar{e}^{0}-e^{0} \bar{e}^{1}+e^{0}\bar{e}^{\bar{1}}-e^{\bar{1}}\bar{e}^{0}) \\
|e^{1}|^2 - |e^{\bar{1}}|^2\\
e^{\bar{1}} \bar{e}^{1}  + e^{1} \bar{e}^{\bar{1}}  \\
\frac{1}{\sqrt{3}}( |e^{1}|^2 +|e^{\bar{1}}|^2 -2 |e^{0}|^2 ) \\
i(e^{\bar{1}} \bar{e}^{1}  - e^{1} \bar{e}^{\bar{1}})  \\
\frac{i}{\sqrt{2}}  (e^{1} \bar{e}^{0}-e^{0} \bar{e}^{1}-e^{0}\bar{e}^{\bar{1}}+e^{\bar{1}}\bar{e}^{0}) \\
\frac{1}{\sqrt{2}} (-e^{1} \bar{e}^{0}-e^{0}\bar{e}^{1}+e^{0}\bar{e}^{\bar{1}}+e^{\bar{1}}\bar{e}^{0})
\end{array}
\right)
\label{eq:SQ-e}
\end{align}
where $\langle S^{\sf x} \rangle = \langle {\bf e}| S^{\sf x} | {\bf e} \rangle$.


\subsection{Mean-field ground state}


The mean-field ground state of $\mathcal{H}^{\sf S=1}_{\sf bbq}[{\bf S}]$~[Eq.~\ref{eq:H-BBQ}] can be derived by varying the ${\bf e}$ vectors in $\mathcal{H}^{\sf S=1}_{\sf bbq}[{\bf e}]$ [Eq.~\ref{eq:Hbbq-e}].
Since the motivation for considering the model is to study the partially polarised spin-nematic state, we do not present the entire phase diagram, but instead determine a parameter range in which this state is stable for the full field range $0<h<h_{\sf sat}$.
This is the case for $J_{11}=1$, $J_{12}>0$, $J_{22}>1$, and from now on we concentrate on this region in parameter space.



In the saturated paramagnet the mean-field wavefunction at every site is described by,
\begin{align}
 {\bf e}_{\uparrow}^{\sf mf} =
i\left(1,0,0\right).
\end{align}
Below the saturation magnetic field, 
\begin{align}
h_{\sf sat} = 4(J_{11}+J_{22}),
\label{eq:hsat}
\end{align}
the ${\bf e}$ vectors cant, forming a 2-sublattice state.
Labelling these two sublattices ${\sf A}$ and ${\sf B}$, the mean field ground state can be described by,
\begin{align}
 {\bf e}_{\sf A}^{\sf mf}  = 
{\bf R}(\theta_h) \cdot  {\bf e}_{\uparrow}^{\sf mf} , \quad
 {\bf e}_{\sf B}^{\sf mf}  = 
{\bf R}(-\theta_h) \cdot  {\bf e}_{\uparrow}^{\sf mf} ,
\label{eq:MFgs}
\end{align}
with,
\begin{align}
{\bf R}(\theta_h) =
 \left(
\begin{array}{ccc}
\cos \theta_h & 0 & \sin \theta_h  \\
0 & 1  & 0 \\
-\sin \theta_h & 0 & \cos \theta_h
\end{array}
\right).
\end{align}
The canting angle, $\theta_h$, is given by,
\begin{align}
 \cos 2\theta_h  = \frac{h}{h_{\sf sat}}.
 \label{eq:theta(h)}
\end{align}
For $h=h_{\sf sat}$ the canting angle is $\theta_h = 0$, while for $h=0$ it is $\theta_h = \pi/4$.
The field evolution of the mean-field ground state described by Eq.~\ref{eq:MFgs} is shown in Fig.~\ref{fig:magnetisation}.


\subsection{Linear flavour-wave theory in the saturated paramagnet}


Above a critical magnetic field, $h_{\sf sat}$ [Eq.~\ref{eq:hsat}], the spins align, forming a saturated paramagnet.
The excitation spectrum for $h \geq h_{\sf sat}$ can be calculated using linearised flavour-wave theory\cite{papanicolau84,papanicolau88,onufrieva85,lauchli06,tsunetsugu06}.
From this the imaginary part of the dynamical spin susceptibility can be determined.


The spin operators are written in terms of a pair of boson creation and annihilation operators, labelled $a$ and $b$, according to,
\begin{align}
S_j^{\sf z} &= 1-2 a_j\dg a_j  - b_j\dg b_j \nonumber \\
S_j^+ &= -\sqrt{2} \left( \sqrt{1 - a_j\dg a_j  - b_j\dg b_j} \ b_j\pdg +b_j\dg a_j  \right) \nonumber \\
S_j^- &= -\sqrt{2} \left( b_j\dg \sqrt{1 - a_j\dg a_j  - b_j\dg b_j}  +a_j\dg b_j  \right),
\label{eq:satPM-S}
\end{align}
where $S^{\pm}=S^{\sf x} \pm iS^{\sf y}$.
The operator $b_j\dg$ creates an excitation with $\Delta S^{\sf z}_j=1$, while $a_j\dg$ creates an excitation with $\Delta S^{\sf z}_j=2$.
It follows from Eq.~\ref{eq:Q} and Eq.~\ref{eq:satPM-S} that,
\begin{align}
&Q^{\sf x^2-y^2} =  a\dg \sqrt{1 - a\dg a  - b\dg b}  + \sqrt{1 - a\dg a  - b\dg b} \ a\pdg \nonumber \\
&Q^{\sf 3z^2-r^2} = \frac{1}{\sqrt{3}} \left(1 -3 b\dg b \right) \nonumber \\
&Q^{\sf xy} = i\left( a\dg \sqrt{1 - a\dg a  - b\dg b}  - \sqrt{1 - a\dg a  - b\dg b} \ a\pdg \right) \nonumber \\
&Q^{\sf yz} = \frac{i}{\sqrt{2}} \left(  \sqrt{1 - a\dg a  - b\dg b} \ b  \right. \nonumber \\
&\left.\qquad \qquad \qquad \qquad - b\dg \sqrt{1 - a\dg a  - b\dg b}+a\dg b-b\dg a \right) \nonumber \\
&Q^{\sf xz} = -\frac{1}{\sqrt{2}} \left(  \sqrt{1 - a\dg a  - b\dg b} \ b   \right. \nonumber \\
&\left.\qquad \qquad \qquad \qquad + b\dg \sqrt{1 - a\dg a  - b\dg b}-a\dg b-b\dg a \right).
\label{eq:satPM-Q}
\end{align}
%


The dispersion of magnetic excitations can be calculated within linear flavour-wave theory.
This involves rewriting $\mathcal{H}^{\sf S=1}_{\sf bbq}[{\bf S},{\bf Q}]$~[Eq.~\ref{eq:H-BBQ-SQ}] in terms of the boson operators defined in Eq.~\ref{eq:satPM-S} and Eq.~\ref{eq:satPM-Q}, and retaining only terms up to bilinear order.
The resulting dispersion relation has two branches,
\begin{align}
\omega_{{\bf k},h}^{\sf a} =& 
-4J_{11} (1-\gamma_{\bf k}^{(1)}) 
+4J_{12} (1-\gamma_{\bf k}^{(2)}) \nonumber \\
& \quad -4J_{22} (1+\gamma_{\bf k}^{(2)})
+2h \nonumber \\
\omega_{{\bf k},h}^{\sf b} =&
-4J_{11} (1-\gamma_{\bf k}^{(1)}) 
+4J_{12} (1-\gamma_{\bf k}^{(2)})
+h
\label{eq:omega-satPM}
\end{align} 
where,
 \begin{align}
\gamma_{\bf k}^{(1)} = \frac{1}{2}(\cos k_{\sf x} +\cos k_{\sf y}), \quad 
\gamma_{\bf k}^{(2)} =\cos k_{\sf x} \cos k_{\sf y}.
\label{eq:gamma}
\end{align}


The imaginary part of the dynamical spin susceptibilty, $\Im m \chi^{\alpha\beta}({\bf q}, \omega)$ [Eq.~\ref{eq:im-chi}], can be calculated within the flavour-wave approach, yielding,
\begin{align}
\Im m  \chi^{\sf xx}({\bf q},\omega)  &= \Im m  \chi^{\sf yy}({\bf q},\omega) 
= \frac{\pi}{2} \delta(\omega - \omega_{\bf q}^{\sf b}) \nonumber \\
\Im m  \chi^{\sf zz}({\bf q},\omega)  &=0.
\end{align} 
The dominant feature is a band of 1-magnon excitations in the perpendicular channel, as shown in Fig.~\ref{fig:bbq-chiqomh}(a) and (b).
This is gapped at all wavevectors, and has uniform spectral weight. 
There is also a band of 2-magnon excitations in the longitudinal channel with zero spectral weight. 
For $h> h_{\sf sat}$ this is gapped at all ${\bf k}$, but exactly at $h= h_{\sf sat}$ the gap closes at ${\bf k}=0$.
This signifies the onset of spin-nematic long-range order.
The spin-dipole remains parallel to the applied magnetic field, and a director order parameter appears in the plane perpendicular to $h$ [see Fig.~\ref{fig:Spin1vsSpin12}].


\subsection{Linear flavour-wave theory in the partially polarised spin nematic}
\label{sec:flvwv-nem}


The dispersion of magnetic excitations in the partially polarised spin-nematic state, which occurs at \mbox{$0<h<h_{\sf sat}$}, is now calculated using linear flavour-wave theory.
This is done by considering small fluctuations around the mean field ground state given in Eq.~\ref{eq:MFgs}.
%


The spin operators can be rewritten as,
\begin{align}
&S_i^{\sf x} = -\frac{1}{\sqrt{2}}  \left[ (\cos \theta_h -e^{i{\bf k}_{\sf M}\cdot {\bf r}_i} \sin \theta_h)  \left(  \sqrt{1 - a\dg a  - b\dg b} \ b   \right. \right. \nonumber \\
&  \left. \left. \hspace{-3mm}
+ b\dg \sqrt{1 - a\dg a  - b\dg b} \right)
+(\cos \theta_h + e^{i{\bf k}_{\sf M}\cdot {\bf r}_i} \sin \theta_h) \left( a\dg b+b\dg a \right)  \right] 
\nonumber \\
&S_i^{\sf y} = \frac{i}{\sqrt{2}}  \left[ (\cos \theta_h + e^{i{\bf k}_{\sf M}\cdot {\bf r}_i} \sin \theta_h)  \left(  \sqrt{1 - a\dg a  - b\dg b} \ b   \right. \right. \nonumber \\
&  \left. \left. \hspace{-3mm}
- b\dg \sqrt{1 - a\dg a  - b\dg b} \right)
+(\cos \theta_h - e^{i{\bf k}_{\sf M}\cdot {\bf r}_i} \sin \theta_h) \left( b\dg a -a\dg b \right)  \right] 
\nonumber \\
&S_i^{\sf z} = \cos 2\theta_h \left(1 -2a\dg a- b\dg b \right) \nonumber \\
& + e^{i{\bf k}_{\sf M}\cdot {\bf r}_i} \sin 2\theta_h \left(  \sqrt{1 - a\dg a  - b\dg b} \ a  + a\dg \sqrt{1 - a\dg a  - b\dg b} \right) 
 \nonumber \\
 \end{align}
while the quadrupole operators are given by,
\begin{align}
&Q_i^{\sf x^2-y^2} =  - e^{i{\bf k}_{\sf M}\cdot {\bf r}_i} \sin 2\theta_h \left(1 -2a\dg a- b\dg b \right) \nonumber \\
&+\cos 2\theta_h  \left(  \sqrt{1 - a\dg a  - b\dg b} \ a  + a\dg \sqrt{1 - a\dg a  - b\dg b} \right) 
\nonumber \\
&Q_i^{\sf 3z^2-r^2} = \frac{1}{\sqrt{3}} \left(1 -3 b\dg b \right) 
\nonumber \\
&Q_i^{\sf xy} = i\left( a\dg \sqrt{1 - a\dg a  - b\dg b}  - \sqrt{1 - a\dg a  - b\dg b} \ a\pdg \right) 
\nonumber \\
&Q_i^{\sf yz} = \frac{i}{\sqrt{2}} \left[  (\cos \theta_h - e^{i{\bf k}_{\sf M}\cdot {\bf r}_i}\sin \theta_h)\left( \sqrt{1 - a\dg a  - b\dg b} \ b   \right. \right. \nonumber \\
& \left.  \left. 
\hspace{-3mm} - b\dg \sqrt{1 - a\dg a  - b\dg b} \right)
+ (\cos \theta_h + e^{i{\bf k}_{\sf M}\cdot {\bf r}_i} \sin \theta_h) \left( a\dg b-b\dg a \right) \right] 
\nonumber \\
&Q_i^{\sf xz} = \frac{1}{\sqrt{2}} \left[  -(\cos \theta_h + e^{i{\bf k}_{\sf M}\cdot {\bf r}_i} \sin \theta_h)\left( \sqrt{1 - a\dg a  - b\dg b} \ b  \right. \right. \nonumber \\
& \left. \left.
\hspace{-3mm} + b\dg \sqrt{1 - a\dg a  - b\dg b} \right)
+ (\cos \theta_h - e^{i{\bf k}_{\sf M}\cdot {\bf r}_i} \sin \theta_h) \left( a\dg b+b\dg a \right) \right] 
\label{eq:SAQA-hintermediate}
\end{align}
where ${\bf k}_{\sf M}=(\pi,\pi)$ is the 2-sublattice ordering vector [see Fig.~\ref{fig:bz}] and $e^{i{\bf k}_{\sf M} \cdot {\bf r}_i} =\pm 1$ depending on whether $i$ is within the ${\sf A}$ or ${\sf B}$ sublattice.


%
\begin{figure*}[ht]
\includegraphics[width=0.98\textwidth]{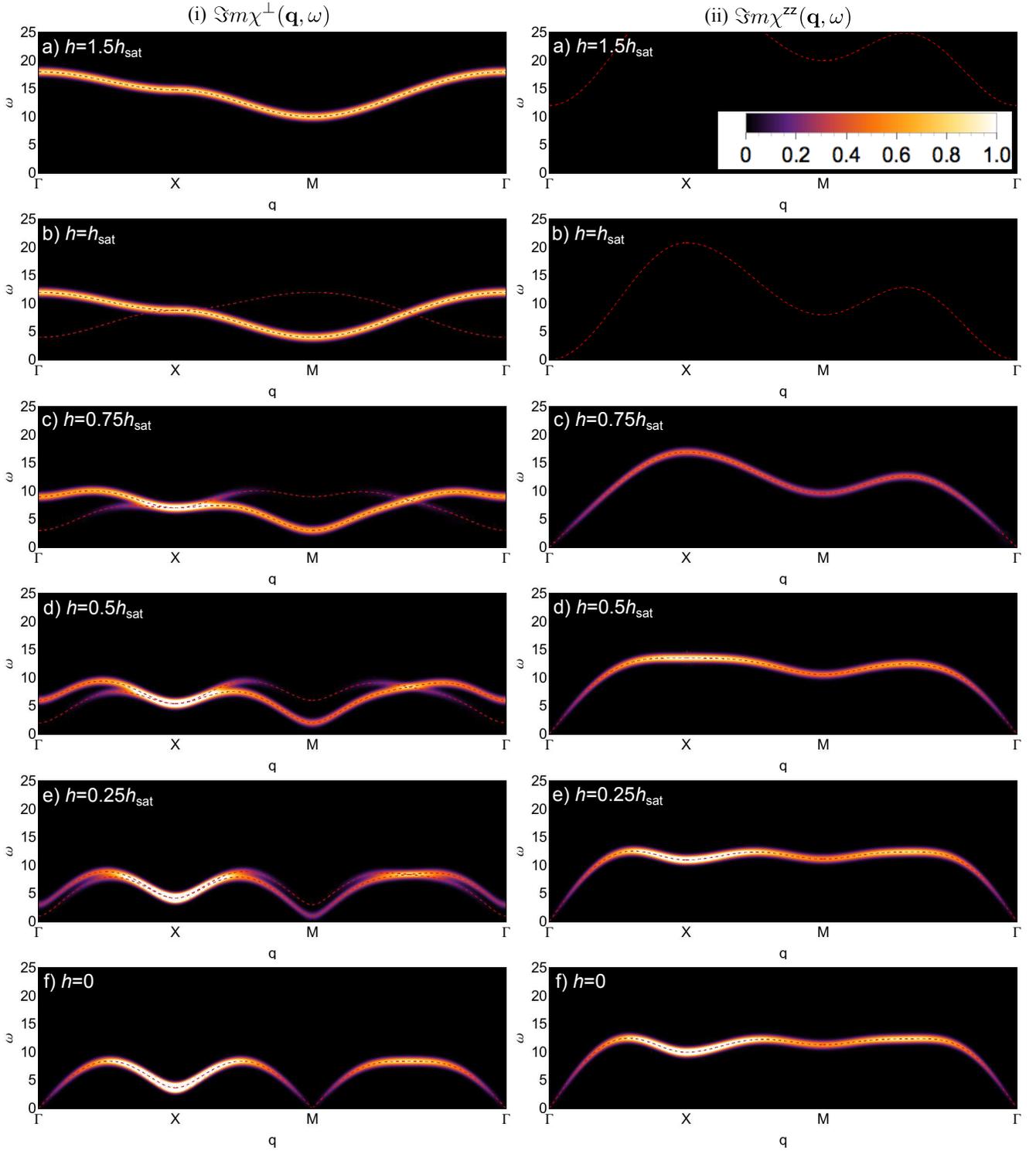}
\caption{\footnotesize{(Color online). 
Flavour-wave predictions for the imaginary part of the dynamic spin susceptibility of a spin-1, partially-polarised, 2-sublattice, spin-nematic state in applied magnetic field.
These follow from considering $\mathcal{H}^{\sf S=1}_{\sf bbq}[{\bf S}]$~[Eq.~\ref{eq:H-BBQ}] with $J_{11}=1$, $J_{12}=0.1$ and $J_{22}=2$.
The external magnetic field is gradually reduced from a) $h=1.5h_{\sf sat}$ to f) $h=0$.
(i) The transverse susceptibility $\Im m   \chi^{\perp}({\bf q},\omega) $ [Eq.~\ref{eq:bbq-chiqomh}].
Dashed red lines show $\omega_{{\bf q},h}^{\sf b}$ at all $h$ and $\omega_{{\bf q}+{\bf q}_{\sf M},h}^{\sf b}$ for $h\leq h_{\sf sat}$ [see Eq.~\ref{eq:omega-satPM}, Eq.~\ref{eq:omegakhb}].
The associated intensity is shown by the colour scale inset in panel (ii)a.
(ii) The longitudinal susceptibility $\Im m   \chi^{\sf zz}({\bf q},\omega) $ [Eq.~\ref{eq:bbq-chiqomh}].
Dashed red lines show $\omega_{{\bf q}+{\bf q}_{\sf M},h}^{\sf a}$ [see Eq.~\ref{eq:omega-satPM}, Eq.~\ref{eq:omegakha}].
The associated intensity is shown by the colour scale inset in panel (ii)a.
All predictions have been convoluted with a gaussian to mimic experimental resolution.
The circuit $\Gamma$-{\sf X}-{\sf M}-$\Gamma$ in the bond-centred Brillouin zone is shown in Fig.~\ref{fig:bz}.
An animated version of this figure is shown in the supplemental material\cite{SM-spin1chiqomh}.
}}
\label{fig:bbq-chiqomh}
\end{figure*}
%


Performing linear flavour-wave theory results in a magnetic dispersion with two branches.
The first has the excitation spectrum,
 \begin{align}
\omega_{{\bf k},h}^{\sf a} =\sqrt{  A_{{\bf k},h}^2-B_{{\bf k},h}^2},
\label{eq:omegakha}
\end{align}
with,
 \begin{align}
&A_{{\bf k},h} = 4J_{11} \sin^2 2\theta_h
-4J_{11} (1-\gamma_{\bf k}^{(1)}) \cos^2 2\theta_h
 \nonumber \\
&  +4J_{12}(1-\gamma_{\bf k}^{(2)})
+4J_{22} \sin^2 2\theta_h
 -4J_{22} (1+\gamma_{\bf k}^{(2)})  \cos^2 2\theta_h \nonumber \\
& +2h \cos 2\theta_h \nonumber \\
& B_{{\bf k},h} = -4J_{11}\gamma_{\bf k}^{(1)} \sin^2 2\theta_h
+4J_{22}\gamma_{\bf k}^{(2)}  \sin^2 2\theta_h,
 \label{eq:flvwav-AkBk}
\end{align}
where $\gamma_{\bf k}^{(1)}$ and $\gamma_{\bf k}^{(2)}$ are defined in Eq.~\ref{eq:gamma}.
The second branch has,
\begin{align}
\omega_{{\bf k},h}^{\sf b}  = &
\frac{1}{2} \left( C_{{\bf k},h}-C_{{\bf k}+{\bf k}_{\sf M},h} \right) \nonumber \\
&+ \frac{1}{2}  \sqrt{(C_{{\bf k},h}+C_{{\bf k}+{\bf k}_{\sf M},h})^2-4D_{{\bf k},h}^2},
\label{eq:omegakhb}
\end{align}
where,
\begin{align}
C_{{\bf k},h} =& -4J_{11} \cos^2 2\theta_h 
+4J_{11} \gamma_{\bf k}^{(1)}  \cos 2\theta_h \nonumber \\
&+4J_{12}(1-\gamma_{\bf k}^{(2)})
 +4J_{22} \sin^2 2\theta_h
 +h \cos 2\theta_h \nonumber \\
 D_{{\bf k},h} =&  -4 J_{22}  \gamma_{\bf k}^{(2)} \sin 2\theta_h .
 \label{eq:flvwav-CkDk}
\end{align}
It follows that the imaginary part of the dynamic spin susceptibility is given by,
 \begin{align}
& \Im m   \chi^{\perp}({\bf q},\omega) = \Im m   \chi^{\sf xx}({\bf q},\omega)  +  \Im m  \chi^{\sf yy}({\bf q},\omega)  =  \nonumber \\
&\qquad  \pi
\left(  u_{{\bf q},h}^{\sf b} \cos \theta_h - v_{{\bf q},h}^{\sf b} \sin \theta_h  \right)^2 
\delta(\omega - \omega_{\bf q}^{\sf b}) \nonumber \\
&\qquad +   \pi
\left(  u_{{\bf q},h}^{\sf b} \sin \theta_h - v_{{\bf q},h}^{\sf b} \cos \theta_h  \right)^2
\delta(\omega - \omega_{{\bf q}+{\bf q}_{\sf M}}^{\sf b}) \nonumber \\
&\Im m  \chi^{\sf zz}({\bf q},\omega)  =  
  \pi \sin^2 2\theta_h \nonumber \\
&\qquad \qquad \qquad (u_{{\bf q}+{\bf q}_{\sf M},h}^{\sf a}+v_{{\bf q}+{\bf q}_{\sf M},h}^{\sf a})^2
\delta(\omega - \omega_{{\bf q}+{\bf q}_{\sf M}}^{\sf a}).
\label{eq:bbq-chiqomh}
\end{align}
where,
\begin{align}
(u_{{\bf k},h}^{\sf a})^2 &=
\frac{A_{{\bf k},h}}{2\sqrt{A_{{\bf k},h}^2-B_{{\bf k},h}^2}}
+\frac{1}{2} \nonumber \\
(v_{{\bf k},h}^{\sf a})^2 &=
\frac{A_{{\bf k},h}}{2\sqrt{A_{{\bf k},h}^2-B_{{\bf k},h}^2}}
-\frac{1}{2} \nonumber \\
2 u_{{\bf k},h}^{\sf a} v_{{\bf k},h}^{\sf b} &= 
\frac{-B_{{\bf k},h}}{\sqrt{A_{{\bf k},h}^2-B_{{\bf k},h}^2}}.
\end{align}
and,
\begin{align}
(u_{{\bf k},h}^{\sf b})^2 &=
\frac{C_{{\bf k},h}+C_{{\bf k}+{\bf k}_{\sf M},h}}{2\sqrt{(C_{{\bf k},h}+C_{{\bf k}+{\bf k}_{\sf M},h})^2-4D_{{\bf k},h}^2}}
+\frac{1}{2} \nonumber \\
(v_{{\bf k},h}^{\sf b})^2 &=
\frac{C_{{\bf k},h}+C_{{\bf k}+{\bf k}_{\sf M},h}}{2\sqrt{(C_{{\bf k},h}+C_{{\bf k}+{\bf k}_{\sf M},h})^2-4D_{{\bf k},h}^2}}
-\frac{1}{2} \nonumber \\
2 u_{{\bf k},h}^{\sf b} v_{{\bf k},h}^{\sf b} &= 
\frac{-2D_{{\bf k},h}}{\sqrt{(C_{{\bf k},h}+C_{{\bf k}+{\bf k}_{\sf M},h})^2-4D_{{\bf k},h}^2}}.
\end{align}
The imaginary part of the dynamical spin susceptibility [Eq.~\ref{eq:bbq-chiqomh}] is plotted in Fig.~\ref{fig:bbq-chiqomh}. 
An animated version of Fig.~\ref{fig:bbq-chiqomh} with continuously varying magnetic field is provided as supplementary material\cite{SM-spin1chiqomh}.


For \mbox{$h>h_{\sf sat}$} [Fig.~\ref{fig:bbq-chiqomh}(a)] the only visible feature is a uniformly bright band of $\Delta S^{\sf z}=1$ excitations in the transverse channel.
The only other excitations of the saturated paramagnet have $\Delta S^{\sf z}=2$ and therefore do not contribute to the spin susceptibility.

At $h=h_{\sf sat}$ [Fig.~\ref{fig:bbq-chiqomh}(b)] the gap to these $\Delta S^{\sf z}=2$ excitations closes, signalling the onset of spin-nematic order. 

For $0<h < h_{\sf sat}$ it is no longer possible to assign an integer value of $\Delta S^{\sf z}$ to a particular excitation. 
In consequence spin and quadrupolar fluctuations are mixed, except for at special points in the Brillouin zone.
Of particular interest is the nature of the low-energy modes, and we consider each of these separately.

The spin-nematic phase possesses a gapless Goldstone mode, which can be seen at ${\bf q}=(0,0)$ in the longitudinal spin susceptibility, $ \Im m   \chi^{\sf zz}({\bf q},\omega)$ [Eq.~\ref{eq:bbq-chiqomh}, Fig.~\ref{fig:bbq-chiqomh}(ii)(b-f)].
This is due to spontaneous breaking of the ${\sf U(1)}$ symmetry of $\mathcal{H}^{\sf S=1}_{\sf bbq}[{\bf S}]$~[Eq.~\ref{eq:H-BBQ}] in field by the spin-nematic state.
The Goldstone mode is associated with rotations of the quadrupoles in the plane perpendicular to the field, and shows up at ${\bf q}=(\pi,\pi)$ in the $\omega_{{\bf q},h}^{\sf a}$ [Eq.~\ref{eq:omegakha}] branch of the dispersion.
Close to ${\bf q}=(\pi,\pi)$ the predominantly quadrupolar fluctuation induces a small spin fluctuation.
These spin fluctuations are parallel to the ${\sf z}$ direction and in phase between the two sublattices, and hence are seen at ${\bf q}=(0,0)$ in the spin susceptibility.
Their magnitude goes to zero approaching ${\bf q}=(0,0)$, which can be seen from the fact that the spectral weight in the spin susceptibility disappears linearly with $|{\bf q}|$ approaching ${\bf q}=(0,0)$.

The longitudinal spin susceptibility, $ \Im m   \chi^{\sf zz}({\bf q},\omega)$ [Eq.~\ref{eq:bbq-chiqomh}, Fig.~\ref{fig:bbq-chiqomh}(ii)(b-f)], also has a gapped mode at ${\bf q}=(\pi,\pi)$.
This is brightest at $h=0$ and gradually fades as the field is increased towards $h=h_{\sf sat}$.
The mode can be thought of as a dynamical spin-density wave, in which spin fluctuations occur parallel to the field direction, but in antiphase between the two sublattices.
Intriguingly, a number of quasi 1-dimensional materials that are good candidates for realising the spin-nematic state show spin-density wave order over large field ranges\cite{svistov10}.
It would be interesting to see if at different values of the interaction parameters this mode condensed.

The transverse spin susceptibility, $ \Im m   \chi^{\perp}({\bf q},\omega)$ [Eq.~\ref{eq:bbq-chiqomh}, Fig.~\ref{fig:bbq-chiqomh}(i)(b-f)], has a single excitation branch at $h=h_{\sf sat}$.
As the field is reduced below $h=h_{\sf sat}$ it splits into two, corresponding to $\omega_{{\bf q},h}^{\sf b}$ and $\omega_{{\bf q}+{\bf q}_{\sf M},h}^{\sf b}$ [Eq.~\ref{eq:omegakhb}].
At a given ${\bf q}$ there is a bright feature at $\omega = \omega_{{\bf q},h}^{\sf b}$, which is predominantly due to a transverse fluctuation of the partially polarised moment.
There is also a weaker feature at $\omega = \omega_{{\bf q}+{\bf q}_{\sf M},h}^{\sf b}$, which is due to the quadrupolar part of the fluctation at ${\bf q}= {\bf q}+{\bf q}_{\sf M}$ inducing a spin fluctuation at ${\bf q}$.
Quadrupolar fluctuations at ${\bf q}$ also induce spin fluctuations at ${\bf q}$, but, unless $h=0$, these are hidden by the transverse fluctuation of the polarised moment.
At $h=0$ there is no polarised moment and all spin fluctuations are induced from quadrupole rotations.
At $h=0$ there are Goldstone mode excitations at ${\bf q}=(\pi,\pi)$ and ${\bf q}=(0,0)$ in the transverse channel, making a total of 3 Goldstone modes in the system.
The spin-nematic state spontaneously breaks the ${\sf SU(2)}$ symmetry of $\mathcal{H}^{\sf S=1}_{\sf bbq}[{\bf S}]$~[Eq.~\ref{eq:H-BBQ}], and the Goldstone modes correspond to 3-dimensional rotations of the quadrupole order parameter [see Appendix~\ref{App:FT-h=0}].


\section{Continuum approach to the spin-1, 2-sublattice spin nematic in applied magnetic field}
\label{sec:fieldtheory}


Here we develop a continuum approach to understanding the fluctuations of the spin-nematic state.
We derive a continuum model directly from the microscopic spin-1 model $\mathcal{H}^{\sf S=1}_{\sf bbq}[{\bf S}]$~[Eq.~\ref{eq:H-BBQ}]. 
After linearising the quantum fields, we show that this exactly reproduces the flavour-wave results presented in Section~\ref{sec:flvwave} at long wavelength.
The advantage of the continuum approach is that it is not tied to a microscopic model, but is instead a theory of the order parameter symmetry.
Thus it can be parametrised from any microscopic model that supports a 2-sublattice spin-nematic state.
In Section~\ref{sec:neutrons} we will parametrise it from $\mathcal{H}^{\sf S=1/2}_{\sf J_1-J_2}$ [Eq.~\ref{eq:HJ1J2}].


\subsection{Continuum theory}


A continuum field theory can be derived directly from $\mathcal{H}^{\sf S=1}_{\sf bbq}[{\bf e}]$ [Eq.~\ref{eq:Hbbq-e}].
The spin states on a site are represented using spin-coherent states, and the overcompleteness of the spin-coherent state basis results in a geometrical phase term in the action.
Assuming that the system at least locally realises a 2-sublattice, partially-polarised, spin-nematic state, $\mathcal{H}^{\sf S=1}_{\sf bbq}[{\bf e}]$ [Eq.~\ref{eq:Hbbq-e}] can be expanded in terms of derivatives of a set of continuum fields, which are defined at the centre of square plaquettes.
These fields are collected in the complex vectors ${\bf e}_{\sf A} ({\bf r},\tau)$ and ${\bf e}_{\sf B} ({\bf r},\tau)$, which are the continuum versions of ${\bf e}_j$ [see Eq.~\ref{eq:ebasis}] on the {\sf A} and {\sf B} sublattices.
The derivation follows in spirit Ref.~[\onlinecite{smerald13}], in which a continuum theory is formulated for a 3-sublattice spin-nematic state on the triangular lattice in the absence of magnetic field. 


The action for the 2-sublattice, partially-polarised spin nematic is given by,
\begin{align}
\mathcal{S}_{\sf 2SL} =  \frac{1}{2}\int dt d^2r \mathcal{L}_{\sf 2SL},
\label{eq:S2SL}
\end{align} 
where we set the lattice constant $b=1$ and,
\begin{align}
 \mathcal{L}_{\sf 2SL} =  \mathcal{L}_{\sf 2SL} ^{\sf kin} - \mathcal{L}_{\sf 2SL} ^{\mathcal{H}}.
 \label{eq:L2SL}
\end{align} 
The kinetic term is given by,
\begin{align}
\mathcal{L}_{\sf 2SL} ^{\sf kin} =
i\bar{{\bf e}}_{\sf A}.\partial_t {\bf e}_{\sf A} 
+i \bar{{\bf e}}_{\sf B}.\partial_t {\bf e}_{\sf B},
\end{align} 
and the Hamiltonian term can be written as,
\begin{align}
&\mathcal{L}_{\sf 2SL} ^{\mathcal{H}} = 
4J_{11} |{\bf e}_{\sf A} \cdot \bar{{\bf e}_{\sf B}}|^2
-2J_{22} |2e^{(1)}_{\sf A} e^{(\bar{1})}_{\sf A} - (e^{(0)}_{\sf A})^2|^2 \nonumber \\
& -2J_{22} |2e^{(1)}_{\sf B} e^{(\bar{1})}_{\sf B} - (e^{(0)}_{\sf B})^2|^2
-h\left( |e^{(1)}_{\sf A}|^2 - |e^{(\bar{1})}_{\sf A}|^2 \right) \nonumber \\
& -h\left( |e^{(1)}_{\sf B}|^2 - |e^{(\bar{1})}_{\sf B}|^2 \right) 
 + \sum_{\lambda={\sf x,y}} \left\{ \right. \nonumber \\
& -J_{11} \left[ ({\bf e}_{\sf A} \cdot \bar{{\bf e}_{\sf B}})( \partial_\lambda \bar{{\bf e}_{\sf A}} \cdot   \partial_\lambda {\bf e}_{\sf B})
 +({\bf e}_{\sf B} \cdot \bar{{\bf e}_{\sf A}})( \partial_\lambda \bar{{\bf e}_{\sf B}} \cdot   \partial_\lambda {\bf e}_{\sf A}) \right]  \nonumber \\
&  +2J_{12} \left[  (\partial_\lambda{\bf e}_{\sf A} \cdot \partial_\lambda\bar{{\bf e}_{\sf A}})
+ (\partial_\lambda{\bf e}_{\sf B} \cdot \partial_\lambda\bar{{\bf e}_{\sf B}}) \right] \nonumber \\
&   +J_{22} [2e^{(1)}_{\sf A} e^{(\bar{1})}_{\sf A} - (e^{(0)}_{\sf A})^2] [2 \partial_\lambda e^{(1)}_{\sf A} \partial_\lambda e^{(\bar{1})}_{\sf A} - (\partial_\lambda e^{(0)}_{\sf A})^2] \nonumber \\
&  +J_{22} [2\bar{e}^{(1)}_{\sf A} \bar{e}^{(\bar{1})}_{\sf A} - (\bar{e}^{(0)}_{\sf A})^2] [2 \partial_\lambda \bar{e}^{(1)}_{\sf A} \partial_\lambda \bar{e}^{(\bar{1})}_{\sf A} - (\partial_\lambda \bar{e}^{(0)}_{\sf A})^2] \nonumber \\
&    +J_{22} [2e^{(1)}_{\sf B} e^{(\bar{1})}_{\sf B} - (e^{(0)}_{\sf B})^2] [2 \partial_\lambda e^{(1)}_{\sf B} \partial_\lambda e^{(\bar{1})}_{\sf B} - (\partial_\lambda e^{(0)}_{\sf B})^2]  \nonumber \\
& \left.   +J_{22} [2\bar{e}^{(1)}_{\sf B} \bar{e}^{(\bar{1})}_{\sf B} - (\bar{e}^{(0)}_{\sf B})^2] [2 \partial_\lambda \bar{e}^{(1)}_{\sf B} \partial_\lambda \bar{e}^{(\bar{1})}_{\sf B} - (\partial_\lambda \bar{e}^{(0)}_{\sf B})^2] 
\right\}.
\end{align} 
It is understood that the constraints $|{\bf e}_{\sf A} |^2=1$ and $|{\bf e}_{\sf B} |^2=1$ have to be enforced.


When working at $h=0$, it is natural to divide the fluctuations into two sets, with one set describing Goldstone modes and other low-energy fluctuations, while a conjugate set describes high-energy fluctuations.
The action considerably simplifies if the high-energy fluctuations are integrated out\cite{smerald13} [see Appendix~\ref{App:FT-h=0}].
However, for fields close to $h=h_{\sf sat}$, it is no longer possible to partition the fluctuations in this way.
Conjugate pairs of fluctuations become degenerate at $h=h_{\sf sat}$, and therefore there is no low-energy fluctuation to integrate out.

\begin{widetext}


\subsection{Linearising the continuum theory}
\label{sec:linfieldtheory}


In order to calculate the dispersion and dynamical susceptibility within the continuum theory, it is first useful to linearise $ \mathcal{L}_{\sf 2SL}$ [Eq.~\ref{eq:L2SL}].
This can be accomplished by writing ${\bf e}_{\sf A} ({\bf r},\tau) $ and ${\bf e}_{\sf B} ({\bf r},\tau) $ in terms of 8 scalar fields,
\begin{align}
& {\bf e}_{\sf A} =
i {\bf R}(\theta_h) \cdot \left(
\begin{array}{c}
1 - \frac{1}{2} \left[ (\psi^{\sf b}_1+\psi^{\sf b}_3)^2+(\psi^{\sf a}_1+\psi^{\sf a}_3)^2+(\psi^{\sf a}_2+\psi^{\sf a}_4)^2+(\psi^{\sf b}_2+\psi^{\sf b}_4)^2 \right] \\
\psi^{\sf b}_1+i\psi^{\sf b}_2 +\psi^{\sf b}_3+i\psi^{\sf b}_4 \\
\psi^{\sf a}_1+i\psi^{\sf a}_2 +\psi^{\sf a}_3+i\psi^{\sf a}_4
\end{array}
\right) \nonumber \\
& {\bf e}_{\sf B} = 
i {\bf R}(-\theta_h) \cdot \left(
\begin{array}{c}
1 - \frac{1}{2} \left[ (\psi^{\sf b}_1-\psi^{\sf b}_3)^2+(\psi^{\sf a}_1-\psi^{\sf a}_3)^2+(\psi^{\sf a}_2-\psi^{\sf a}_4)^2+(\psi^{\sf b}_2-\psi^{\sf b}_4)^2 \right] \\
\psi^{\sf b}_1+i\psi^{\sf b}_2 -\psi^{\sf b}_3-i\psi^{\sf b}_4 \\
\psi^{\sf a}_1+i\psi^{\sf a}_2 -\psi^{\sf a}_3-i\psi^{\sf a}_4
\end{array}
\right).
\label{eq:efields-linearised}
\end{align}
The $\psi^{\sf a}$ fields are associated with the $a$ bosons in the flavour-wave theory and the $\psi^{\sf b}$ fields with the $b$ bosons (see Section~\ref{sec:flvwv-nem}).
%


The $\psi$ fields are substituted into $ \mathcal{L}_{\sf 2SL}$ [Eq.~\ref{eq:L2SL}] and terms up to quadratic order are retained.
After Fourier transform using,
\begin{align}
\psi({\bf r},t) =\frac{1}{(2\pi)^3}\int d\omega   d^2k  \ e^{i({\bf k}\cdot{\bf r}-\omega t)} \psi({\bf k},\omega),
\end{align} 
$\mathcal{S}_{\sf 2SL}$ [Eq.~\ref{eq:S2SL}] can be rewritten as,
\begin{align}
\mathcal{S}^{\sf lin}_{\sf 2SL} =
\frac{1}{2} \frac{1}{(2\pi)^3}\int d\omega   d^2k \
\mathcal{L}({\bf k},\omega),
\label{eq:S2SL-kom}
\end{align}
where,
\begin{align}
\mathcal{L}({\bf k},\omega) =
\sum_{i=a,b} \Psi^i({\bf k},\omega) \cdot
{\bf L}^i({\bf k},\omega) \cdot
\Psi^i(-{\bf k},-\omega).
\label{eq:lagkom}
\end{align}
Here,
\begin{align}
\Psi^i({\bf k},\omega) &=
\left( \psi_1^i({\bf k},\omega),\psi_2^i({\bf k},\omega),\psi_3^i({\bf k},\omega),\psi_4^i({\bf k},\omega)  \right),
\end{align}
\begin{align}
&{\bf L}^{\sf a}({\bf k},\omega) =
4 \left(
\begin{array}{cccc}
\tilde{A}_{{\bf k},h}+ \tilde{B}_{{\bf k},h} & -i \omega & 0 & 0  \\
i \omega & \tilde{A}_{{\bf k},h} -  \tilde{B}_{{\bf k},h} & 0 & 0  \\
0 & 0 & \tilde{A}_{{\bf k}+{\bf k}_{\sf M},h}+ \tilde{B}_{{\bf k}+{\bf k}_{\sf M},h} & -i \omega  \\
0 & 0 & i \omega  & \tilde{A}_{{\bf k}+{\bf k}_{\sf M},h} - \tilde{B}_{{\bf k}+{\bf k}_{\sf M},h}
\end{array}
\right)
\label{eq:La}
\end{align}
and,
\begin{align}
{\bf L}^{\sf b}({\bf k},\omega) &=
4\left(
\begin{array}{cccc}
\tilde{C}_{{\bf k},h}  & -i\omega & \tilde{D}_{{\bf k},h}& 0  \\
i\omega & \tilde{C}_{{\bf k},h}  & 0 & -\tilde{D}_{{\bf k},h}  \\
\tilde{D}_{{\bf k},h} & 0 & \tilde{C}_{{\bf k}+{\bf k}_{\sf M},h} & -i\omega  \\
0 & -\tilde{D}_{{\bf k},h} & i\omega  & \tilde{C}_{{\bf k}+{\bf k}_{\sf M},h}
\end{array}
\right),
\label{eq:Lb}
\end{align}
with,
\begin{align}
\tilde{A}_{{\bf k},h} &=  
4J_{11}(1+\cos^22\theta_h) -J_{11}\cos^22\theta_h \ k^2
 +2J_{12} \ k^2  + 4 J_{22} \sin^22\theta_h  + 2J_{22}\cos^22\theta_h \ k^2 \nonumber \\
\tilde{B}_{{\bf k},h} &= -4J_{11} \left(1-\frac{k^2}{4}\right)\sin^22\theta_h
+ 4J_{22} \left(1-\frac{k^2}{2}\right)\sin^22\theta_h \nonumber \\
\tilde{A}_{{\bf k}+{\bf k}_{\sf M},h} &=  
4J_{11}\sin^22\theta_h + J_{11}\cos^22\theta_h \ k^2
 +2J_{12} \ k^2   + 4 J_{22} \sin^22\theta_h  + 2J_{22}\cos^22\theta_h \ k^2 \nonumber \\
 \tilde{B}_{{\bf k}+{\bf k}_{\sf M},h} &= 4J_{11} \left(1-\frac{k^2}{4}\right)\sin^22\theta_h
+ 4J_{22} \left(1-\frac{k^2}{2}\right)\sin^22\theta_h \nonumber \\
\tilde{C}_{{\bf k},h} &=  4J_{11} \left(1-\frac{k^2}{4} \right) \cos 2\theta_h +2J_{12}k^2  + 4 J_{22} \nonumber \\
\tilde{C}_{{\bf k}+{\bf k}_{\sf M},h} &=  -4J_{11} \left(1-\frac{k^2}{4} \right) \cos 2\theta_h +2J_{12}k^2  + 4 J_{22} \nonumber \\
\tilde{D}_{{\bf k},h} &= -4 J_{22} \left(1-\frac{k^2}{2} \right)\sin 2\theta_h
\end{align}
These are just the small ${\bf k}$ expansions of Eq.~\ref{eq:flvwav-AkBk} and Eq.~\ref{eq:flvwav-CkDk}.


Diagonalisation of $\mathcal{S}^{\sf lin}_{\sf 2SL}$ [Eq.~\ref{eq:S2SL-kom}] is accomplished using the unitary matrices,
\begin{align}
&{\bf U}_{\sf a} = \left(
\begin{array}{cccc}
i\tilde{v}^{\sf a}({\bf k},\omega)  & i\tilde{u}^{\sf a}({\bf k},\omega) & 0  & 0  \\
-\tilde{u}^{\sf a}({\bf k},\omega)   & \tilde{v}^{\sf a}({\bf k},\omega)  & 0 & 0 \\
0  & 0  & i\tilde{v}^{\sf a}({\bf k}+{\bf k}_{\sf M},\omega)  & i\tilde{u}^{\sf a}({\bf k}+{\bf k}_{\sf M},\omega)  \\
0   & 0   & -\tilde{u}^{\sf a}({\bf k}+{\bf k}_{\sf M},\omega) & \tilde{v}^{\sf a}({\bf k}+{\bf k}_{\sf M},\omega)
\end{array}
\right),
\end{align}
and,
\begin{align}
&{\bf U}_{\sf b} =
\frac{1}{\sqrt{2}} \left(
\begin{array}{cccc}
i\tilde{v}^{\sf b}({\bf k},\omega)  & i\tilde{u}^{\sf b}({\bf k},\omega) & i\tilde{u}^{\sf b}({\bf k}+{\bf k}_{\sf M},\omega)  & i\tilde{v}^{\sf b}({\bf k}+{\bf k}_{\sf M},\omega)  \\
-\tilde{v}^{\sf b}({\bf k},\omega)   & -\tilde{u}^{\sf b}({\bf k},\omega)  & \tilde{u}^{\sf b}({\bf k}+{\bf k}_{\sf M},\omega) & \tilde{v}^{\sf b}({\bf k}+{\bf k}_{\sf M},\omega)  \\
i\tilde{u}^{\sf b}({\bf k},\omega)  & -i\tilde{v}^{\sf b}({\bf k},\omega)  & i\tilde{v}^{\sf b}({\bf k}+{\bf k}_{\sf M},\omega)  & -i\tilde{u}^{\sf b}({\bf k}+{\bf k}_{\sf M},\omega)  \\
\tilde{u}^{\sf b}({\bf k},\omega)   & -\tilde{v}^{\sf b}({\bf k},\omega)   & -\tilde{v}^{\sf b}({\bf k}+{\bf k}_{\sf M},\omega) & \tilde{u}^{\sf b}({\bf k}+{\bf k}_{\sf M},\omega)
\end{array}
\right),
\end{align}
with,
\begin{align}
\tilde{u}^{\sf a}({\bf k},\omega) &= 
\frac{1}{\sqrt{2}} 
\sqrt{1+\frac{\tilde{B}_{{\bf k},h}}
{\sqrt{\omega^2 +\tilde{B}_{{\bf k},h}^2} }}, \quad
\tilde{v}^{\sf a}({\bf k},\omega) = 
\frac{1}{\sqrt{2}} 
\sqrt{1-\frac{\tilde{B}_{{\bf k},h}}
{\sqrt{\omega^2 +\tilde{B}_{{\bf k},h}^2} }},
\end{align}
and,
\begin{align}
\tilde{u}^{\sf b}({\bf k},\omega) &= 
\frac{1}{\sqrt{2}} 
\sqrt{1+\frac{\frac{\tilde{C}_{{\bf k},h}-\tilde{C}_{{\bf k}+{\bf k}_{\sf M},h}}{2}+\omega}
{\sqrt{\left(\frac{\tilde{C}_{{\bf k},h}-\tilde{C}_{{\bf k}+{\bf k}_{\sf M},h}}{2}+\omega \right)^2 +\tilde{D}_{{\bf k},h}^2} }},
\quad
\tilde{v}^{\sf b}({\bf k},\omega) = 
\frac{1}{\sqrt{2}} 
\sqrt{1- \frac{\frac{\tilde{C}_{{\bf k},h}-\tilde{C}_{{\bf k}+{\bf k}_{\sf M},h}}{2}+\omega}
{\sqrt{\left(\frac{\tilde{C}_{{\bf k},h}-\tilde{C}_{{\bf k}+{\bf k}_{\sf M},h}}{2}+\omega \right)^2 +\tilde{D}_{{\bf k},h}^2} }}.
\end{align}
In consequence one finds,
\begin{align}
\mathcal{L}({\bf k},\omega) =
\sum_{i=a,b} \tilde{\Psi}^i({\bf k},\omega) \cdot
[{\bf G}^i({\bf k},\omega)]^{-1} \cdot 
\tilde{\Psi}^i(-{\bf k},-\omega),
\end{align}
with,
\begin{align}
\tilde{\Psi}^i(-{\bf k},-\omega) = {\bf U}_i\dg \cdot \Psi^i(-{\bf k},-\omega)
\label{eq:psi-psibara}
\end{align}
Here the diagonal Green's function matrices are given by,
\begin{align}
&[{\bf G}^{\sf a}({\bf k},\omega)]^{-1} = 
4{\bf U}_{\sf a}\dg \cdot {\bf L}^{\sf a}({\bf k},\omega) \cdot {\bf U}_{\sf a}
= 4 \left(
\begin{array}{cccc}
[G^{{\sf a}-}({\bf k},\omega)]^{-1}  & 0 & 0 & 0  \\
0   & [G^{{\sf a}+}({\bf k},\omega)]^{-1}  & 0 & 0  \\
0  & 0 & [G^{{\sf a}-}({\bf k}+{\bf k}_{\sf M},\omega)]^{-1} & 0 \\
0   & 0  & 0 & [G^{{\sf a}+}({\bf k}+{\bf k}_{\sf M},\omega)]^{-1} 
\end{array}
\right)
\end{align}
with,
\begin{align}
[G^{{\sf a}\pm}({\bf k},\omega)]^{-1} =
\tilde{A}_{{\bf k},h}
 \pm \sqrt{ \tilde{B}_{{\bf k},h}^2 +\omega^2 },
  \label{eq:Gfunca}
\end{align}
and,
\begin{align}
[{\bf G}^{\sf b}({\bf k},\omega)]^{-1} &= 
4{\bf U}_{\sf b}\dg \cdot {\bf L}^{\sf b}({\bf k},\omega) \cdot {\bf U}_{\sf b} = 
4\left(
\begin{array}{cccc}
[G^{{\sf b}-}({\bf k}+{\bf k}_{\sf M},\omega)]^{-1}  & 0 & 0 & 0  \\
0   & [G^{{\sf b}+}({\bf k}+{\bf k}_{\sf M},\omega)]^{-1}  & 0 & 0  \\
0  & 0 & [G^{{\sf b}-}({\bf k},\omega)]^{-1} & 0 \\
0   & 0  & 0 & [G^{{\sf b}+}({\bf k},\omega)]^{-1} 
\end{array}
\right),
\end{align}
with,
\begin{align}
[G^{{\sf b}\pm}({\bf k},\omega)]^{-1} =
 \frac{1}{2}(\tilde{C}_{{\bf k},h}+\tilde{C}_{{\bf k}+{\bf k}_{\sf M},h})
 \pm  \frac{1}{2}\sqrt{(\tilde{C}_{{\bf k},h}-\tilde{C}_{{\bf k}+{\bf k}_{\sf M},h})^2 - 4\omega(\tilde{C}_{{\bf k},h}-\tilde{C}_{{\bf k}+{\bf k}_{\sf M},h}) +4 \tilde{D}_{{\bf k},h}^2 +4\omega^2 }.
 \label{eq:Gfuncb}
\end{align}

\end{widetext}


It is useful to calculate the imaginary part of the Green's functions given in Eq.~\ref{eq:Gfunca} and Eq.~\ref{eq:Gfuncb}.
One finds,
\begin{align}
\Im m G^{{\sf a}+}({\bf k},\omega) &= 0 \nonumber \\
\Im m G^{{\sf a}-}({\bf k},\omega) &=  \frac{1}{4} g^{\sf a}_{{\bf k},h} \ \delta(\omega -  \tilde{\omega}^{\sf a}_{{\bf k},h}),
\end{align}
where,
\begin{align}
g^{\sf a}_{{\bf k},h} = \frac{ \tilde{A}_{{\bf k},h} }{\sqrt{\tilde{A}_{{\bf k},h}^2 -  \tilde{B}_{{\bf k},h}^2} } ,
\end{align}
 only positive frequency contributions have been retained, and,
\begin{align}
 \tilde{\omega}^{\sf a}_{{\bf k},h} = \sqrt{\tilde{A}_{{\bf k},h}^2 -  \tilde{B}_{{\bf k},h}^2}.
 \label{eq:omtildekha}
\end{align}
Similarly,
\begin{align}
\Im m G^{{\sf b}+}({\bf k},\omega) &= 0 \nonumber \\
\Im m G^{{\sf b}-}({\bf k},\omega) &=   \frac{1}{4} g^{\sf b}_{{\bf k},h} \   \delta(\omega -  \tilde{\omega}^{\sf b}_{{\bf k},h}),
\end{align}
where,
\begin{align}
g^{\sf b}_{{\bf k},h}  = \frac{ \tilde{C}_{{\bf k},h}+\tilde{C}_{{\bf k}+{\bf k}_{\sf M},h} }{\sqrt{(\tilde{C}_{{\bf k},h}+\tilde{C}_{{\bf k}+{\bf k}_{\sf M},h})^2 - 4 \tilde{D}_{{\bf k},h}^2} } ,
\end{align}
and,
\begin{align}
 \tilde{\omega}^{\sf b}_{{\bf k},h} &= \frac{1}{2}(\tilde{C}_{{\bf k},h}-\tilde{C}_{{\bf k}+{\bf k}_{\sf M},h}) \nonumber \\
& +\frac{1}{2}\sqrt{(\tilde{C}_{{\bf k},h}+\tilde{C}_{{\bf k}+{\bf k}_{\sf M},h})^2 - 4 \tilde{D}_{{\bf k},h}^2}.
  \label{eq:omtildekhb}
\end{align}
It is clear that $ \tilde{\omega}^{\sf a}_{{\bf k},h}$ [Eq.~\ref{eq:omtildekha}] and $ \tilde{\omega}^{\sf b}_{{\bf k},h}$ [Eq.~\ref{eq:omtildekhb}] are equivalent at long wavelength to the flavour-wave dispersion relations $ \omega^{\sf a}_{{\bf k},h}$ [Eq.~\ref{eq:omegakha}] and $ \omega^{\sf b}_{{\bf k},h}$ [Eq.~\ref{eq:omegakhb}].    
%


\subsection{Dynamical spin susceptibility}


%
\begin{figure*}[ht]
\includegraphics[width=0.98\textwidth]{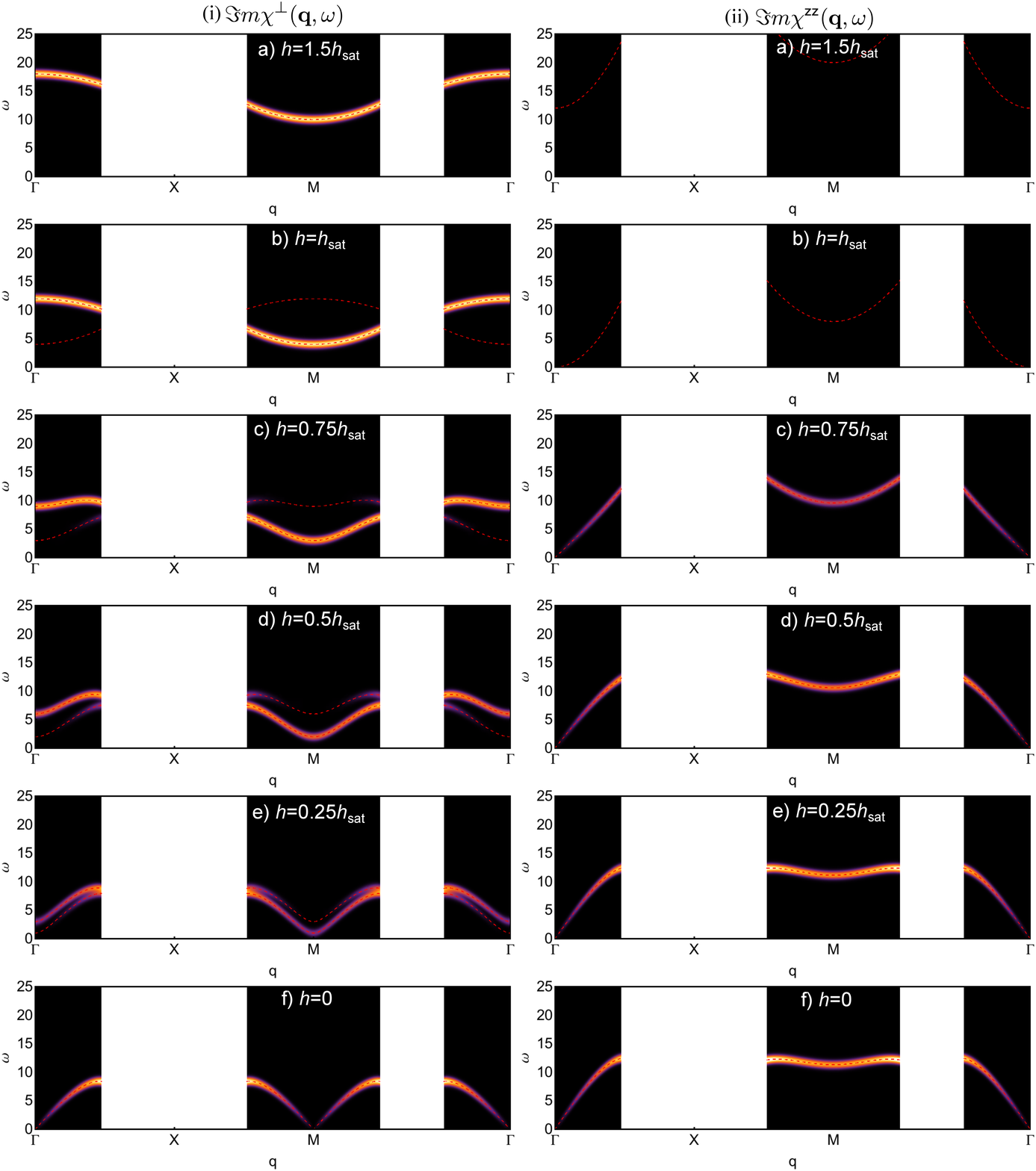}
\caption{\footnotesize{(Color online). 
Continuum theory predictions for the imaginary part of the dynamic spin susceptibility of a spin-1, partially-polarised, 2-sublattice, spin-nematic state in applied magnetic field.
The action $\mathcal{S}^{\sf lin}_{\sf 2SL}$ [Eq.~\ref{eq:S2SL-kom}] is parametrised using $J_{11}=1$, $J_{12}=0.1$ and $J_{22}=2$.
The external magnetic field is gradually reduced from a) $h=1.5h_{\sf sat}$ to f) $h=0$.
(i) The transverse susceptibility $\Im m   \chi^{\perp}({\bf q},\omega) $ [Eq.~\ref{eq:FT-chiomh}].
Dashed red lines show $\tilde{\omega}_{{\bf q},h}^{\sf b}$ at all $h$ and $\tilde{\omega}_{{\bf q}+{\bf q}_{\sf M},h}^{\sf b}$ for $h\leq h_{\sf sat}$ [see Eq.~\ref{eq:omtildekhb}].
(ii) The longitudinal susceptibility $\Im m   \chi^{\sf zz}({\bf q},\omega) $ [Eq.~\ref{eq:FT-chiomh}].
Dashed red lines show $\tilde{\omega}_{{\bf q}+{\bf q}_{\sf M},h}^{\sf a}$ [see Eq.~\ref{eq:omtildekha}].
All predictions have been convoluted with a gaussian to mimic experimental resolution.
The circuit $\Gamma$-{\sf X}-{\sf M}-$\Gamma$ in the bond-centred Brillouin zone is shown in Fig.~\ref{fig:bz}.
The same linear, normalised colour intensity scale is used as in Fig.~\ref{fig:bbq-chiqomh}.
}}
\label{fig:FTspin1-chiqomh}
\end{figure*}

The imaginary part of the dynamical spin susceptibility can be calculated within the linearised theory.
Using Eq.~\ref{eq:SQ-e} and Eq.~\ref{eq:efields-linearised} the spin fields can be written as,
\begin{align}
S^{\sf x}_{\sf A}({\bf r},t) &= -\sqrt{2}(\cos\theta_h - \sin \theta_h) (\psi^{\sf b}_1 + \psi^{\sf b}_3) \nonumber \\
S^{\sf y}_{\sf A}({\bf r},t) &= \sqrt{2}(\cos\theta_h + \sin \theta_h) (\psi^{\sf b}_2 + \psi^{\sf b}_4) \nonumber \\
S^{\sf z}_{\sf A}({\bf r},t) &= \cos 2\theta_h +2 \sin 2\theta_h (\psi^{\sf a}_1 + \psi^{\sf a}_3) \nonumber \\
S^{\sf x}_{\sf B}({\bf r},t) &= -\sqrt{2}(\cos\theta_h + \sin \theta_h) (\psi^{\sf b}_1 - \psi^{\sf b}_3) \nonumber \\
S^{\sf y}_{\sf B}({\bf r},t) &= \sqrt{2}(\cos\theta_h - \sin \theta_h) (\psi^{\sf b}_2 - \psi^{\sf b}_4) \nonumber \\
S^{\sf z}_{\sf B}({\bf r},t) &= \cos 2\theta_h -2 \sin 2\theta_h (\psi^{\sf a}_1 - \psi^{\sf a}_3).
\label{eq:spinfields}
\end{align}
Using the results of Section~\ref{sec:linfieldtheory}, as well as the definition of the spin susceptibility given in Eq.~\ref{eq:im-chi}, one finds,
\begin{align}
&\Im m \chi^{\perp}({\bf q},\omega) = \Im m \chi^{\sf xx}({\bf q},\omega) + \Im m \chi^{\sf yy}({\bf q},\omega) = \nonumber \\
& \ \pi
\left[( \tilde{u}^{\sf b}({\bf q}+{\bf q}_{\sf M},\omega) \cos\theta_h -  \tilde{v}^{\sf b}({\bf q}+{\bf q}_{\sf M},\omega)\sin\theta_h )^2  \right] \nonumber \\
& \qquad \times g^{\sf b}_{\bf q} \ \delta(\omega - \tilde{\omega}^{\sf b}_{\bf q}) \nonumber \\
 & \ +\pi
\left[( \tilde{v}^{\sf b}({\bf q},\omega)\cos\theta_h - \tilde{u}^{\sf b}({\bf q},\omega)\sin\theta_h ) ^2 \right] \nonumber \\
&\qquad \times g^{\sf b}_{\bf q} \ \delta(\omega - \tilde{\omega}^{\sf b}_{{\bf q}+{\bf q}_{\sf M}}) \nonumber \\
& \Im m \chi^{\sf zz}({\bf q},\omega) =
 \pi \sin^2 2\theta_h \ [\tilde{u}^{\sf a}({\bf q}+{\bf q}_{\sf M},\omega)+\tilde{v}^{\sf a}({\bf q}+{\bf q}_{\sf M},\omega)]^2  \nonumber \\
  & \qquad \qquad \qquad \times g^{\sf a}_{{\bf q}+{\bf q}_{\sf M}} \ \delta(\omega - \tilde{\omega}^{\sf a}_{{\bf q}+{\bf q}_{\sf M}}).
  \label{eq:FT-chiomh}
\end{align}
where either ${\bf q} \approx 0$ or ${\bf q} \approx {\bf q}_{\sf M}$.
The summed dynamical spin susceptibility is plotted in Fig.~\ref{fig:FTspin1-chiqomh}.
It can be seen that, at long wavelength, this is exactly equivalent to Fig.~\ref{fig:bbq-chiqomh}, which depicts the flavour-wave prediction for the imaginary part of the dynamical spin susceptibility [Eq.~\ref{eq:bbq-chiqomh}].
Alternatively, this equivalence is clear from directly comparing Eq.~\ref{eq:bbq-chiqomh} and Eq.~\ref{eq:FT-chiomh}.
%


\subsection{Hydrodynamic parametrisation}


We now parametrise the linearised field theory $\mathcal{S}^{\sf lin}_{\sf 2SL}$ [Eq.~\ref{eq:S2SL-kom}]  in terms of hydrodynamic parameters.
This frees the theory from any particular microscopic model, and thus allows it to be applied to any partially-polarised, 2-sublattice AFQ state.
We concentrate in particular on the case $h\approx h_{\sf sat}$.

The action $\mathcal{S}^{\sf lin}_{\sf 2SL}$ [Eq.~\ref{eq:S2SL-kom}] contains all symmetry allowed terms for a 2-sublattice, partially-polarised AFQ state on the square lattice at a linear level.
This is not the case for the non-linear action $\mathcal{S}_{\sf 2SL}$ [Eq.~\ref{eq:S2SL}], which describes the long-wavelength fluctuations of the spin-nematic state found in $\mathcal{H}^{\sf S=1}_{\sf bbq}[{\bf S}] $ [Eq.~\ref{eq:H-BBQ}].
For a general 2-sublattice spin-nematic state, it may be necessary to include other relevant terms in the non-linear action.
However, the only effect these will have on the linear theory is to change the hydrodynamic parameters.

In consequence it is possible to write a general, hydrodynamically-parametrised, linearised Lagrangian for a 2-sublattice, partially-polarised AFQ state,
\begin{align}
\mathcal{L}_{\sf hyd}({\bf k},\omega) &=
\frac{2}{\chi_h^{\sf Q,z} \omega_{{\bf k},h}^{\sf Q,z} (\omega - \omega_{{\bf k},h}^{\sf Q,z}) } | \tilde{\Psi}^{\sf Q,z}({\bf k},\omega)|^2 \nonumber \\
&+\frac{2}{\chi_h^{\sf S,z} \omega_{{\bf k},h}^{\sf S,z} (\omega - \omega_{{\bf k},h}^{\sf S,z}) } | \tilde{\Psi}^{\sf S,z}({\bf k},\omega)|^2  \nonumber \\
&+ \frac{1}{ \omega - \omega_{{\bf k},h}^{\sf xy,\pi} } | \tilde{\Psi}^{\sf xy,\pi}({\bf k},\omega)|^2 \nonumber \\
&+ \frac{1}{ \omega - \omega_{{\bf k},h}^{\sf xy,0} } | \tilde{\Psi}^{\sf xy,0}({\bf k},\omega)|^2.
\label{eq:Lhydro}
\end{align}
This describes 4 modes, and these have all been mapped onto ${\bf k} \approx 0$.
The dispersion relations of the 4 modes are given by,
\begin{align}
\omega_{{\bf k},h}^{\sf Q,z} &= \sqrt{ (v_h^{\sf Q,z})^2  \ {\bf k}^2 +\sigma_{h_{\sf sat}}^2 {\bf k}^4  } \nonumber \\
\omega_{{\bf k},h}^{\sf S,z} &= \sqrt{ (\Delta_{h_{\sf sat}}^{\sf S,z})^2 + (v_{h_{\sf sat}}^{\sf S,z})^2 {\bf k}^2 } \nonumber \\
 \omega_{{\bf k},h}^{\sf xy,\pi} &= \sqrt{ (\Delta_{h}^{\sf xy,\pi})^2 + (v_{h_{\sf sat}}^{\sf xy,\pi})^2 {\bf k}^2 } \nonumber \\
 \omega_{{\bf k},h}^{\sf xy,0} &= \sqrt{ (\Delta_{h}^{\sf xy,0})^2 + (v_{h_{\sf sat}}^{\sf xy,0})^2 {\bf k}^2 }.
 \label{eq:omhydro}
\end{align}
For the hydrodynamic parameters that depend strongly on field we write,
\begin{align}
(v_h^{\sf Q,z})^2 &= (1-h/h_{\sf sat}) (v^{\sf Q,z})^2 \nonumber \\
\Delta_{h}^{\sf xy,\pi} &= \Delta_{h_{\sf sat}}^{\sf xy,\pi} \frac{h}{h_{\sf sat}} \nonumber \\
\Delta_{h}^{\sf xy,0} &= h ,
\label{eq:hydrohdep}
\end{align}
while the others are only expected to change weakly with varying magnetic field close to $h=h_{\sf sat}$.

The hydrodynamic parameters can be taken from any microscopic model supporting a partially polarised, 2-sublattice AFQ state.
Table~\ref{tab:Jtohydroconversion} shows the parametrisation from $\mathcal{H}^{\sf S=1}_{\sf bbq}[{\bf S}] $ [Eq.~\ref{eq:H-BBQ}].

The $\omega_{{\bf k},h}^{\sf Q,z}$ [Eq.~\ref{eq:omhydro}] mode is gapless and associated with breaking {\sf U(1)} symmetry.
This is the Goldstone mode and describes rotations of the quadrupolar order parameter in the plane perpendicular to the applied field.
For $h < h_{\sf sat}$ the mode has a linear dispersion in the vicinity of  ${\bf k} = 0$.
As $h\to h_{\sf sat}$ from below, $(v_h^{\sf Q,z})^2 \to 0$ and therefore at $h = h_{\sf sat}$ the mode has a quadratic dispersion, as expected for a saturated paramagnet.

The $\omega_{{\bf k},h}^{\sf S,z}$ [Eq.~\ref{eq:omhydro}] mode is associated with spin fluctuations parallel to the magnetic field. 
These can be thought of as a dynamic spin-density wave.
The mode is gapped and both the gap, $\Delta_{h}^{\sf S,z} \approx \Delta_{h_{\sf sat}}^{\sf S,z}$, and the velocity, $v_{h}^{\sf S,z} \approx v_{h_{\sf sat}}^{\sf S,z}$, are only weakly field dependent close to $h=h_{\sf sat}$, and thus we approximate them with their values at $h=h_{\sf sat}$.

The $ \omega_{{\bf k},h}^{\sf xy,\pi}$ [Eq.~\ref{eq:omhydro}] mode is associated with spin fluctuations transverse to the field direction and antiparallel on the two sublattices.
The gap, $\Delta_{h}^{\sf xy,\pi}$, is linearly dependent on the field [see Eq.~\ref{eq:hydrohdep}], but the velocity $v_{h}^{\sf xy,\pi} \approx v_{h_{\sf sat}}^{\sf xy,\pi}$ is approximately field independent.

The $ \omega_{{\bf k},h}^{\sf xy,0}$ [Eq.~\ref{eq:omhydro}] mode is associated with spin fluctuations transverse to the field direction and parallel on the two sublattices.
The gap, $\Delta_{h}^{\sf xy,0}$, is linearly dependent on the field [see Eq.~\ref{eq:hydrohdep}], but the velocity $v_{h}^{\sf xy,0} \approx v_{h_{\sf sat}}^{\sf xy,0}$ is approximately field independent.

\begin{table}
\begin{center}
\footnotesize
  \begin{tabular}{| c | c |}
    \hline
Hydro-   & Spin-1 model   \\ 
dynamic  & $\mathcal{H}^{\sf S=1}_{\sf bbq}[{\bf S}]$~[Eq.~\ref{eq:H-BBQ}]  \\ \hline 
\multirow{2}{*}{$(\Delta_h^{\sf S,z})^2$} & \multirow{2}{*}{$64 J_{11}(J_{22} \sin^2 2\theta_h + J_{11} \cos^2 2\theta_h)$} \\
&\\ \hline
\multirow{2}{*}{$(\Delta_h^{\sf xy,\pi})^2$} & \multirow{2}{*}{$16 (J_{11}-J_{22})^2 \cos^2 2\theta_h$} \\
&\\ \hline
\multirow{2}{*}{$(\Delta_h^{\sf xy,0})^2$} & \multirow{2}{*}{$16 (J_{11}+J_{22})^2 \cos^2 2\theta_h$} \\
&\\ \hline
\multirow{2}{*}{$(v_h^{\sf Q,z})^2$} & \multirow{2}{*}{$8(J_{11}+J_{22}) (J_{11}+2J_{12}+2J_{22}) \sin^2 2\theta_h$} \\
&\\ \hline
\multirow{2}{*}{$(v_h^{\sf S,z})^2$} & $16 J_{12}(J_{11}+J_{22})+8(J_{11}-J_{22})(J_{11}-2J_{22}) $ \\
& $+ 8\cos^2 2\theta_h [2J_{12}(J_{11}-J_{22})-(3J_{11}-J_{22})(J_{11}-2J_{22})]$  \\ \hline
\multirow{2}{*}{$(v_h^{\sf xy,\pi})^2$} & $16(J_{22}-J_{11})(J_{12}+J_{22})\sin^2 2\theta_h $ \\
& $+8 J_{11}(J_{22}-J_{11})\cos^2 2\theta_h $  \\ \hline
\multirow{2}{*}{$(v_h^{\sf xy,0})^2$} & $16(J_{11}+J_{22})(J_{12}+J_{22})\sin^2 2\theta_h $ \\
& $-8 J_{11}(J_{11}+J_{22})\cos^2 2\theta_h $  \\ \hline
\multirow{2}{*}{$\sigma_h$} & $ (J_{11} + 2 (J_{12} + J_{22}))/3  $ \\
& $\times (-J_{11} + 6 J_{12} - 
   J_{22} + (4 J_{11} + 7 J_{22}) \cos 4\theta_h) $  \\ \hline
\multirow{2}{*}{$\chi_h^{\sf Q,z}$} & \multirow{2}{*}{$[4(J_{11}+J_{22})]^{-1}$} \\
&\\ \hline
\multirow{2}{*}{$\chi_h^{\sf S,z}$} & \multirow{2}{*}{$[4J_{11}]^{-1}$} \\
&\\ \hline
\multirow{2}{*}{$\rho^{\sf Q,z}_h$} & \multirow{2}{*}{$2(J_{11}+2J_{12}+2J_{22})\sin^2 2\theta_h$} \\
&\\ \hline
    \end{tabular}
\end{center} 
\caption{\footnotesize{
Relationship between the hydrodynamic parameters appearing in the continuum field theory, $\mathcal{L}_{\sf hyd}({\bf k},\omega)$ [Eq.~\ref{eq:Lhydro}], and the parameters of the microscopic model $\mathcal{H}^{\sf S=1}_{\sf bbq}[{\bf S}] $ [Eq.~\ref{eq:H-BBQ}].
}}
\label{tab:Jtohydroconversion}
\end{table}

In order to calculate the imaginary part of the dynamic spin susceptibility, it is first necessary to determine expressions for the spin moments.
To do this we use $\mathcal{H}^{\sf S=1}_{\sf bbq}[{\bf S}] $ [Eq.~\ref{eq:H-BBQ}] as a guide, and re-express the spin fields appearing in Eq.~\ref{eq:spinfields} in terms of the fields and hydrodynamic parameters appearing in $\mathcal{L}_{\sf hyd}({\bf k},\omega)$ [Eq.~\ref{eq:Lhydro}], making use of the results presented in Section.~\ref{sec:linfieldtheory}. 
It follows that,
\begin{align}
&\delta S^{\sf x}({\bf q},\omega) = 
-\sin \theta_h \ \tilde{\Psi}^{\sf xy,\pi}(-{\bf q},-\omega) 
+\cos \theta_h \ \tilde{\Psi}^{\sf xy,0}(-{\bf q},-\omega)  \nonumber \\
&\delta S^{\sf x}({\bf q}_{\sf M}+{\bf q},\omega) = 
\cos \theta_h \ \tilde{\Psi}^{\sf xy,\pi}(-{\bf q},-\omega) \nonumber \\
&\qquad \qquad \qquad \qquad -\sin \theta_h \ \tilde{\Psi}^{\sf xy,0}(-{\bf q},-\omega)  \nonumber \\
&\delta S^{\sf z}({\bf q},\omega) = 
- \frac{1}{2}  \sqrt{\rho^{\sf Q,z}_h {\bf q}^2 \chi_h^{\sf Q,z} }
\ \tilde{\Psi}^{\sf Q,z}(-{\bf q},-\omega) \nonumber \\
&\delta S^{\sf z}({\bf q}_{\sf M}+{\bf q},\omega) = 
- \sin 2\theta_h
\ \tilde{\Psi}^{\sf S,z}(-{\bf q},-\omega),
\end{align}
where ${\bf q} \approx 0$.
The imaginary part of the dynamical spin susceptibility is thus given by,
\begin{align}
&\Im m \chi_{\sf bc}^{\perp}({\bf q},\omega) = \Im m \chi_{\sf bc}^{\sf xx}({\bf q},\omega) + \Im m \chi_{\sf bc}^{\sf yy}({\bf q},\omega) = \nonumber \\
& \qquad  \pi \left[ \cos^2 \theta_h \ \delta(\omega -  \omega_{{\bf q},h}^{\sf xy,0}) +  \sin^2 \theta_h \ \delta(\omega - \omega_{{\bf q},h}^{\sf xy,\pi}) \right] \nonumber \\
&\Im m \chi_{\sf bc}^{\perp}({\bf q}_{\sf M}+{\bf q},\omega) = \nonumber \\
&\qquad \Im m \chi_{\sf bc}^{\sf xx}({\bf q}_{\sf M}+{\bf q},\omega) + \Im m \chi_{\sf bc}^{\sf yy}({\bf q}_{\sf M}+{\bf q},\omega) = \nonumber \\
& \qquad  \pi \left[ \cos^2 \theta_h \ \delta(\omega -  \omega_{{\bf q},h}^{\sf xy,\pi}) +  \sin^2 \theta_h \ \delta(\omega - \omega_{{\bf q},h}^{\sf xy,0}) \right] \nonumber \\
&\Im m \chi_{\sf bc}^{\sf zz}({\bf q},\omega) = \frac{\pi}{2} \frac{\rho^{\sf Q,z}_h {\bf q}^2}{ \omega_{{\bf q},h}^{\sf Q,z}}
\delta(\omega - \omega_{{\bf q},h}^{\sf Q,z}) \nonumber \\
&\Im m \chi_{\sf bc}^{\sf zz}({\bf q}_{\sf M}+{\bf q},\omega) = 2\pi \sin^2 2\theta_h \frac{1}{\chi_h^{\sf S,z}  \omega_{{\bf q},h}^{\sf S,z}}
\delta(\omega - \omega_{{\bf q},h}^{\sf S,z}),
\label{eq:chihydro-spin1}
\end{align}
where ${\bf q} \approx 0$.
 We have anticipated the mapping made in Section~\ref{sec:spinmapping} and introduced the subscript {\sf bc} which stands for bond-centred.


\section{Mapping from spin-1 site-based nematic to spin-1/2 bond nematic}
\label{sec:spinmapping}


%
\begin{figure*}[ht]
\includegraphics[width=0.98\textwidth]{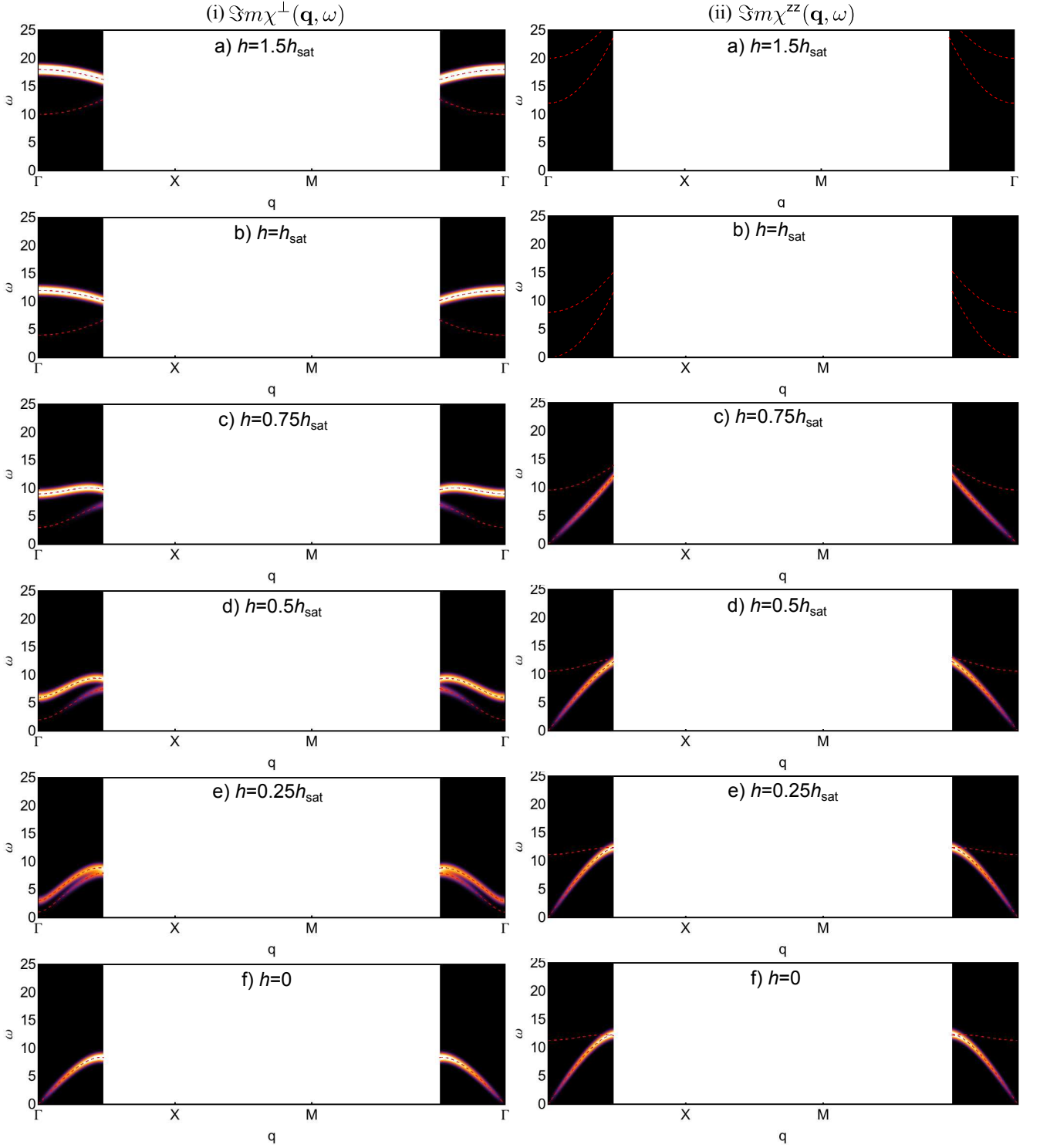}
\caption{\footnotesize{(Color online). 
Predictions for the dynamic spin susceptibility of a spin-1, partially-polarised, 2-sublattice, spin-nematic state in applied magnetic field (Fig.~\ref{fig:FTspin1-chiqomh}) mapped onto the site-centred lattice
The relationship between the bond- and site-centred lattices is shown in Fig.~\ref{fig:bz}, and the mapping of the dynamical susceptibility is performed using Eq.~\ref{eq:chi-sitecentred}.
The external magnetic field is gradually reduced from a) $h=1.5h_{\sf sat}$ to f) $h=0$.
(i) The transverse susceptibility $\Im m   \chi^{\perp}_{\sf sc}({\bf q},\omega) $ [Eq.~\ref{eq:chi-sitecentred}].
Dashed red lines show $\tilde{\omega}_{{\bf q},h}^{\sf b}$ and $\tilde{\omega}_{{\bf q}+{\bf q}_{\sf M},h}^{\sf b}$ [see Eq.~\ref{eq:omtildekhb}].
(ii) The longitudinal susceptibility $\Im m   \chi^{\sf zz}({\bf q},\omega) $ [Eq.~\ref{eq:chi-sitecentred}].
Dashed red lines show $\tilde{\omega}_{{\bf q},h}^{\sf a}$ and $\tilde{\omega}_{{\bf q}+{\bf q}_{\sf M},h}^{\sf a}$ [see Eq.~\ref{eq:omtildekha}].
All predictions have been convoluted with a gaussian to mimic experimental resolution.
The circuit $\Gamma$-{\sf X}-{\sf M}-$\Gamma$ in the site-centred Brillouin zone is shown in Fig.~\ref{fig:bz}.
The same linear, normalised colour intensity scale is used as in Fig.~\ref{fig:bbq-chiqomh}.
}}
\label{fig:FTspinhalf-chiqomh}
\end{figure*}

We now show how the imaginary part of the long-wavelength, dynamical spin susceptibility can be calculated in a bond-nematic phase, focusing on the 2-sublattice state realised in $\mathcal{H}^{\sf S=1/2}_{\sf J_1-J_2}$ [Eq.~\ref{eq:HJ1J2}] for $J_1<0$, $J_2/|J_1| >0.4$ and $h\approx h_{\sf sat}$ [see Fig.~\ref{fig:J1J2h-nematic}].
In such a state the order parameter is bond-centred, but the physical spins, the fluctuations of which are measured in inelastic neutron scattering experiments, live on the sites of the square lattice.

At low energy, one can view the bond-nematic state in terms of an effective spin-1 degree of freedom that lives on the bonds\cite{shannon06}.
Fluctuations of these spin-1 degrees of freedom are described by $\mathcal{L}_{\sf hyd}({\bf k},\omega)$ [Eq.~\ref{eq:Lhydro}], and, in terms of the original spin-1/2 degrees of freedom, correspond to changing the mix of triplet states on a bond.
However, inelastic neutron scattering measures the fluctuations of the site-centred spin-1/2, and it is therefore necessary to determine the mapping that needs to be applied to $\Im m \chi({\bf q},\omega)$ [Eq.~\ref{eq:chihydro-spin1}] in order to make experimentally relevant predictions.

We consider two lattices, the original square lattice of the spin-1/2 degrees of freedom, with lattice constant $a$, and a bond-centred lattice with lattice constant \mbox{$b=a/\sqrt{2}$} [see Fig.~\ref{fig:Spin1vsSpin12}].
While in the rest of the article the lattice constants have been absorbed into the definition of the wavevector, rendering it dimensionless, for clarity in this section the lattice constants are explicitly included in the calculations.
In the site-centred lattice we label the real-space coordinates by the vector ${\bf x}$, and in the bond-centred lattice by ${\bf r}$.
In reciprocal space we use ${\bf p}$ and ${\bf q}$.
The relations between these coordinates are,
  \begin{align}
x_{\sf x} =\frac{1}{\sqrt{2}}(r_{\sf x}+r_{\sf y}), \qquad
x_{\sf y} =-\frac{1}{\sqrt{2}}(r_{\sf x}-r_{\sf y}),
\end{align} 
 and,
  \begin{align}
p_{\sf x} =\frac{1}{\sqrt{2}}(q_{\sf x}+q_{\sf y}), \qquad
p_{\sf y} =-\frac{1}{\sqrt{2}}(q_{\sf x}-q_{\sf y}),
\end{align} 
which can be seen from Fig.~\ref{fig:bz}.

We assume that the spin at a site, $S^{\sf sc}$, ({\sf sc} stands for site centred) can be written as the sum of the quasi-spin, $S^{\sf bc}$, degrees of freedom on the 4 neighbouring bond centres, which leads to,
  \begin{align}
S^{\sf sc}_{{\bf x}_i} = S^{\sf bc}_{{\bf x}_i+\frac{{\bf e}_{\sf x}}{2}} +S^{\sf bc}_{{\bf x}_i-\frac{{\bf e}_{\sf x}}{2}} + S^{\sf bc}_{{\bf x}_i+\frac{{\bf e}_{\sf y}}{2}} + S^{\sf bc}_{{\bf x}_i-\frac{{\bf e}_{\sf y}}{2}},
\end{align} 
where ${\bf e}_{\sf x}=(a,0)$ and ${\bf e}_{\sf y}=(0,a)$.
Taking the Fourier transform results in,
  \begin{align}
S^{\sf sc}_{{\bf p}} &= \sum_{{\bf q}} 
\delta\left( p_{\sf x}-\frac{q_{\sf x}+q_{\sf y}}{\sqrt{2}} \pm \frac{2m\pi}{a}\right) \nonumber \\
&\times \delta\left( p_{\sf y}+\frac{q_{\sf x}-q_{\sf y}}{\sqrt{2}} \pm \frac{2n\pi}{a}\right) 
\cos\frac{q_{\sf x} b}{2} \cos\frac{q_{\sf y} b}{2}
S^{\sf bc}_{{\bf q}},
\end{align} 
where $m$ and $n$ are integers.
This allows the imaginary part of the dynamical spin susceptibility to be calculated in the site-centred coordinate system as,
  \begin{align}
&\Im m \chi_{\sf sc}^{\sf \alpha \alpha}({\bf p},\omega)=  \nonumber \\ 
&\left[ \cos^2 \left[ \frac{(p_{\sf x}-p_{\sf y})a}{4} \right] 
\cos^2 \left[ \frac{(p_{\sf x}+p_{\sf y})a}{4} \right] 
\Im m \chi_{\sf bc}^{\sf \alpha \alpha}({\bf p},\omega)  \right.\nonumber \\
& \left. +\sin^2 \left[ \frac{(p_{\sf x}-p_{\sf y})a}{4} \right] 
\sin^2 \left[ \frac{(p_{\sf x}+p_{\sf y})a}{4}  \right] 
\Im m \chi_{\sf bc}^{\sf \alpha \alpha}({\bf p}_{\sf M}+{\bf p},\omega) \right]
\label{eq:chi-sitecentred}
\end{align} 
where \mbox{$(p_{\sf x}+p_{\sf y})^2/2+(p_{\sf x}-p_{\sf y})^2/2 = {\bf p}^2$} has been used, ${\bf p} \approx 0$ and $\Im m \chi_{\sf bc}^{\sf \alpha \alpha}$ is given in Eq.~\ref{eq:chihydro-spin1}.

The result of mapping the spin-1 dynamical susceptibility predictions shown in Fig.~\ref{fig:FTspin1-chiqomh} onto the site-centred lattice using Eq.~\ref{eq:chi-sitecentred} is shown in Fig.~\ref{fig:FTspinhalf-chiqomh}.
It can be seen that the mapping from the bond-centred to site-centred lattice results in all the low-energy modes appearing at the $\Gamma$ point, which is expected, since the 2-sublattice AFQ state does not break the translational symmetry of the site-centred lattice.
For \mbox{${\bf q}\approx 0$}, one has \mbox{$\cos^2 [ (q_{\sf x}-q_{\sf y})a/4] \approx 1$}, \mbox{$\cos^2 [ (q_{\sf x}+q_{\sf y})a/4] \approx 1$}, \mbox{$\sin^2 [ (q_{\sf x}-q_{\sf y})a/4] \approx 0$} and  \mbox{$\sin^2 [ (q_{\sf x}+q_{\sf y})a/4] \approx 0$} and therefore the contribution from $\Im m \chi_{\sf bc}^{\sf \alpha \alpha}({\bf q},\omega)$ dominates over the contribution from $\Im m \chi_{\sf bc}^{\sf \alpha \alpha}({\bf q}_{\sf M}+{\bf q},\omega)$.
In consequence, when making experimental predictions in Section~\ref{sec:neutrons} below, we concentrate on the modes $\omega_{{\bf k},h}^{\sf Q,z}$ and $ \omega_{{\bf k},h}^{\sf xy,0}$ [Eq.~\ref{eq:omhydro}], since these will determine the dominant experimental signatures.


\section{Microscopic theory of a spin-1/2 bond nematic}
\label{sec:parametrisation}


In order to make quantitative experimental predictions for materials it is necessary to determine the hydrodynamic parameters appearing in $\mathcal{L}_{\sf hyd}({\bf k},\omega)$ [Eq.~\ref{eq:Lhydro}].
Here we consider $\mathcal{H}^{\sf S=1/2}_{\sf J_1-J_2}$ [Eq.~\ref{eq:HJ1J2}] and microscopically calculate the hydrodynamic parameters in the vicinity of $h=h_{\sf sat}$.
In Section~\ref{sec:neutrons} these will be fed into $\mathcal{L}_{\sf hyd}({\bf k},\omega)$ [Eq.~\ref{eq:Lhydro}], allowing quantitative predictions to be made for inelastic neutron scattering experiments.

All magnets have a saturation magnetic field, above which the spins are aligned parallel to the field direction.
For systems with an antiferromagnetic ground state at $h=0$, the simplest way to connect the zero-field and high-field states is via a canting of the ordered moment towards the field direction.
As field is reduced from above saturation, there is a phase transition from the fully-polarised state to the canted antiferromagnet at the saturation field.
This can be understood in terms of the condensation of magnons out of the fully-saturated ``vacuum'' state\cite{batyev84,batyev85}.

In frustrated magnets the condensation of single magnons may not be the first instability of the fully-saturated state as magnetic field is lowered.
One possibility is that the 1-magnon instability is preceded by a 2-magnon instability, in which bound pairs of magnons condense\cite{shannon06}.
If this occurs, then this corresponds to a quadrupolar ordering of the spin degrees of freedom just below the saturation magnetic field, and hence the creation of a spin-nematic state.
We show below that this is the case for $\mathcal{H}^{\sf S=1/2}_{\sf J_1-J_2}$ [Eq.~\ref{eq:HJ1J2}] over a wide range of parameters.

In the saturated paramagnet it is useful to use the hardcore boson representation, in which spin operators are replaced according to,
\begin{align}
S^z_l=-1/2+a^\dagger_l a_l \; , \;\;
S_l^+ =a_l^\dagger \; , \; \;  S_l^- =a_l \; .
\label{eq:hardcoreSpin}
\end{align}
Thus $\mathcal{H}^{\sf S=1/2}_{\sf J_1-J_2}$ [Eq.~\ref{eq:HJ1J2}] can be rewritten as,
\begin{align}
\mathcal{H}^{\sf S=1/2}_{\sf J_1-J_2} &= \sum_{\bf q}[\omega (\bolq) - \mu(h)] a^\dagger_{\bolq}a_{\bolq} \nonumber \\
& \qquad +\frac{1}{2N}  \sum_{\bolq,\bolk,\bolk^\prime}  V_{\bolq} 
a_{\bolk+\bolq}^\dagger a_{\bolk^\prime-\bolq}^\dagger
a_{\bolk}a_{\bolk^\prime},
\label{eq:HJ1J2boson}
\end{align}
where,
\begin{align}
&\omega(\bolq) =\epsilon(\bolq)-\epsilon_{\sf min}, \quad
\mu(h)=h_{\sf c1}-h , \nonumber \\
& h_{\sf c1} = \epsilon({\bf 0})-\epsilon_{\sf min}, \quad
V_{\bolq} =2(\epsilon(\bolq)+U) ,
\label{eq:omq-muh-hc1}
\end{align}
$U\to \infty$ is the on-site repulsion and,
\begin{align}
\epsilon({\bf q}) &= 
J_1 (\cos q_{\sf x} + \cos q_{\sf y}) 
+ 2J_2 \cos q_{\sf x}  \cos q_{\sf y} .
\end{align}
In the parameter range \mbox{$-2\leq J_1/J_2\leq2$} with \mbox{$J_2>0$}, 
\mbox{ $\epsilon_{\sf min}=\epsilon(\bolQ_{\pm})=-2J_2$},
where \mbox{$\bolQ_{+}=(\pi,0)$} and \mbox{$\bolQ_{-}=(0,\pi)$}.
The saturation field for 1-magnon condensation is thus given by,
\begin{align}
h_{\sf c1}=2J_1+4J_2.
\label{eq:hc1}
\end{align}

For $\mu(h)<0$ ($h > h_{\sf c1}$) the system is fully-polarised unless the 1-magnon instability is preceded by another instability.
We denote the fully-saturated state by $\ket{\Omega}$, where $a_l\ket{\Omega}=0$.
Since the interaction term in Eq.~\ref{eq:HJ1J2boson} is normal ordered, there is no self-energy term in the 1-magnon dispersion relation, and it is exactly given by,
\begin{align}
\omega^{1-m}_{{\bf k},h} = \omega(\bolk)-\mu(h) .
\label{eq:omkh1-m}
\end{align}
Expanding $\omega(\bolk)$ near the minima $\bolQ_{\pm}$ leads to,
\begin{align}
\omega(k_x,k_y+\pi)=\frac{k_x^2}{2m_{1}^{(1)}}+\frac{k_y^2}{2m_{2}^{(1)}}+O(k^4)\ ,\ 
\end{align}
where,
\begin{align}
m^{(1)}_{1} =\frac{1}{-J_1 + 2J_2}, \quad
m^{(1)}_{2} =\frac{1}{J_1 + 2J_2},
\end{align}
and,
\begin{align}
\omega(k_x+\pi,k_y)
=\frac{k_x^2}{2m_{2}^{(1)}}+\frac{k_y^2}{2m_{1}^{(1)}}+O(k^4).
\end{align}

In the fully-saturated state it is very simple to calculate the imaginary part of the dynamical spin susceptibility.
It is exactly given by,
\begin{align}
\Im m\chi_{\sf sat}^{\perp}(\omega, \bolk) &= \Im m\chi_{\sf sat}^{\sf xx}(\omega, \bolk) +
\Im m\chi_{\sf sat}^{\sf yy}(\omega, \bolk) = \nonumber \\
& \pi  \delta(\omega-\omega^{1-m}_{{\bf k},h}) \nonumber \\
\Im m\chi_{\sf sat}^{\sf zz}(\omega, \bolk) &= 0.
\label{eq:chisat}
\end{align}
This describes a sharp band of 1-magnon excitations, with equal weight at all wavevectors.

\begin{figure}[t]
\centering
\includegraphics[width=0.49\textwidth]{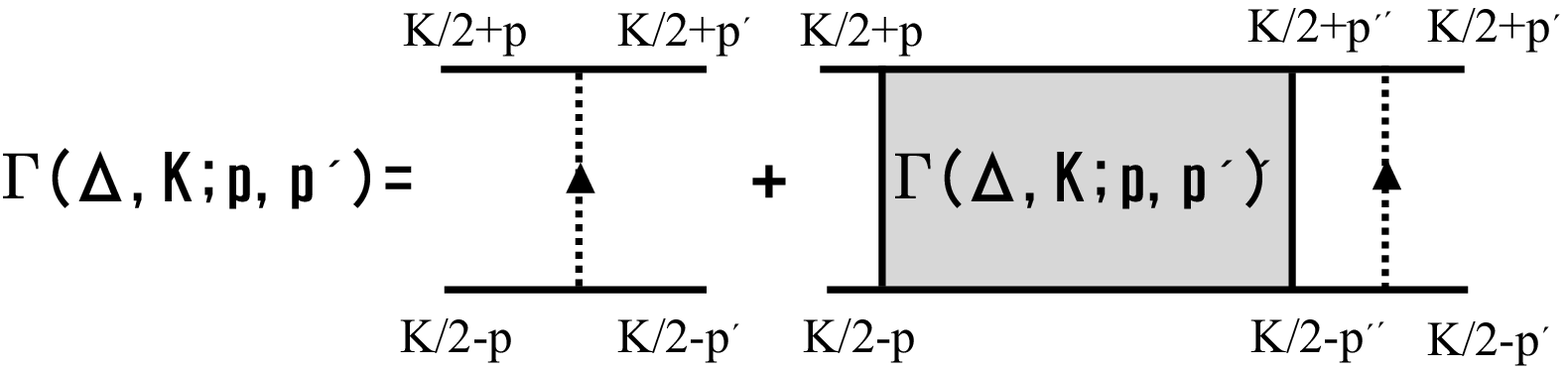}
\caption{\footnotesize{(Color online). 
Diagrammatic calculation of 2-magnon binding energy for $\mathcal{H}^{\sf S=1/2}_{\sf J_1-J_2}$ [Eq.~\ref{eq:HJ1J2}].
The exact scattering amplitude $\Gamma(\Delta,\bolK;\bolp,\bolp^\prime)$ [Eq.~\ref{eq:ladder}] is calculated from a ladder diagram.
$\bolK$ is the centre of mass momenta and $\Delta$ the binding energy of magnon pairs.
}}
\label{fig:ladder}
\end{figure}

We next consider the condensation of bound pairs of magnons out of the fully-saturated state.
The 2-particle Green's function can be calculated exactly using the ladder diagram shown in Fig.~\ref{fig:ladder}.
The scattering amplitude is given by\cite{batyev84,batyev85,nikuni95,ueda13,ueda-arXiv},
\begin{equation}
\begin{split}
&\Gamma(\Delta,\bolK;\bolp,\bolp^\prime)
=V_{\bolp^\prime-\bolp}+V_{-\bolp^\prime-\bolp}\\
&-\frac{1}{2}\int \frac{d^2 p^{\prime\prime}}{(2\pi)^2}
\frac{\Gamma(\Delta,\bolK;\bolp,\bolp^{\prime\prime})
[V_{\bolp^\prime-\bolp^{\prime\prime}}
+V_{-\bolp^\prime-\bolp^{\prime\prime}}]}
{\omega(\bolK/2+\bolp^{\prime\prime})+\omega(\bolK/2-\bolp^{\prime\prime})+\Delta-i0^+},
\end{split}
\label{eq:ladder}
\end{equation}
where $\Delta$ and $\bolK$ are respectively the binding energy
and the center-of-mass momentum of the bound state.
This integral equation is exactly soluble\cite{batyev84,batyev85}.


\begin{figure}[t]
\centering
\includegraphics[width=0.45\textwidth]{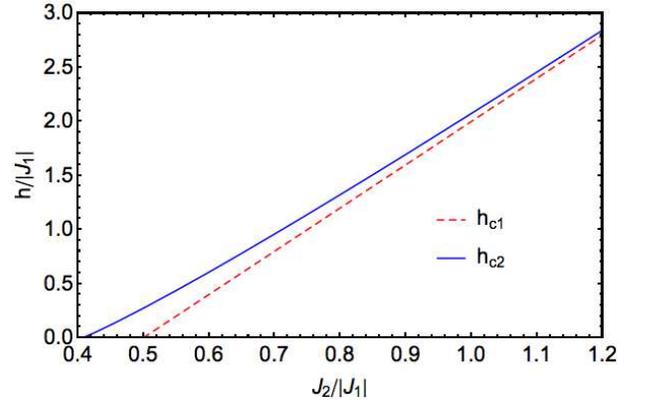}
\caption{\footnotesize{(Color online). 
First instability of the fully-polarised state on lowering magnetic field.
The critical field for one magnon condensation [$h_{\sf c1}$, Eq.~\ref{eq:hc1}] is shown as a red, dashed line and the critical field for condensation of bound-magnon pairs [$h_{\sf c2}$, Eq.~\ref{eq:hc2}] as a blue line.
For $0.4\lesssim |J_1/J_2|$ the 2-magnon instability precedes the 1-magnon instability ($h_{\sf c2}>h_{\sf c1}$). 
}}
\label{fig:Hc1Hc2}
\end{figure}

\begin{figure}[ht]
\centering
\includegraphics[width=0.45\textwidth]{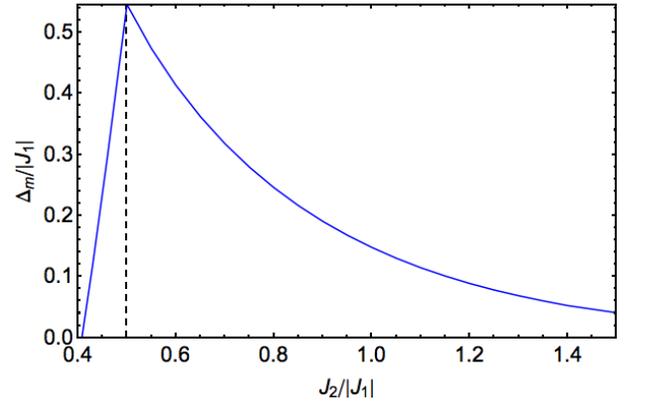}
\caption{\footnotesize{(Color online). 
The 2-magnon binding energy in the fully-saturated state of $\mathcal{H}^{\sf S=1/2}_{\sf J_1-J_2}$ [Eq.~\ref{eq:HJ1J2}].
The binding energy, \mbox{$\Delta_m=\Delta_B(\bolK=(0,0))$}, is calculated from $\Gamma(\Delta,\bolK;\bolp,\bolp^\prime)$ [Eq.~\ref{eq:ladder}, Fig.~\ref{fig:ladder}].
It determines the saturation field for 2-magnon condensation, $h_{\sf c2} = h_{\sf c1} + \Delta_m/2$.
For $\Delta_m>0$ [see Fig.~\ref{fig:Hc1Hc2}], $h_{\sf c2}>h_{\sf c1}$ and the 2-magnon condensate occurs at higher field than the 1-magnon condensate.
}}
\label{fig:BEnergy}
\end{figure}

Divergence of $\Gamma(\Delta,\bolK;\bolp,\bolp^\prime)$ [Eq.~\ref{eq:ladder}] implies the existence of a stable bound state below the 2-magnon continuum\cite{nakanishi69,ueda13}.
The $\bolK$-dependent binding energy is denoted as $\Delta_B(\bolK)$.
The wavefunction of the bound state can be determined from the residue of the divergence of $\Gamma(\Delta,\bolK;\bolp,\bolp^\prime)$ [Eq.~\ref{eq:ladder}], and this is considered in more detail below.
When lowering magnetic field, if the bound-magnon gap closes before the 1-magnon gap then the bound state condenses and the spin-nematic state appears.
The critical field for 2-magnon condensation is given by,
\begin{align}
h_{\sf c2} = h_{\sf c1} + \Delta_{\sf m}/2 ,
\label{eq:hc2}
\end{align}
where $\Delta_m$ is the maximum value of the binding energy.
For $0.4\lesssim |J_1/J_2|\lesssim 5$ there is a stable bound state at $\bolK=(0,0)$, and this is the leading instability on lowering magnetic field [see Fig.~\ref{fig:Hc1Hc2}].
The binding energy, \mbox{$\Delta_m=\Delta_B(\bolK=(0,0))$}, as a function of $J_2/|J_1|$  is shown in Fig.~\ref{fig:BEnergy}.


\begin{figure}[ht]
\centering
\includegraphics[width=0.49\textwidth]{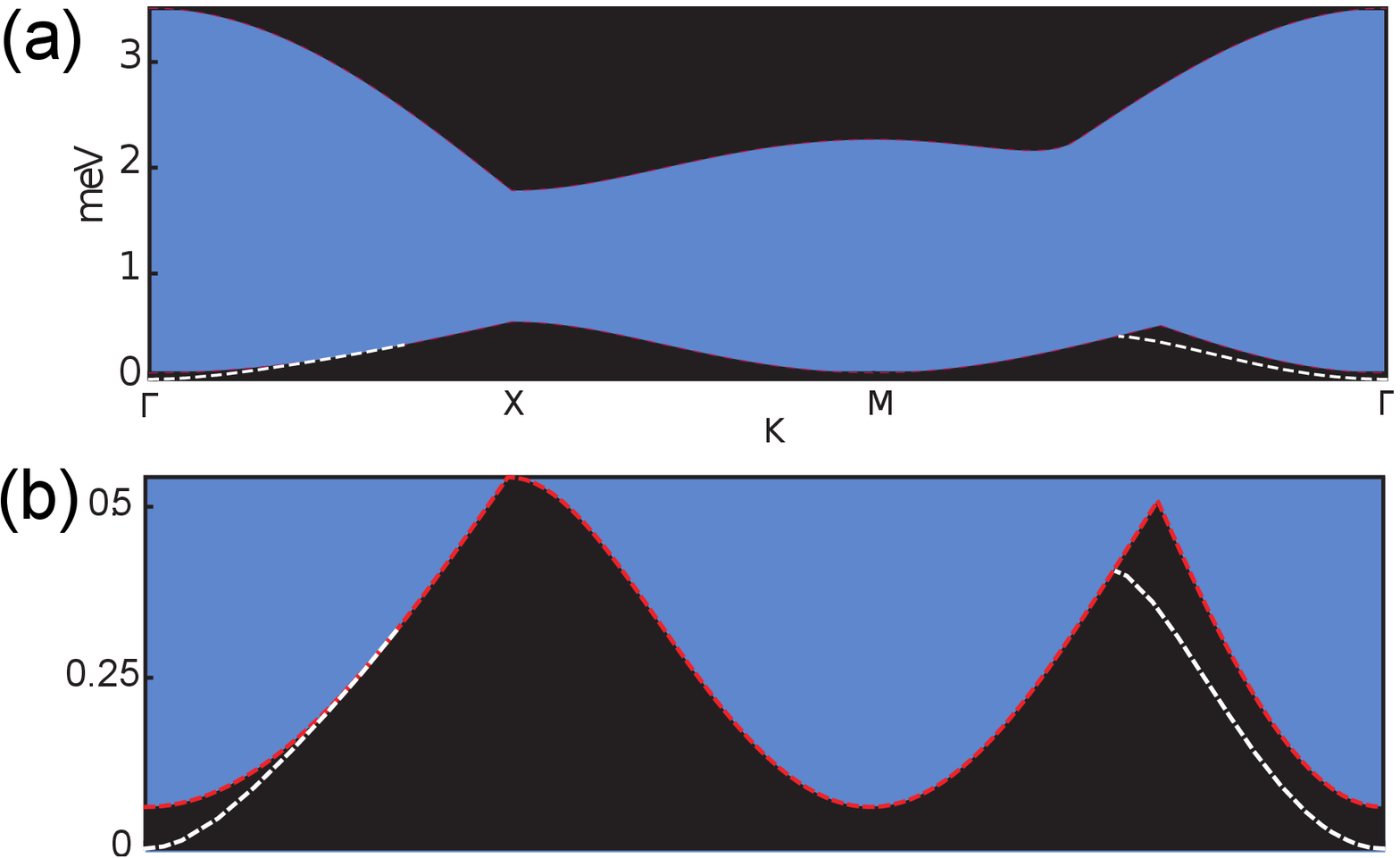}
\caption{\footnotesize{(Color online). 
2-magnon dispersion of $\mathcal{H}^{\sf S=1/2}_{\sf J_1-J_2}$ [Eq.~\ref{eq:HJ1J2}] at saturation.
The 2-magnon continuum, defined by \mbox{$\omega(\bolK/2+\bolp)+\omega(\bolK/2-\bolp)+\Delta_m$}, is shown as a blue region.
The dispersion of the 2-magnon bound state, $\omega^{2-m}_{{\bf K},h}$ [Eq.~\ref{eq:omega-2mag}], is shown as a white dashed line for the region of ${\bf K}$ in which it lies below the 2-magnon continuum.
(a) shows the full extent of the continuum, while (b) shows a detailed view of the low-energy spectrum.
The exchange parameters are taken as $J_1=-3.6$K and $J_2=3.2$K and the field as $h=h_{\sf c2}$ [Eq.~\ref{eq:hc2}].
}}
\label{fig:ContBM_plot}
\end{figure}


In the fully-polarised phase with $h>h_{\sf c2}$, 
the bound-magnon dispersion relation is given by,
\begin{equation}
\omega^{2-m}_{{\bf K},h} = \Delta_m-\Delta_B(\bolK)-\mu_2(h) \approx \frac{K^2}{2m^{(2)}}-\mu_2(h)+O(K^4),
\label{eq:omega-2mag}
\end{equation}
where,
\begin{equation}
\mu_2(h)=2(h_{\sf c2}-h) .
\end{equation}
This dispersion, $\omega^{2-m}_{{\bf K},h}$ [Eq.~\ref{eq:omega-2mag}], is shown in Fig.~\ref{fig:ContBM_plot} in relation to the 2-magnon continuum.

Slightly below the saturation field, for small \mbox{$\mu_2(h)>0$}, one can view the system as a dilute gas of bound magnons.
We consider an effective Hamiltonian,
\begin{equation}
\begin{split}
\mathcal{H}_{\text{eff}} = &\sum_{|\bolK|<\Lambda}
\left\{
\frac{(K)^2}{2m^{(2)}}-
\mu_2(h)
\right\}
b^\dagger_{\bolK}b_{\bolK}\\
&+\frac{\Gamma^{(2)}}{4N}\sum_{\bolK_1,\bolK_2,\bolq}
b^\dagger_{\bolK_1+\bolq}b^\dagger_{\bolK_2-\bolq}b_{\bolK_1}b_{\bolK_2} 
+ \cdots \ ,
\end{split}
\label{eq:bmH}
\end{equation}
where the bound-state creation operator is,
\begin{align}
b_\bolK^\dagger=\sum_{\bolp}\chi_\bolK(\bolp)a^\dagger_{\bolK/2+\bolp}a^\dagger_{\bolK/2-\bolp}, 
\label{eq:boundstate-wfn}
\end{align}
$\Gamma^{(2)}$ is the renormalized interaction between the low-energy bound magnons\cite{ueda-arXiv}, $\Lambda$ is a momentum cutoff, and the dots represent higher-order interaction terms that can be neglected in the dilute limit. 
The effective free energy of the condensed phase is given by,
\begin{equation}
\frac{E}{N} =\frac{1}{4}
\Gamma^{(2)}\rho_2^2-\mu_2(h) \rho_2\ ,
\label{eq:EA}
\end{equation}
where $\VEV{b_{\bolK=0}}=\sqrt{\rho_2}e^{i\theta_0}$. 
Minimizing this free energy, results in,
\begin{equation}
\rho_2=\frac{2\mu_2(h)}{\Gamma^{(2)}} .
\end{equation}

Next we consider the Goldstone mode of the spin-nematic state in the dilute limit at zero temperature.
Taking the cutoff $\Lambda\rightarrow \infty$, the effective action is given by,
\begin{align}
\mathcal{S}_{\sf eff} &=
\int d^2x\left[
\frac{i}{2} b({\bf x})^\ast \partial_t b({\bf x})-b({\bf x}) \partial_t b({\bf x})^\ast)
-\frac{|\nabla b({\bf x})|^2}{2m^{(2)}} \right. \nonumber \\
& \left. \qquad +\mu_2(h) |b({\bf x})|^2 
-\frac{\Gamma^{(2)}}{4}|b({\bf x})|^4 
\right] , 
\end{align}
where \mbox{$b({\bf x})=\frac{1}{N}\int d^2K b_{\bolK}
e^{i \bolK \cdot {\bf x}}$}. 
Substituting,
\begin{align}
b({\bf x})=\sqrt{\rho_2+\delta \rho_2({\bf x})} \ e^{i\theta({\bf x})},
 \end{align}
into $\mathcal{S}_{\sf eff}$ [Eq.~\ref{eq:bmH}] and integrating out the high-energy mode $\delta \rho_2$, results in,
\begin{equation}
\mathcal{S}_{\sf eff}=\int dx^2 \frac{1}{\Gamma^{(2)}}(\partial_t \theta({\bf x}))^2
-\frac{\rho_2}{2m^{(2)}}(\nabla \theta({\bf x}))^2 .
\label{eq:Seff2}
\end{equation}
Hence, the disperison of the Goldstone mode at small $\bolK$ within the dilute limit is given by\cite{ueda-arXiv2},
\begin{equation}
\omega^{2-m}_{{\bolK},h}=\sqrt{\frac{\Gamma^{(2)} \rho_2  \bolK^2}{2m^{(2)}}}
=\sqrt{\frac{\mu_2(h)}{m^{(2)}}}|\bolK |.
\label{eq:omegak2M-nem}
\end{equation}


\begin{figure}[ht]
\centering
\includegraphics[width=0.45\textwidth]{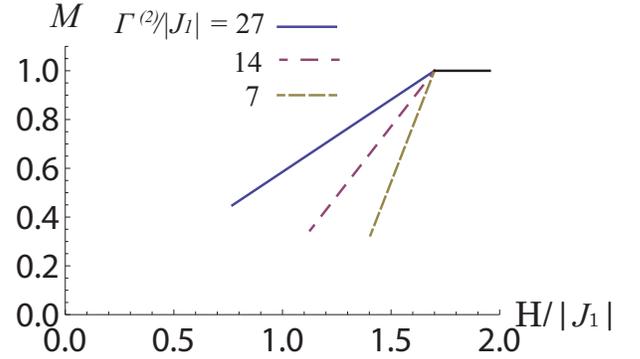}
\caption{\footnotesize{(Color online). 
Magnetisation of the spin-nematic state found in $\mathcal{H}^{\sf S=1/2}_{\sf J_1-J_2}$ [Eq.~\ref{eq:HJ1J2}] just below the saturation field as a function of magnetic field $h$.
M(h) [Eq.~\ref{eq:Mh}] is plotted at several values of the effective interaction $\Gamma^{(2)}$.
Comparison of $\partial M(h)/\partial h$ at $h\to h_{\sf sat}$ [Eq.~\ref{eq:dM/dh}] with experimental data can be used to estimate $\Gamma^{(2)}$ for a given material.
}}
\label{fig:MagCurve_Gamm2}
\end{figure}


In 2d, the parameter $\Gamma^{(2)}$ tends to zero approaching the saturation field.
However, even a very small interlayer coupling will cutoff this approach to zero, and $\Gamma^{(2)}$ remains finite at $h=h_{\sf c2}$ 
\footnote{
Here we make a comment on the dimensionality.
For a small interlayer coupling $J_\perp$, $\Gamma^{(2)}=O(J/(|\log J_\perp/J_1|+\cdots))\rightarrow 0$ as 
$J_\perp \rightarrow 0$.
In 2d, the suppression of $\Gamma^{(2)}$ implies a steep magnetisation curve just below the saturation field.
In quasi-2d materials, the logarithmic divergence is cut-off by a small $J_\perp$, and even with $J_\perp/|J_1| \sim 10^{-2}$ one finds $\log J_\perp/|J_1| \sim O(1)$.
Hence in any real material $\Gamma^{(2)}$ will be finite.
}.
A good way to estimate $\Gamma^{(2)}$ for a given material is from the gradient of the experimental magnetisation curve close to saturation.
One finds theoretically for the magnetisation,
\begin{align}
M(h\approx h_{\sf sat})=\frac{2}{N} \sum_l \VEV{S^z_l}=1-\frac{16(h_{\sf c2}-h)}{\Gamma^{(2)}},
\label{eq:Mh}
\end{align}
and therefore,
\begin{align}
\left. \frac{\partial M}{\partial h}\right|_{h=h_{\sf sat}} = \frac{16}{\Gamma^{(2)}}
\label{eq:dM/dh}
\end{align}
Magnetisation curves for several values of $\Gamma^{(2)}$ are shown in Fig.~\ref{fig:MagCurve_Gamm2}.

Finally, we will consider the bound-state wavefunction $\chi_\bolK(\bolp)$ [Eq.~\ref{eq:boundstate-wfn}] and demonstrate that this describes the same 2-sublattice bond-nematic state considered in Section~\ref{sec:flvwave} and Section~\ref{sec:fieldtheory}.

From taking the residue of the divergent scattering amplitude $\Gamma(\Delta,\bolK;\bolp,\bolp^\prime)$ [Eq.~\ref{eq:ladder}], one finds,
\begin{equation}
\chi_{\bolK=0}(\bolp) = \chi_{0}(\bolp) \propto \frac{\cos p_x - \cos p_y}{2\omega(\bolp)+\Delta_m} ,
\label{eq:nem-wfn}
\end{equation}
where the minimum of $\omega(\bolp)$ [Eq.~\ref{eq:omq-muh-hc1}] is zero.
The wavefunction is normalised by requiring $\sum_\bolp |\chi_{0}(\bolp)|^2=1$, where 
the summation over $\bolp$ is taken in half of the Brillouin zone 
as $\bolp$ and $-\bolp$ gives the same bound state.
This wavefunction has d-wave symmetry, since interchange of the coordinates $x$ and $y$ leads to $\chi_{0}(\bolp)\rightarrow -\chi_{0}(\bolp)$ while $\bolK=0\rightarrow 0$.

The bound-magnon condensed phase is described by the coherent state\cite{zhitomirsky10},
\begin{align}
|\text{Nem}\rangle 
=C_1\exp (\phi\sum_\bolp \chi_0(\bolp)a^\dagger_\bolp a^\dagger_{-\bolp})
\ket{\Omega},
\label{eq:Nemket}
\end{align}
where $C_1=\Pi_\bolp \sqrt{1-|\phi \chi_0(\bolp)|^2}$ and 
$\phi=\sqrt{\rho_2}e^{i\theta_0}$.
In this phase we consider the expectation value of the bond-operator,
\begin{align}
Q^{--}(\bolr) &= 
S^-_{\bf l} S^-_{{\bf l}+{\bf r}}
= \frac{1}{2} \left( Q^{\sf xx}_{{\bf l},{\bf l}+{\bf r}} - Q^{\sf yy}_{{\bf l},{\bf l}+{\bf r}} \right) - i Q^{\sf xy}_{{\bf l},{\bf l}+{\bf r}} \nonumber \\
&=  Q^{\sf x^2-y^2}_{{\bf l},{\bf l}+{\bf r}} - i Q^{\sf xy}_{{\bf l},{\bf l}+{\bf r}},
\end{align}
where $Q^{\alpha \beta}_{ij}$ is defined in Eq.~\ref{eq:OP} and \mbox{$Q^{\sf x^2-y^2}_{ij} = 1/2(Q^{\sf xx}_{ij} - Q^{\sf yy}_{ij})$} is defined in analogy with Eq.~\ref{eq:Q} for the spin-1 theory.
2-sublattice AFQ order of the type considered throughout this article corresponds to a non-zero expectation value $\VEV{Q^{--}(\bolr)}$ on nearest-neighbour bonds, with a change of sign between vertical and horizontal bonds.
Calculating the expectation value of this bond operator with respect to the condensed phase, $|\text{Nem}\rangle$ [Eq.~\ref{eq:Nemket}] gives,
\begin{align}
\VEV{Q^{--}(\bolr)}_{\sf nem} &=\VEV{S^-_{\bf l} S^-_{{\bf l}+{\bf r}}}_{\sf nem} =
\sum_{\bolp} 
\VEV{a_{\bolp}a_{-\bolp}}_{\sf nem} \exp(i \bolp\cdot \bolr) \nonumber \\
&=\sum_\bolp \frac{\phi\chi_0(\bolp)}{1-|\phi\chi_0(\bolp)|^2}\exp(i \bolp\cdot \bolr) \nonumber \\
&=\phi \sum_\bolp \chi_0(\bolp)\exp(i \bolp\cdot \bolr)+O(\phi^2) .
\label{eq:Q--}
\end{align}
The permutation ${\bf r}=(r_x,r_y)\rightarrow (r_y,r_x)$ 
shows the d-wave symmetry of this bond operator,
\begin{equation}
\VEV{Q^{--}(r_x,r_y)}_{\sf nem} = -\VEV{Q^{--}(r_y,r_x)}_{\sf nem},
\end{equation}
which, on nearest-neighbour bonds, exactly corresponds to the 2-sublattice AFQ phase shown in Fig.~\ref{fig:J1J2h-nematic} and Fig.~\ref{fig:Spin1vsSpin12}.

Next, we consider the asymptotic behaviour 
of $\VEV{Q^{--}(\bolr)}_{\sf nem}$ [Eq.~\ref{eq:Q--}] for the low-density case $\phi\ll 1$. 
For large $\bolr$, the oscillation of the wave function is fast, and 
the dominant integrand comes from $\bolp$ close to $(0,0)$, $(0,\pi)$ 
and $(\pi,0)$.
Considering $\bolr=(r,0)$ for simplicity, one can show for $r\gg 1$,
\begin{align}
&\VEV{Q^{--}(r,0)}_{\sf nem}  \approx 
c_2\frac{\exp (- r/\xi_0)}{\sqrt{r/\xi_0}} ,
\end{align}
where $\xi_0\approx 1/\sqrt{2m_1^{(1)} \Delta_m}$ is a measure of the size of the bound state 
and $c_{2}$ is a constant.
The larger the value of the binding energy $\Delta_m$ [see Fig.~\ref{fig:BEnergy}], the more localised the bound state.

As an example, we take $J_2/|J_1|=0.9$ and show that the nearest-neighbour bonds gives the dominant contribution to $\VEV{Q^{--}(\bolr)}$ [Eq.~\ref{eq:Q--}].
One finds,
\begin{align}
\VEV{Q^{--}(2,0)}_{\sf nem}/\VEV{Q^{--}(1,0)}_{\sf nem} &=0.15 \nonumber \\
\VEV{Q^{--}(3,0)}_{\sf nem}/\VEV{Q^{--}(1,0)}_{\sf nem} &=0.33 \nonumber \\
\VEV{Q^{--}(4,0)}_{\sf nem}/\VEV{Q^{--}(1,0)}_{\sf nem} &=0.10 \nonumber \\
\VEV{Q^{--}(0,0)}_{\sf nem}=\VEV{Q^{--}(1,1)}_{\sf nem} &=0
\end{align}
to first order in $\phi$.
In Section~\ref{sec:flvwave}, we have in effect considered the nearest-neighbour order parameter $\VEV{Q^{--}(1,0)}_{\sf nem}$, and the above analysis shows that this approximation becomes better the larger the value of $\Delta_m$.
In the limit where $\Delta_m \to \infty$ the wavefunction $\chi_{\bolK=0}(\bolp)$ [Eq.~\ref{eq:nem-wfn}] is only non-zero on nearest-neighbour bonds to first order in $\phi$, and we write,
\begin{equation}
\chi_{n.n.}(\bolp)= \sqrt{\frac{2}{N}}(\cos p_x - \cos p_y).
\end{equation}
In consequence the only non-zero bond-operator expectation values are,
\begin{equation}
\VEV{Q^{--}(1,0)}_{\sf nem} = -\VEV{Q^{--}(0,1)}_{\sf nem} = \phi .
\end{equation}

While the mapping to the spin-1 model [Section~\ref{sec:flvwave}] needs a nearest-neighbour bond order, 
this is not the case for the continuum theory [Section~\ref{sec:fieldtheory}].
The sole requirement is $\bolK\ll 1/\xi_0$, and therefore the continuum theory is valid even for small $\Delta_m$.

In summary, we have considered in this section the ground state and low-energy excitation spectrum of $\mathcal{H}^{\sf S=1/2}_{\sf J_1-J_2}$ [Eq.~\ref{eq:HJ1J2}] for magnetic fields close to saturation.
At the saturation field, $h_{\sf sat} = h_{c2}$, bound-magnon pairs condense to form a spin-nematic state.
In the limit that the density of bound pairs is dilute, the excitation spectrum can be calculated from the microscopic model.
This allows the hydrodynamic parameters appearing in $\mathcal{L}_{\sf hyd}({\bf k},\omega)$ [Eq.~\ref{eq:Lhydro}] to be determined, and we show an explicit example of this in the following section.


\section{Predictions for inelastic neutron scattering experiments in a spin-1/2 bond nematic}
\label{sec:neutrons}


Spin-nematic order does not break time-reversal symmetry, and therefore does not produce an internal magnetic field.
This makes the spin-nematic state essentially invisible to most common probes of magnetism, such as elastic scattering of neutrons, Knight shift of the NMR spectra and the asymmetry of oscillations in $\mu$sr spectra\cite{andreev84}.
However, excitations of the quadrupolar order parameter mix a spin-dipole component into the wavefunction, and this can, in principle, be detected by dynamic probes of magnetism.

Here we make predictions that demonstrate how a spin-nematic state could be identified via inelastic scattering of neutrons.
We consider in particular the class of materials that are well described by $\mathcal{H}^{\sf S=1/2}_{\sf J_1-J_2}$ [Eq.~\ref{eq:HJ1J2}].
For definiteness we show predictions relevant to BaCdVO(PO$_4$)$_2$, where fits to magnetic susceptibility give the exchange parameters $J_1=-3.6$K and $J_2=3.2$K [\onlinecite{nath08}]. 
We take a square lattice with lattice constant $a=4.5$\AA.
The saturation field of BaCdVO(PO$_4$)$_2$ has been measured to be $h_{\sf sat} \approx 4.2$T, which is easily achievable in a neutron scattering experiment.

\begin{figure}[t]
\centering
\includegraphics[width=0.49\textwidth]{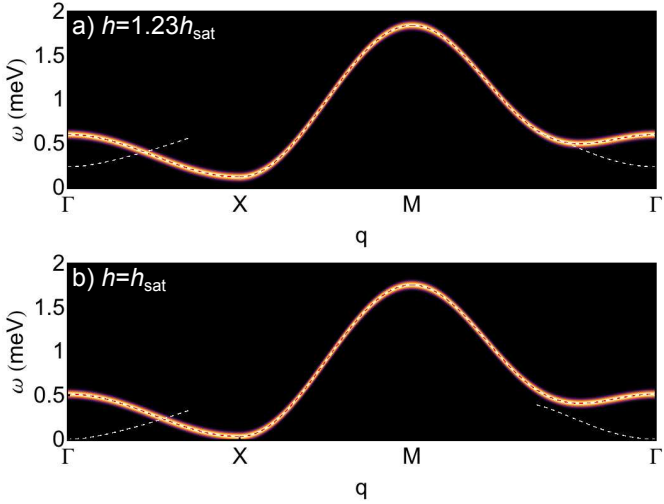}
\caption{\footnotesize{(Color online). 
Predictions for inelastic neutron scattering from a system described by $\mathcal{H}^{\sf S=1/2}_{\sf J_1-J_2}$ [Eq.~\ref{eq:HJ1J2}] above the saturation field $h_{\sf sat}=h_{\sf c2}$ [Eq.~\ref{eq:hc2}]. 
The imaginary part of the dynamic spin susceptibility, $\Im m\chi_{\sf sat}^{\perp}(\omega, {\bf q})$ [Eq.~\ref{eq:chisat}], is calculated exactly using the parameters $J_1=-3.6$K and $J_2=3.2$K, which are believed to describe BaCdVO(PO$_4$)$_2$ \cite{nath08}.
The 1-magnon dispersion, $\omega^{1-m}_{{\bf q},h}$ [Eq.~\ref{eq:omkh1-m}], is shown by a red dashed line and the dispersion of the 2-magnon bound state, $\omega^{2-m}_{{\bf q},h}$ [Eq.~\ref{eq:omega-2mag}], by a white dashed line.
a) At $h=1.23 h_{\sf sat}$ the dispersion of the 2-magnon bound state is gapped for all ${\bf q}$.
b) At $h=h_{\sf sat}$ the gap to the 2-magnon bound state closes at ${\bf q}=(0,0)$ while 1-magnon excitations remains gapped for all ${\bf q}$. 
The circuit $\Gamma$-{\sf X}-{\sf M}-$\Gamma$ in the site-centred Brillouin zone is shown in Fig.~\ref{fig:bz}.
The same linear, normalised colour intensity scale is used as in Fig.~\ref{fig:bbq-chiqomh}.
}}
\label{fig:chiJ1J2sat}
\end{figure}

\begin{figure*}[ht]
\centering
\includegraphics[width=0.9\textwidth]{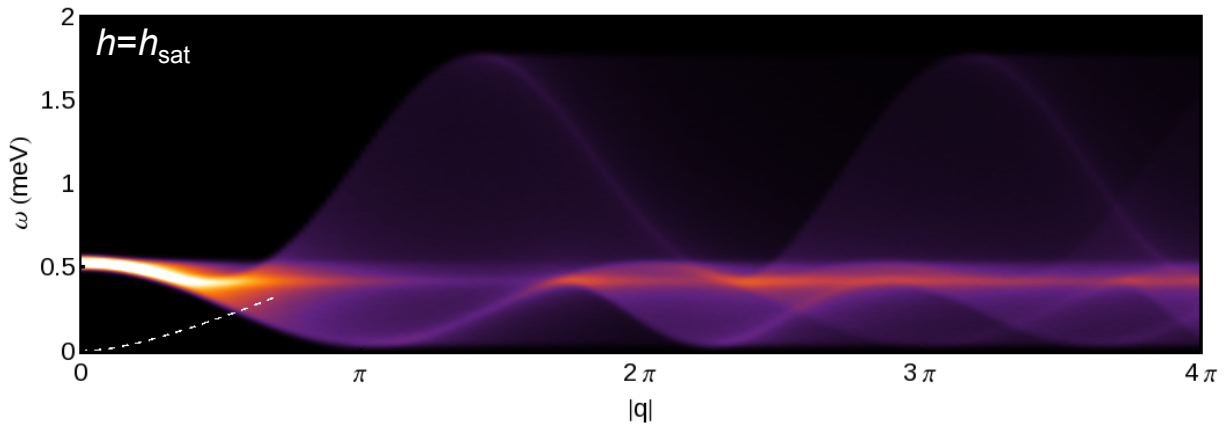}
\caption{\footnotesize{(Color online). 
Predictions for angle-integrated inelastic neutron scattering experiments on a spin-1/2 frustrated ferromagnet 
with incipient spin-nematic order, at the saturation field $h=h_{\sf sat}$.
The imaginary part of the dynamic spin susceptibility [Eq.~\ref{eq:chisat}] has been integrated over $4\pi$ solid angle in order to mimic a powder sample.
The Hamiltonian considered is $\mathcal{H}^{\sf S=1/2}_{\sf J_1-J_2}$ [Eq.~\ref{eq:HJ1J2}], and the parameters $J_1=-3.6$K and $J_2=3.2$K are believed to describe BaCdVO(PO$_4$)$_2$ \cite{nath08}.
The white dashed line shows the dispersion of the 2-magnon continuum, $\omega^{2-m}_{{\bf q},h}$ [Eq.~\ref{eq:omega-2mag}], which becomes gapless at the saturation magnetic field $h=h_{\sf sat}$.
At this value of magnetic field, spectral weight resides in the 1-magnon excitation, which is gapped for all momentum transfers, ${\bf q}$.
The same linear, normalised colour intensity scale is used as in Fig.~\ref{fig:bbq-chiqomh}.
}}
\label{fig:angres}
\end{figure*}

In order to make quantitative predictions for experiment it is necessary to determine the hydrodynamic parameters appearing in $\Im m \chi_{\sf sc}^{\sf \alpha \alpha}({\bf p},\omega)$ [Eq.~\ref{eq:chi-sitecentred}].
All momentum transfers are now relabelled as ${\bf q}$, since this is commonly used.
We concentrate on the 1-magnon mode with dispersion $ \omega_{{\bf q},h}^{\sf xy,0}$ [Eq.~\ref{eq:omhydro}] and the 2-magnon mode with dispersion $\omega_{{\bf q},h}^{\sf Q,z}$ [Eq.~\ref{eq:omhydro}], as these will be the most experimentally visible.
%


 \begin{figure*}[h]
\centering
\includegraphics[width=0.85\textwidth]{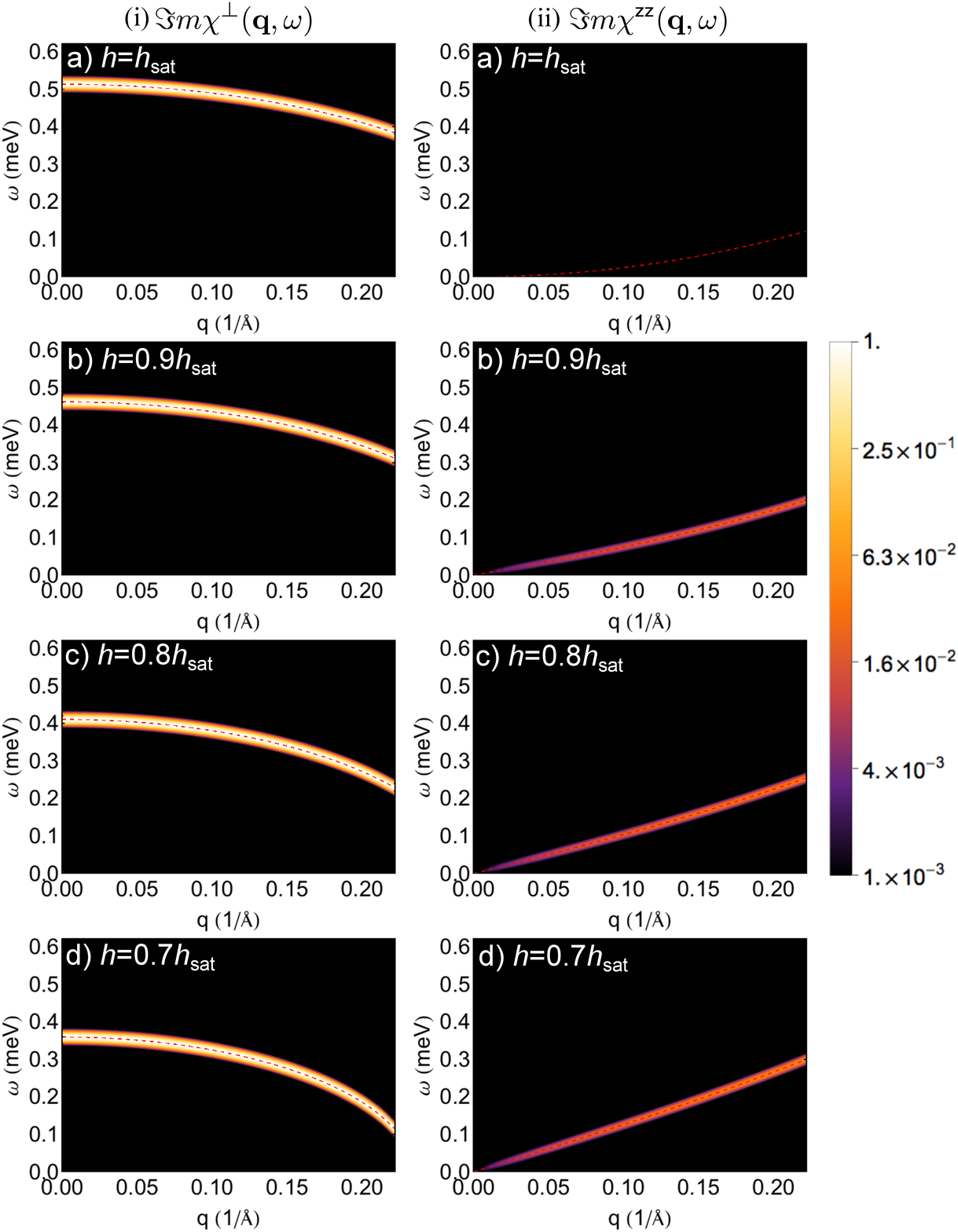}
\caption{\footnotesize{(Color online). 
Predictions for inelastic neutron scattering from a single-crystal sample 
of a spin-1/2 frustrated ferromagnet exhibiting two-sublattice, bond-centered
spin-nematic order in applied magnetic field.
(a)---(d) Predictions at small ${\bf q}$, for magnetic field $h$ ranging from the 
saturation value $h_{\sf sat}$ to $h = 0.7h_{\sf sat}$.
(i) The majority of spectral weight resides in a gapped spin-wave mode, 
visible in the transverse part of the dynamical susceptibility 
$\chi^{\perp}({\bf q},\omega)$.
(ii) The onset of spin-nematic order for $h < h_{\sf sat}$ is heralded by the 
emergence of a ghostly, linearly-dispersing Goldstone mode in the longitudinal 
susceptibility $\chi^{zz}({\bf q},\omega)$.
Predictions for $\chi^{\alpha\beta}({\bf q},\omega)$ 
were calculated for a material described by the \mbox{spin-1/2} $J_1$--$J_2$ 
model $\mathcal{H}^{\sf S=1/2}_{\sf J_1-J_2}$~[Eq.~\ref{eq:HJ1J2}],  
as described in Section~\ref{sec:neutrons} of this article, for parameters $J_1=-3.6\ \text{K}$ 
and $J_2=3.2\ \text{K}$ relevant to BaCdVO(PO$_4$)$_2$ \cite{nath08}.
All predictions have been convoluted with a gaussian of standard deviation $ 0.006\ \text{meV}$ 
to mimic experimental resolution.
Equivalent results for a powder sample are shown in Fig.~\ref{fig:chih0_8angint}.
}}
\label{fig:chiJ1J2}
\end{figure*}


Comparing $\omega_{{\bf q},h}^{\sf Q,z}$ [Eq.~\ref{eq:omhydro}] to $\omega^{2-m}_{{\bf q},h}$ [Eq.~\ref{eq:omega-2mag} and Eq.~\ref{eq:omegak2M-nem}] and using the coefficient of $\mathcal{S}_{\sf eff}$ [Eq.~\ref{eq:Seff2}] to set $\rho^{\sf Q,z}_h$, one finds for the 2-magnon mode,
 \begin{align}
\sigma_{h_{\sf sat}} &= 
\frac{1}{2m^{(2)}} \nonumber \\
(v_h^{\sf Q,z})^2 &= 
\frac{\mu_2(h)}{m^{(2)}} =
 \frac{2h_{\sf sat}}{m^{(2)}} \left( 1 - \frac{h}{h_{\sf sat}} \right) \nonumber \\
\rho^{\sf Q,z}_h &=
\frac{\mu_2(h)}{\Gamma^{(2)} m^{(2)}} =
\frac{2h_{\sf sat}}{\Gamma^{(2)} m^{(2)}} \left( 1 - \frac{h}{h_{\sf sat}} \right) \nonumber \\
\chi_h^{\sf Q,z} &= \frac{\rho^{\sf Q,z}_h}{(v_h^{\sf Q,z})^2} = \frac{1}{\Gamma^{(2)}},
\label{eq:hydro-BaCd}
\end{align}
 where $h_{\sf sat}= h_{\sf c2}$.
For the 1-magnon mode one can compare $ \omega_{{\bf q},h}^{\sf xy,0}$ [Eq.~\ref{eq:omhydro}] to $\omega^{1-m}_{{\bf k},h}$ [Eq.~\ref{eq:omkh1-m}] to find,
 \begin{align}
\Delta_{h}^{\sf xy,0} &= h \nonumber \\
(v_{h_{\sf sat}}^{\sf xy,0})^2 &=  -\frac{J_1+2J_2}{2}.
\end{align}
For $J_2/|J_1|=3.2/3.6 \approx 0.89$, one can use the results of Section~\ref{sec:parametrisation} to show,
\begin{align}
\frac{h_{\sf c1}}{|J_1|} &= 1.56, \quad
\frac{\Delta_{\sf m}}{|J_1|} =0.196, \nonumber \\
\frac{h_{\sf c2}}{|J_1|} &= 1.65, \quad
m^{(2)} |J_1|  =1.27.
\end{align}
Furthermore, from fitting the experimentally measured magnetisation curve for BaCdVO(PO$_4$)$_2$\cite{nath08} with \mbox{$M(h\approx h_{\sf sat})$} [Eq.~\ref{eq:Mh}], one can estimate,
 \begin{align}
\frac{\Gamma^{(2)}}{|J_1|} \approx 10.
\end{align}


For $h\geq h_{\sf sat}$ it is possible to exactly calculate both $\Im m\chi_{\sf sat}^{\sf \alpha\alpha}(\omega, {\bf q})$ [see Eq.~\ref{eq:chisat}] and the 2-magnon dispersion relation, $\omega^{2-m}_{{\bf q},h}$ [Eq.~\ref{eq:omega-2mag}].
In Fig.~\ref{fig:chiJ1J2sat} we show predictions for inelastic neutron scattering at $h=1.23h_{\sf sat}$ and $h=h_{\sf sat}$.
The only signal is a sharp and uniformly intense band of 1-magnon excitations. 
Also shown in Fig.~\ref{fig:chiJ1J2sat} is the 2-magnon dispersion, which is gapped for $h> h_{\sf sat}$, and softens at ${\bf q}=0$ for $h=h_{\sf sat}$, but is invisible to inelastic neutron scattering experiments.
In Fig.~\ref{fig:angres} these predictions are integrated over $4\pi$ of solid angle in order to mimic inelastic neutron scattering from a powder sample.

 
For $h<h_{\sf sat}$ we calculate predictions for inelastic neutron scattering using Eq.~\ref{eq:chi-sitecentred}.
These predictions are shown in Fig.~\ref{fig:chiJ1J2} at a range of magnetic field values and in the vicinity of ${\bf q}=0$.
%


As field is reduced below $h=h_{\sf sat}$ intensity appears in the Goldstone mode excitation, $\omega_{{\bf q},h}^{\sf Q,z}$ [Eq.~\ref{eq:omhydro}], for ${\bf q} \neq 0$.
For small ${\bf q}$ and fixed $h$, the intensity of this mode grows linearly with ${\bf q}$, as can be seen from Eq.~\ref{eq:chi-sitecentred}.
For fixed, small ${\bf q}$ and small $h_{\sf sat}-h$ the intensity grows as $\sqrt{h_{\sf sat}-h}$.
The velocity of this mode goes as $v_h^{\sf Q,z} \propto \sqrt{h_{\sf sat}-h}$ [Eq.~\ref{eq:hydro-BaCd}], and therefore the dispersion becomes steeper as field is reduced.


The 1-magnon excitation has a gap $\Delta_{h}^{\sf xy,0} = h$ that slowly reduces with lowering field.
To a first approximation the velocity is constant, and the intensity in this mode is uniform over ${\bf q}$ and does not vary with $h$.


The relative intensity of the two modes can be estimated from,
\begin{align}
\frac{I_{\sf Q,z}({\bf q},h)}{I_{\sf xy,0}({\bf q},h)} \approx \frac{\rho^{\sf Q,z}_h {\bf q}^2}{ \omega_{{\bf q},h}^{\sf Q,z}},
\end{align}
where $I_{\sf Q,z}$ is the intensity of the Goldstone mode and $I_{\sf xy,0}$ the intensity of the gapped 1-magnon mode.
%


In Fig.~\ref{fig:chih0_8angint} we show predictions for inelastic neutron scattering from a powder sample, calculated by taking the predictions shown in Fig.~\ref{fig:chiJ1J2} and averaging over $4\pi$ of solid angle. 
We show predictions for $h=0.8h_{\sf sat}$.
Also shown is a constant-${\bf q}$ cut at ${\bf q}=0.18$\AA$^{-1}$, showing the relative intensity of the two modes. 
At these values of ${\bf q}$ and $h$, the peak intensity in the Goldstone mode is about $3\%$ of the peak intensity in the 1-magnon mode. 


There will also be a small contribution to the inelastic scattering from the 2-magnon continuum [see Fig.~\ref{fig:ContBM_plot}].
This is spread over a large region of ${\bf q}$ and $\omega$ space, and at leading order the intensity grows as $h_{\sf sat}-h$ as the field is reduced.
Thus the contribution to the scattering will be considerably smaller than from the Goldstone mode excitation, which is sharp and has an intensity growing as $\sqrt{h_{\sf sat}-h}$, and can be safely ignored.


While the predictions shown in Fig.~\ref{fig:chiJ1J2sat}, Fig.~\ref{fig:chiJ1J2} and Fig.~\ref{fig:chih0_8angint} are specific to BaCdVO(PO$_4$)$_2$, a very similar analysis can be made for any compound described by $\mathcal{H}^{\sf S=1/2}_{\sf J_1-J_2}$ [Eq.~\ref{eq:HJ1J2}].
In fact, $\mathcal{L}_{\sf hyd}({\bf q},\omega)$ [Eq.~\ref{eq:Lhydro}]  can be applied to any system with a partially polarised, 2-sublattice spin-nematic order parameter.


\section{Discussion and conclusions}
\label{sec:conclusion}


In this article we have explored how inelastic neutron scattering can be used to probe for 
the existence of a spin-nematic state in applied magnetic field.
Following the philosophy detailed in Refs.~[\onlinecite{smerald13,smerald-thesis,smerald-arXiv}], 
we suggest that a good way to recognise this state experimentally is via the excitation spectrum, 
since the ground state is essentially invisible to common probes of magnetism.
To this end, we have developed a general theory of the magnetic excitations 
of a two-sublattice, bond-centered spin-nematic state in applied magnetic field.
We parameterise this theory from the microscopic model believed to describe the 
spin-1/2 frustrated-ferromagnet BaCdVO(PO$_4$)$_2$ [\onlinecite{nath08}], a promising 
candidate for spin-nematic order, and make predictions for inelastic scattering of neutrons 
from this material.
We also introduce an effective spin-1 model supporting the same form of 
spin-nematic order, and use it to explore the evolution of magnetic excitations
for a wide range of magnetic field.
The main experimental predictions of this article are summarised in Fig.~\ref{fig:chih0_8angint} 
and Fig.~\ref{fig:chiJ1J2}.


The starting point was to first derive a phenomenological theory of a 2-sublattice, partially-polarised spin-nematic state.
This involved constructing a spin-1 bilinear-biquadratic model [$\mathcal{H}^{\sf S=1}_{\sf bbq}[{\bf S}]$,~Eq.~\ref{eq:H-BBQ}] with both the desired symmetries and a spin-nematic ground state.
We note that if spin-1 compounds with large biquadratic coupling can be synthesised, this model may become experimentally relevant in its own right. 
Alternatively, it could be realised in molecular condensates of cold atoms.
In this article, the spin-1 model served as a guide to the derivation of a continuum field theory description of the long-wavelength excitations of the spin-nematic state [$\mathcal{L}_{\sf hyd}({\bf q},\omega)$, Eq.~\ref{eq:Lhydro}].
We found four low-energy modes, including a Goldstone mode associated with the breaking of {\sf U(1)} symmetry and predominantly describing quadrupole fluctuations of the order parameter, and three gapped modes describing mixed spin and quadrupole fluctuations.

One of the most experimentally promising places to search for the spin-nematic state is in square lattice, spin-1/2 frustrated ferromagnets described by $\mathcal{H}^{\sf S=1/2}_{\sf J_1-J_2}$ [Eq.~\ref{eq:HJ1J2}].
Close to saturation the spin-nematic state is expected to be realised for a large range of $J_2/J_1$ values.
In spin-1/2 nematic states the quadrupolar order parameter lives on the bonds of the lattice and not on the sites.
The continuum theory thus describes effective bond-centred spin fluctuations, and in order to accurately describe experiments, it was necessary to make a mapping onto the site-centred lattice.
After this procedure it became clear that the experimental response is dominated by only 2 modes, the Goldstone mode and a gapped mode that can be identified with single-magnon condensation out of the fully-saturated state.

The continuum theory contains a number of hydrodynamic parameters that have to be calculated from microscopic considerations.
In order to do this we considered $\mathcal{H}^{\sf S=1/2}_{\sf J_1-J_2}$ [Eq.~\ref{eq:HJ1J2}] close to the saturation magnetic field, and used exact diagrammatic calculations to determine the saturation magnetic field, assuming condensation of first single magnons and then bound-magnon pairs.
For a wide range of $J_2/J_1$ values the bound-magnon pairs condense at higher field than single-magnon excitations, forming a spin-nematic state.
Using similar diagrammatic calculations, the continuum theory was fully parametrised.

This allowed us to make predictions for inelastic neutron scattering experiments, focusing on the material BaCdVO(PO$_4$)$_2$.
%
%
In Fig.~\ref{fig:chih0_8angint} and Fig.~\ref{fig:chiJ1J2} we show predictions for small momentum transfers ${\bf q}$ and fields $h \lesssim h_{\sf sat}$, where the dominant feature in the spectrum is the gapped 1-magnon excitation band.
However, we showed that there is also spectral weight in the Goldstone mode excitation and this grows with increasing ${\bf q}$ and decreasing $h$.
Experimental detection of this excitation would be strong evidence for the existence of a spin-nematic state.

The predictions we make for BaCdVO(PO$_4$)$_2$ are quantitative, and this allows one to determine whether it is realistic to expect to see the Goldstone mode experimentally.
The saturation field has been measured as $h_{\sf sat}\approx 4.2$T, and therefore the full field range $0<h<h_{\sf sat}$ can be accessed in inelastic neutron scattering experiments.
Assuming this material is well described by $\mathcal{H}^{\sf S=1/2}_{\sf J_1-J_2}$ [Eq.~\ref{eq:HJ1J2}], one would expect a phase transition at intermediate field between the spin-nematic state (higher field) and a canted antiferromagnet (lower field).
Since the frustration parameter $J_2/|J_1|\approx 0.9$ is relatively close to the highly frustrated point $J_2/|J_1|\approx 0.5$, one would expect that the spin-nematic state would be the ground state over a sizeable field range.
The lower the magnetic field in which the spin-nematic can be measured, the more intense the scattering from the Goldstone mode.
Taking a relatively conservative value of $h=0.8h_{\sf sat}$ [see Fig.~\ref{fig:chih0_8angint}] one can see that an energy resolution of better than about 0.1meV is needed at a momentum transfer of about $0.15-0.2$\AA$^{-1}$ --  achievable values in neutron scattering experiments. 
The intensity of the Goldstone mode at these values of field and momentum transfer is expected to be about 3\% of the 1-magnon excitation.
This is comparable to measuring an ordered moment of about $0.2\mu_{\sf B}$ -- and this is possible experimentally.

Finally we would like to emphasise the generality of these results for understanding spin-nematic phases, and end with the hope that the ghostly Goldstone mode will be revealed through the machine of inelastic neutron scattering.


{\it Acknowledgments.}   
We thank Tsutomu Momoi, Karlo Penc, Markos Skoulatos and Burkhard Schmidt for useful discussions.
H.T.U is grateful for support from JSPS KAKENHI Grant No. 26800209.
A. S. acknowledges the Swiss National Science Foundation and its SINERGIA network ``Mott physics beyond the Heisenberg model'' for financial support.
This work was supported by the Okinawa Institute for Science and Technology
Graduate University.


\appendix



\section{Continuum theory of the 2-sublattice antiferroquadrupolar state at h=0}
\label{App:FT-h=0}


At $h=0$ the continuum field theory developed in Section~\ref{sec:fieldtheory} for the 2-sublattice antiferroquadrupolar (AFQ) spin-nematic state can be considerably simplified.
This allows comparison to lattice gauge theory calculations used to understand the $h=0$ spin-nematic state in $\mathcal{H}^{\sf S=1/2}_{\sf J_1-J_2}$ [Eq.~\ref{eq:HJ1J2}] \cite{shindou09,shindou11,shindou13}.

One simplification arises from the increased symmetry at $h=0$.
In the case $h\neq 0$ the symmetry of $\mathcal{S}_{\sf 2SL}$ [Eq.~\ref{eq:S2SL}] is {\sf U(1)}, and breaking this symmetry via the formation of an AFQ state results in a single Goldstone mode, as shown in Fig.~\ref{fig:FTspin1-chiqomh}b-e.
For $h=0$, the symmetry of the action is increased to {\sf SU(2)}. 
As a result, there are three Goldstone mode excitations, as can be seen in Fig.~\ref{fig:FTspin1-chiqomh}f)
There is a degenerate pair of Goldstone modes associated with rotations of the quadrupolar order parameter out of the ordering plane (Fig.~\ref{fig:FTspin1-chiqomh}(i)f), and a third associated with rotations within the ordering plane (Fig.~\ref{fig:FTspin1-chiqomh}(ii)f).
There is also a gapped mode that can be interpreted as a dynamical spin-density wave excitation.

A second simplification arises from a natural division at $h=0$ of conjugate pairs of fields into those associated with high- and low-energy fluctuations.
This is not the case as $h\to h_{\sf sat}$.
At $h=0$, fields associated with high-energy fluctuations can be eliminated by a Gaussian integral.
The resulting continuum theory is based on an {\sf SU(3)} generalisation of the non-linear sigma model (nl$\sigma$m).
A similar treatment of the 3-sublattice AFQ state on the triangular lattice was presented in Refs.~[\onlinecite{smerald13,smerald-thesis}], and the development in this Appendix closely follows these references.

\begin{figure}[t]
\centering
\includegraphics[width=0.45\textwidth]{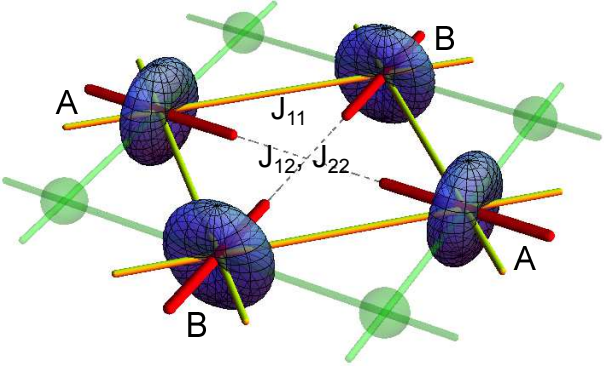}
\caption{\footnotesize{(Color online). 
2-sublattice antiferroquadrupolar (AFQ) spin-nematic state on the site- and bond-centred lattices at $h=0$.
Green spheres represent spin-1/2 degrees of freedom at the vertices of a square lattice (site-centred lattice).
The bond-centred, nematic order parameter is represented by red cylinders, and the associated distribution of spin fluctuations by a blue surface.
A bond-centred lattice is introduced and shown in yellow.
Long wavelength fluctuations of the order parameter are described by, $\mathcal{S}_{\sf nl\sigma m}[{\bf U}] $ [Eq.~\ref{eq:S-of-U-square}].
}}
\label{fig:Spin1vsSpin12h0}
\end{figure}

In order to demonstrate the validity of the approach, we first derive the continuum theory from a simple spin-1 model, $\mathcal{H}^{\sf S=1}_{\sf bbq}[{\bf S}]$ [Eq.~\ref{eq:H-BBQ}].
This is useful, since one can check that the resulting action reproduces the results of flavour-wave theory [see Section~\ref{sec:flvwave}] at long wavelength.
Using the results of Section~\ref{sec:spinmapping}, the continuum model can be mapped onto the site-centred lattice.
Thus we can describe the low-energy fluctuations of the spin-nematic state found to exist in $\mathcal{H}^{\sf S=1/2}_{\sf J_1-J_2}$ [Eq.~\ref{eq:HJ1J2}].

Previous approaches to understanding the spin-nematic state at $h=0$ in $\mathcal{H}^{\sf S=1/2}_{\sf J_1-J_2}$ [Eq.~\ref{eq:HJ1J2}] reformulate the problem in terms of a lattice gauge theory, and solve this using a large-N mean-field approach\cite{shindou09,shindou11,shindou13}.
This gives, in principle, all the excitations of the spin-nematic state.
As such, it is interesting to compare the $h=0$ field theory derived in this Appendix with the lattice gauge theory calculations.


\subsection{Deriving the action}


Before embarking on the calculation, we briefly summarise the mains steps,  following the same logic as in Refs.~[\onlinecite{smerald13,smerald-thesis}].
First, we note that in $\mathcal{H}^{\sf S=1}_{\sf bbq}[{\bf S}]$ [Eq.~\ref{eq:H-BBQ}] there is an {\sf SU(3)}-symmetric point at $J_{22}=0$.
When deriving a continuum theory, we consider small, but non-zero $J_{22}$, and make an expansion around the high-symmetry point.

The starting point of the calculation is the simplest subunit of the lattice, the square plaquette shown in Fig.~\ref{fig:Spin1vsSpin12h0}.
The spin-1 degrees of freedom on this plaquette are described using ${\bf d}$-vectors [Eq.~\ref{eq:dvecs}], and the energy is given by $\mathcal{H}^{\sf S=1}_{\sf bbq}[{\bf d}]$ [Eq.~\ref{eq:Hbbq-d}].
The ${\bf d}$-vectors obey a length constraint ${\bf d}\cdot \bar{\bf d}=1$ and the phase is set using ${\bf d}^2=\bar{\bf d}^2$.
Energy is minimised on a plaquette by selecting real ${\bf d}$ vectors, with orthogonal alignment between 1st neighbours and parallel alignment between 2nd neighbours.
A set of matrices can be defined that act on the four ${\bf d}$-vectors, and allow any configuration to be reached.
When these act on a ground-state configuration of the plaquette of ${\bf d}$-vectors, some of the matrices leave the energy invariant, others lead to an energy increase proportional to $J_{22}$, while still more lead to an energy increase proportional to $J_{11}$, $J_{12}$ or a combination of the two.
The first two sets of matrices will become the low-energy modes in the continuum theory, while the third set are considered high-energy modes and are eliminated by a Gaussian integral.

The square plaquette defines a basic unit from which to build a square lattice, and continuum fields are defined at the centre of plaquettes.
Assuming that the system has at least local 2-sublattice AFQ order,  the plaquettes can be stitched together and an action derived in terms of the continuum fields.
This action also includes a dynamical term, arising from the quantum mechanical overlap of nearby director configurations.

One choice for the ground state of a plaquette is given by,
\begin{align}
{\bf d}_{\sf A}^{\sf gs} = (1,0,0), \quad 
{\bf d}_{\sf B}^{\sf gs} = (0,1,0),
\end{align}
where {\sf A} and {\sf B} label the two sublattices [see Fig.~\ref{fig:Spin1vsSpin12h0}].
Including fluctuations, and assuming the plaquette is close to a ground state configuration, the ${\bf d}$ vectors can be approximated by,
\begin{align}
{\bf d}_{\sf A} &\approx {\bf U}(\boldsymbol{\phi})
\left(
\begin{array}{c}
1-(l_1^{\sf z})^2/2-(l_2^{\sf z})^2/2-(v^{\sf z}_{\sf A})^2/2 \\
l_1^{\sf z}-i l_2^{\sf z} \\
i v^{\sf z}_{\sf A}
\end{array}
\right),
\nonumber \\
{\bf d}_{\sf B}  &\approx {\bf U}(\boldsymbol{\phi})
\left(
\begin{array}{c}
l_1^{\sf z}+i l_2^{\sf z}  \\
1-(l_1^{\sf z})^2/2-(l_2^{\sf z})^2/2-(v^{\sf z}_{\sf B})^2/2 \\
i v^{\sf z}_{\sf B}
\end{array}
\right).
\label{eq:dexpansion}
\end{align}
Here,
\begin{align}
{\bf U}(\boldsymbol{\phi})=\exp\left[{i\sum_{i=1}^4\lambda_i \phi_i}\right],
\end{align}
is a unitary matrix describing low-energy fluctuations in terms of a set of parameters $\boldsymbol{\phi}=(\phi_1,\phi_2,\phi_3,\phi_4)$ and,
 \begin{align}
\lambda_1&=
\left(
\begin{array}{ccc}
0 &  -i & 0  \\
i &0  & 0 \\
0 & 0 &0
\end{array}
\right)
\quad
\lambda_2=
\left(
\begin{array}{ccc}
0 & 0 &i \\
0 & 0 & 0 \\
-i & 0 &0  
\end{array}
\right)
\nonumber \\
\lambda_3&=
\left(
\begin{array}{ccc}
0 & 0 & 0  \\
0 & 0  & -i \\
0 & i &0
\end{array}
\right)
\quad
\lambda_4=
\left(
\begin{array}{ccc}
0 & 1 & 0  \\
1 &0 & 0 \\
0& 0 &0
\end{array}
\right),
\label{eq:generators}
\end{align}
are the relevant subset of {\sf SU(3)} generators.
The parameters $l_1^{\sf z}$, $l_2^{\sf z}$, $v^{\sf z}_{\sf A}$ and $v^{\sf z}_{\sf B}$ describe small cantings of the plaquette configuration away from the low-energy subspace, and have been included in Eq.~\ref{eq:dexpansion} at quadratic order.

Plaquettes are stitched together to form a lattice by promoting the parameters $\boldsymbol{\phi}$, $l_1^{\sf z}$, $l_2^{\sf z}$, $v^{\sf z}_{\sf A}$ and $v^{\sf z}_{\sf B}$ to fields, and these are defined at the centres of the plaquettes.
These fields are allowed to fluctuate in time and space, and the continuum limit is taken under the assumption that the plaquette configurations only vary significantly over long lengthscales.

The action is given by,
\begin{align}
\mathcal{S}_{\sf nl\sigma m}=  \frac{1}{4}\int d\tau d^2r 
\left[ \mathcal{L}^{\sf kin}_{\sf nl\sigma m} +\mathcal{L}^{\sf H}_{\sf nl\sigma m} \right],
\label{eq:S2SL}
\end{align}
where,
 \begin{align}
\mathcal{L}^{\sf kin}_{\sf nl\sigma m} = 
2\bar{{\bf d}}_{\sf A}.\partial_\tau {\bf d}_{\sf A} 
+2 \bar{{\bf d}}_{\sf B}.\partial_\tau {\bf d}_{\sf B},
\end{align}
and,
 \begin{align}
\mathcal{L}^{\sf H}_{\sf nl\sigma m} = 
\langle \mathcal{H} \rangle_{\sf plaq},
\end{align}
where $\langle \mathcal{H} \rangle_{\sf plaq}$ refers to the Hamiltonian on a single plaquette.
After substituting in Eq.~\ref{eq:dexpansion} for the ${\bf d}$ vectors, expanding to quadratic order in the high-energy canting fields, making a gradient expansion around the plaquette centres and rewriting,
\begin{align}
 {\bf U}({\bf r},\tau)= \left(
\begin{array}{ccc}
n_{\sf A}^{\sf x}({\bf r},\tau) & n_{\sf B}^{\sf x}({\bf r},\tau) & n_{\sf C}^{\sf x}({\bf r},\tau)  \\
n_{\sf A}^{\sf y}({\bf r},\tau) & n_{\sf B}^{\sf y}({\bf r},\tau)  & n_{\sf C}^{\sf y}({\bf r},\tau) \\
n_{\sf A}^{\sf z}({\bf r},\tau) & n_{\sf B}^{\sf z}({\bf r},\tau) & n_{\sf C}^{\sf z}({\bf r},\tau)
\end{array}
\right),
\label{eq:Urt}
\end{align}
one arrives at,
\begin{align}
&\mathcal{L}^{\sf kin}_{\sf nl\sigma m} = 
2\bar{{\bf n}}_{\sf A}\partial_\tau{\bf n}_{\sf A} 
+2\bar{{\bf n}}_{\sf B}\partial_\tau{\bf n}_{\sf B}
 +4l_1^{\sf z} (\bar{{\bf n}}_{\sf A}\partial_\tau{\bf n}_{\sf B} -{\bf n}_{\sf A}\partial_\tau\bar{{\bf n}}_{\sf B}) \nonumber \\
&  -4il_2^{\sf z} (\bar{{\bf n}}_{\sf A}\partial_\tau{\bf n}_{\sf B} +{\bf n}_{\sf A}\partial_\tau\bar{{\bf n}}_{\sf B}) 
    +2iv^{\sf z}_{\sf A} (\bar{{\bf n}}_{\sf C}\partial_\tau{\bf n}_{\sf A} +{\bf n}_{\sf C}\partial_\tau\bar{{\bf n}}_{\sf A}) \nonumber \\
&    +2iv^{\sf z}_{\sf B} (\bar{{\bf n}}_{\sf B}\partial_\tau{\bf n}_{\sf C} +{\bf n}_{\sf B}\partial_\tau\bar{{\bf n}}_{\sf C}),
\end{align}
and,
\begin{align}
&\mathcal{L}_{\sf nl\sigma m} = 
32J_{11} \left( (l_1^{\sf z})^2 
+(l_2^{\sf z})^2 \right) 
+32J_{22} (l_2^{\sf z})^2 \nonumber \\
&+16J_{22} \left( (v^{\sf z}_{\sf A})^2 + (v^{\sf z}_{\sf B})^2 \right) 
 -4J_{22}\left( {\bf n}_{\sf A}^2 \bar{\bf n}_{\sf A}^2 +{\bf n}_{\sf B}^2 \bar{\bf n}_{\sf B}^2 \right)
  \nonumber \\
&+ \sum_{\lambda ={\sf x,y}} \left\{
4J_{11}| \bar{{\bf n}}_{\sf A} \partial_\lambda {\bf n}_{\sf B} |^2 \right. \nonumber \\ 
&+2J_{22}\left[ (\partial_\lambda {\bf n}_{\sf A})^2+ (\partial_\lambda \bar{\bf n}_{\sf A})^2
+(\partial_\lambda {\bf n}_{\sf B})^2 + (\partial_\lambda \bar{\bf n}_{\sf B})^2  \right] \nonumber \\
&\left. +4J_{12} \left[ |\partial_\lambda {\bf n}_{\sf A}|^2+|\partial_\lambda {\bf n}_{\sf B}|^2
- | \bar{\bf n}_{\sf A} \partial_\lambda {\bf n}_{\sf A}|^2 - | \bar{\bf n}_{\sf B} \partial_\lambda {\bf n}_{\sf B}|^2 \right] \right\}.
\end{align}
The fields ${\bf n}_{\sf A}$ and ${\bf n}_{\sf B}$ inherit the length and phase constraints of the ${\bf d}$ vectors and are further constrained to be orthogonal, $\bar{\bf n}_{\sf A}\cdot {\bf n}_{\sf B}=0$.
The auxiliary field ${\bf n}_{\sf C} = {\bf n}_{\sf A} \times {\bf n}_{\sf B}$ is introduced as a convenient piece of book-keeping, and is not an independent degree of freedom.

The high-energy, canting fields can be eliminated by a Gaussian integral, giving,
\begin{align}
l_1^{\sf z}&= -\frac{1}{16J_{11}} (\bar{{\bf n}}_{\sf A}\partial_\tau{\bf n}_{\sf B} -{\bf n}_{\sf A}\partial_\tau\bar{{\bf n}}_{\sf B}), \nonumber \\
 l_2^{\sf z} &= \frac{i}{16(J_{11}+J_{22})}  (\bar{{\bf n}}_{\sf A}\partial_\tau{\bf n}_{\sf B} +{\bf n}_{\sf A}\partial_\tau\bar{{\bf n}}_{\sf B}) \nonumber \\
  v^{\sf z}_{\sf A}& = -\frac{i}{16J_{22}}  (\bar{{\bf n}}_{\sf C}\partial_\tau{\bf n}_{\sf A} +{\bf n}_{\sf C}\partial_\tau\bar{{\bf n}}_{\sf A} )
 \nonumber \\
v^{\sf z}_{\sf B} & = -\frac{i}{16J_{22}}  (\bar{{\bf n}}_{\sf B}\partial_\tau{\bf n}_{\sf C} +{\bf n}_{\sf B}\partial_\tau\bar{{\bf n}}_{\sf C}).
\end{align}
It follows that the action is,
\begin{align}
&\mathcal{S}_{\sf nl\sigma m}[{\bf n}_{\sf A},{\bf n}_{\sf B},\bar{\bf n}_{\sf A},\bar{\bf n}_{\sf B}] \approx \frac{1}{4}\int_0^\beta d\tau \int d^2r \{ \nonumber \\
&2\bar{{\bf n}}_{\sf A}\partial_\tau{\bf n}_{\sf A} 
+2\bar{{\bf n}}_{\sf B}\partial_\tau{\bf n}_{\sf B}
 -\frac{1}{8J_{11}} (\bar{{\bf n}}_{\sf A}\partial_\tau{\bf n}_{\sf B} -{\bf n}_{\sf A}\partial_\tau\bar{{\bf n}}_{\sf B})^2\nonumber \\
&  +\frac{1}{8(J_{11}+J_{22})} (\bar{{\bf n}}_{\sf A}\partial_\tau{\bf n}_{\sf B} +{\bf n}_{\sf A}\partial_\tau\bar{{\bf n}}_{\sf B})^2 \nonumber \\
&  + \frac{1}{16J_{22}} (\bar{{\bf n}}_{\sf C}\partial_\tau{\bf n}_{\sf A} +{\bf n}_{\sf C}\partial_\tau\bar{{\bf n}}_{\sf A})^2 \nonumber \\
&    + \frac{1}{16J_{22}}(\bar{{\bf n}}_{\sf B}\partial_\tau{\bf n}_{\sf C} +{\bf n}_{\sf B}\partial_\tau\bar{{\bf n}}_{\sf C})^2
 -4J_{22}\left( {\bf n}_{\sf A}^2 \bar{\bf n}_{\sf A}^2 +{\bf n}_{\sf B}^2 \bar{\bf n}_{\sf B}^2 \right) \nonumber \\
 &+ \sum_{\lambda ={\sf x,y}} 
4J_{11}| \bar{{\bf n}}_{\sf A} \partial_\lambda {\bf n}_{\sf B} |^2 \nonumber \\
&+2J_{22}\left[ (\partial_\lambda {\bf n}_{\sf A})^2+ (\partial_\lambda \bar{\bf n}_{\sf A})^2
+(\partial_\lambda {\bf n}_{\sf B})^2 + (\partial_\lambda \bar{\bf n}_{\sf B})^2  \right] \nonumber \\
& +4J_{12} \left[ |\partial_\lambda {\bf n}_{\sf A}|^2+|\partial_\lambda {\bf n}_{\sf B}|^2
- | \bar{\bf n}_{\sf A} \partial_\lambda {\bf n}_{\sf A}|^2 - | \bar{\bf n}_{\sf B} \partial_\lambda {\bf n}_{\sf B}|^2 \right] \}.
\label{eq:action-n}
\end{align}
This action can be rewritten by reintroducing the matrix $ {\bf U}({\bf r},\tau)$ [Eq.~\ref{eq:Urt}] as,
\begin{align}
&\mathcal{S}_{\sf nl\sigma m}[{\bf U}] 
=  \frac{1}{16} \int_0^\beta d\tau \int d^2r \left\{ \right. \nonumber \\
&\quad 8\mathrm{Tr}[{\bf P}. {\bf U}\dg\cdot\partial_\tau {\bf U}]
-\chi^{\sf S,z}_0 (\Delta^{\sf S,z}_0)^2\left[ 2-\sum_{i=1}^2 \left| [{\bf U}^\mathrm{T}.{\bf U}]_{ii} \right|^2 \right]
\nonumber \\
& \quad 
+\mathrm{Tr}\left[ \boldsymbol{\Lambda}_{\chi}^{\sf Q}. 
\left( {\bf U}\dg\cdot\partial_\tau {\bf U} + {\bf U}^\mathrm{T} \cdot\partial_\tau \bar{\bf U} \right)\dg \right. \nonumber \\
&\qquad \qquad \times \left. \left( {\bf U}\dg\cdot\partial_\tau {\bf U} + {\bf U}^\mathrm{T} \cdot\partial_\tau \bar{\bf U} \right) 
\right]  \nonumber \\  
& \quad -\chi^{\sf S,z}_0 
\left( [{\bf U}\dg\cdot\partial_\tau {\bf U}]_{21} -[{\bf U}\dg\cdot\partial_\tau {\bf U}]_{12} \right)^2 
 \nonumber \\
 & \quad +\sum_{\sf \lambda = x,y}
\mathrm{Tr}\left[ 
\boldsymbol{\Lambda}_{\rho}^{\sf Q}. 
\left( {\bf U}\dg\cdot\partial_\lambda {\bf U} + {\bf U}^\mathrm{T} \cdot\partial_ \lambda \bar{\bf U} \right)\dg \right. \nonumber \\
&\qquad \qquad \qquad \times \left.
\left( {\bf U}\dg\cdot\partial_ \lambda {\bf U} + {\bf U}^\mathrm{T} \cdot\partial_ \lambda \bar{\bf U} \right) \right] \nonumber \\
& \quad \qquad  -\rho^{\sf S,z}_0
\left( [{\bf U}\dg\cdot\partial_ \lambda {\bf U}]_{21} -[{\bf U}\dg\cdot\partial_ \lambda {\bf U}]_{12} \right)^2,
\label{eq:S-of-U-square}
\end{align}
where,
\begin{align}
{\bf P} &=
\left(
\begin{array}{ccc}
1 & 0 & 0  \\
0 & 1 & 0 \\
0 & 0 & 0
\end{array}
\right) \nonumber \\
\boldsymbol{\Lambda}_{\chi}^{\sf Q} &=
\left(
\begin{array}{ccc}
\chi^{\sf Q, z}_0 & 0 & 0  \\
0 & \chi^{\sf Q,\sf z}_0  & 0 \\
0 & 0 & 2\chi^{\sf xy}_0-\chi^{\sf Q, z}_0
\end{array}
\right) \nonumber \\
\boldsymbol{\Lambda}_{\rho}^{\sf Q} &=
 \left(
\begin{array}{ccc}
\rho^{\sf Q, z}_0 & 0 & 0  \\
0 & \rho^{\sf Q, z}_0  & 0 \\
0 & 0 & 2\rho^{\sf Q xy}_0 -\rho^{\sf Q, z}_0
\end{array}
\right).
\end{align}
This formulation makes clear the {\sf SU(2)} symmetry of the order parameter.
In $\mathcal{S}_{\sf nl\sigma m}[{\bf U}] $ [Eq.~\ref{eq:S-of-U-square}] we have introduced a set of hydrodynamic parameters.
The relationship between these and the exchange parameters of $\mathcal{H}^{\sf S=1}_{\sf bbq}[{\bf S}]$ [Eq.~\ref{eq:H-BBQ}] is shown in Table~\ref{table:2-sublattice-dictionary}.
These can be compared to the hydrodynamic parameters presented in Table~\ref{tab:Jtohydroconversion}.
Using the relation $v^2=\rho/\chi$, one can see that these are equivalent at $h=0$, except for the velocity $(v^{\sf S,z}_0)^2$, which differs by a factor proportional to $J_{22}^2$, which is considered a small parameter in this Appendix.


\begin{table}
\begin{center}
  \begin{tabular}{| c | c | }
    \hline
\multirow{2}{*}{ $\mathcal{S}_{\sf nl\sigma m}[{\bf U}]$ [Eq.~\ref{eq:S-of-U-square}] }  &  
\multirow{2}{*}{ $\mathcal{H}^{\sf S=1}_{\sf bbq}[{\bf S}]$ [Eq.~\ref{eq:H-BBQ}] }    \\ 
&   \\ \hline
\multirow{2}{*}{$\chi^{\sf Q,z}_0$}  &  
\multirow{2}{*}{$[4(J_{11}+J_{22})]^{-1}$}    \\ 
&   \\ \hline
\multirow{2}{*}{$\chi^{\sf S,z}_0$}  &  
\multirow{2}{*}{$[4J_{11}]^{-1}$}    \\ 
&   \\ \hline
\multirow{2}{*}{$\chi^{\sf xy}_0$}  &  
\multirow{2}{*}{$[8J_{22}]^{-1}$}    \\ 
&   \\ \hline
\multirow{2}{*}{$\rho^{\sf Q,z}_0$}  &  
\multirow{2}{*}{$2 (J_{11}+2J_{12}+2J_{22})$}    \\ 
&   \\ \hline
\multirow{2}{*}{$\rho^{\sf S,z}_0$}  &  
\multirow{2}{*}{$2 (J_{11}+2J_{12}-2J_{22})$}    \\ 
&   \\ \hline
\multirow{2}{*}{$\rho^{\sf xy}_0$}  &  
\multirow{2}{*}{$2 (J_{11}+J_{22})$}    \\ 
&   \\ \hline
\multirow{2}{*}{$\Delta^{\sf S,z}_0$}  &  
\multirow{2}{*}{$8\sqrt{J_{11}J_{22}}$}    \\ 
&   \\ \hline
    \end{tabular}
\end{center} 
\caption{\footnotesize{
Dictionary for translating between the parameters of the 
continuum field theory for 2-sublattice AFQ order, 
$\mathcal{S}_{\sf nl\sigma m}[{\bf U}]$ [Eq.~\ref{eq:S-of-U-square}],
and the microscopic model 
$\mathcal{H}^{\sf S=1}_{\sf bbq}[{\bf S}]$ [Eq.~\ref{eq:H-BBQ}].
}}
\label{table:2-sublattice-dictionary}
\end{table}



\subsection{Linearising the action}


\begin{figure*}[t]
\centering
\includegraphics[width=0.99\textwidth]{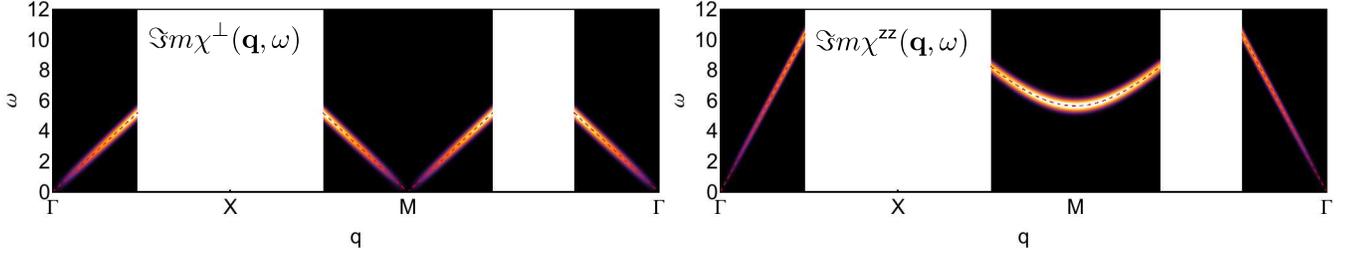}
\caption{\footnotesize{(Color online). 
Predictions for the imaginary part of the dynamical spin susceptibility for a 2-sublattice, spin-1 antiferroquadrupolar (AFQ) state in the absence of magnetic field.
$\Im m \chi^{\perp}_{\sf nl\sigma m}$ and $\Im m \chi^{\sf zz}_{\sf nl\sigma m}$ [Eq.~\ref{eq:chinlsm}] are calculated from $\mathcal{S}_{\sf nl\sigma m}[\boldsymbol{\phi}]$ [Eq.~\ref{eq:nlsm-lin}].
Hydrodynamic parameters are taken from Table~\ref{table:2-sublattice-dictionary} with $J_{11}=1$, $J_{12}=1$ and $J_{22}=0.5$.
Dashed red lines show $\omega^{\sf Q,z}_{{\bf q},0}$, $\omega^{\sf xy}_{{\bf q},0}$ and $\omega^{\sf S,z}_{{\bf q},0}$.
All predictions have been convoluted with a gaussian to mimic experimental resolution.
The circuit $\Gamma$-{\sf X}-{\sf M}-$\Gamma$ in the bond-centred Brillouin zone is shown in Fig.~\ref{fig:bz}.
The same linear, normalised colour intensity scale is used as in Fig.~\ref{fig:bbq-chiqomh}.
}}
\label{fig:appSpin1chi}
\end{figure*}

It is useful to linearise $\mathcal{S}_{\sf nl\sigma m}[{\bf U}]$ to further bring out the physical content of the action.
At linear order one can approximate,
\begin{align}
 {\bf U}({\bf r},\tau)  \approx \left(
\begin{array}{ccc}
1 & \phi_1+i\phi_4 & -\phi_2  \\
 -\phi_1+i\phi_4 & 1  & \phi_3 \\
\phi_2 & -\phi_3 & 1
\end{array}
\right),
\end{align}
and therefore,
\begin{align}
&\mathcal{S}_{\sf nl\sigma m}[\boldsymbol{\phi}] \approx \frac{1}{2} \int_0^\beta d\tau  \int d^2r \{ \nonumber \\
&\chi^{\sf Q,z}_0(\partial_\tau\phi_1)^2+  \sum_{\lambda ={\sf x,y}}  \rho^{\sf Q,z}_0 (\partial_\lambda\phi_1)^2  \nonumber \\
&+\chi^{\sf S,z}_0 (\partial_\tau\phi_4)^2 +  \sum_{\lambda ={\sf x,y}}  \rho^{\sf S,z}_0(\partial_\lambda\phi_4)^2+\chi^{\sf S,z}_0 (\Delta^{\sf S,z}_0)^2 \phi_4^2 \nonumber \\
&+ \chi^{\sf xy}_0 \left[(\partial_\tau\phi_2)^2+(\partial_\tau\phi_3)^2\right] +  \sum_{\lambda ={\sf x,y}} \rho^{\sf xy}_0 [(\partial_\lambda\phi_2)^2 +  (\partial_\lambda\phi_3)^2]
 \}.
 \label{eq:nlsm-lin}
\end{align}
From this one can calculate the dispersion relations of the 4 modes as,
\begin{align}
\omega^{\sf Q,z}_{{\bf k},0} &= \sqrt{ \frac{\rho^{\sf Q,z}_0 }{\chi^{\sf Q,z}_0 }} |{\bf k}|  \nonumber \\
 \omega^{\sf S,z}_{{\bf k},0} &= \sqrt{ (\Delta^{\sf S,z}_0)^2 + \frac{\rho^{\sf S,z}_0 }{\chi^{\sf S,z}_0 } |{\bf k}|^2} \nonumber \\
 \omega^{\sf xy}_{{\bf k},0} &=  \sqrt{ \frac{\rho^{\sf xy}_0 }{\chi^{\sf xy}_0 }} |{\bf k}|,
\label{eq:omega-FT}
\end{align}
where $ \omega^{\sf xy}_{{\bf k},0}$ is twofold degenerate in the 2-sublattice Brillouin zone.
This describes 3 Goldstone modes, $ \omega^{\sf xy}_{{\bf k},0}$ and $ \omega^{\sf Q,z}_{{\bf k},0}$, and 1 gapped mode, $ \omega^{\sf S,z}_{{\bf k},0}$.

The Goldstone mode $\omega^{\sf Q,z}_{{\bf k},0}$ corresponds to real rotations of the ${\bf d}$ vectors in the ordering plane.
This in turn results in rotations of the quadrupolar order parameter in the ordering plane.

The pair of Goldstone modes described by $\omega^{\sf xy}_{{\bf k},0}$ corresponds to real rotations of the ${\bf d}$ vectors out of the ordering plane.
This results in rotations of the quadrupolar order parameter out of the ordering plane.

The gapped mode $ \omega^{\sf S,z}_{{\bf k},0} $ corresponds to an imaginary rotation of the ${\bf d}$ vectors in the ordering plane.
This results in spin fluctuations perpendicular to the ordering plane, which are out of phase on the 2 sublattices and can be thought of as a dynamical spin-density wave.


\subsection{Imaginary part of the dynamical spin susceptibility}


Inelastic neutron scattering measures the imaginary part of the dynamical spin susceptibility, $\Im m \chi^{\alpha\beta}({\bf q}, \omega)$ [Eq.~\ref{eq:im-chi}].
This can be calculated from the linearised action, $\mathcal{S}_{\sf nl\sigma m}[\boldsymbol{\phi}]$ [Eq.~\ref{eq:nlsm-lin}].

\begin{figure*}[t]
\centering
\includegraphics[width=0.8\textwidth]{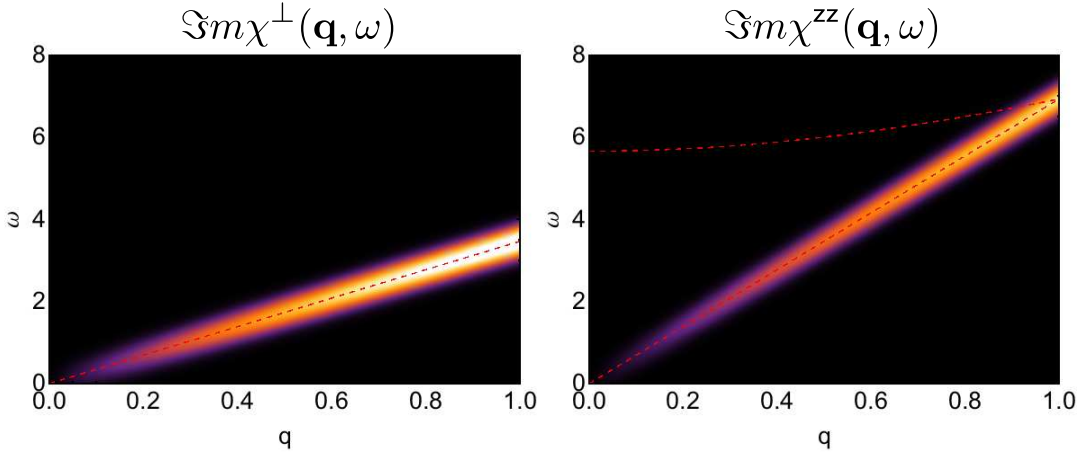}
\caption{\footnotesize{(Color online). 
Predictions for the imaginary part of the dynamical spin susceptibility for a 2-sublattice, 
spin-1/2 antiferroquadrupolar (AFQ) state in the absence of magnetic field.
The mapping described in Section~\ref{sec:spinmapping} is used to transform 
$\Im m \chi^{\perp}_{\sf nl\sigma m}$ and $\Im m \chi^{\sf zz}_{\sf nl\sigma m}$ 
[Eq.~\ref{eq:chinlsm}] into the site-centred Brillouin zone.
An arbitrary set of hydrodynamic parameters have been chosen as 
$(v^{\sf Q,z}_0)^2=48$, $(v^{\sf xy}_0)^2=12$, $(v^{\sf S,z}_0)^2=16$ 
and $(\Delta^{\sf S,z}_0)^2=32$.
Dashed red lines show $\omega^{\sf Q,z}_{{\bf q},0}$, $\omega^{\sf xy}_{{\bf q},0}$ 
and $\omega^{\sf S,z}_{{\bf q},0}$.
All predictions have been convoluted with a gaussian of $\text{FWHM} = xxx$
to mimic experimental resolution.
The same linear, normalised colour intensity scale is used as in Fig.~\ref{fig:bbq-chiqomh}.
The results in this Figure can be compared directly with the predictions 
of the lattice gauge theory given in Ref. \onlinecite{shindou09,shindou11,shindou13}.
}}
\label{fig:appSpinhalfchi}
\end{figure*}

In order to do this it is first necessary to determine how the spin moments are related to the quantum fields.
This can be achieved by dividing the ${\bf d}$-vectors into a real and imaginary part, ${\bf d} = {\bf u}+i{\bf v}$, and noticing that ${\bf S}=2{\bf u}\times {\bf v}$.
To lowest order in each of the spin components one can show\cite{smerald13,smerald-thesis},
\begin{align}
&{\bf S}_{\sf A}= \nonumber \\
&\left( \hspace{-0.5mm}
\begin{array}{c}
-2 \phi_2 \phi_4+2\chi^{\sf xy}_0 \phi_1 \partial_t \phi_2 + \chi^{\sf Q,z}_0 \phi_2 \partial_t \phi_1 +\chi^{\sf S,z}_0\chi^{\sf xy}_0   \partial_t \phi_2 \partial_t \phi_4 \\
2\chi^{\sf xy}_0  \partial_t \phi_2   \\
2\phi_4 -  \chi^{\sf Q,z}_0 \partial_t \phi_1
\end{array} \hspace{-0.5mm}
\right)
\nonumber \\
&{\bf S}_{\sf B}= \nonumber \\
&\left( \hspace{-0.5mm}
\begin{array}{c}
-2\chi^{\sf xy}_0  \partial_t \phi_3   \\
-2 \phi_3 \phi_4+2\chi^{\sf xy}_0 \phi_1 \partial_t \phi_3 - \chi^{\sf Q,z}_0 \phi_3 \partial_t \phi_1  -\chi^{\sf S,z}_0\chi^{\sf xy}_0   \partial_t \phi_3 \partial_t \phi_4 \\
-2\phi_4 -  \chi^{\sf Q,z}_0 \partial_t \phi_1
\end{array} \hspace{-0.5mm}
\right)
\label{eq:nlsm-spincomp}
\end{align}
It follows that,
\begin{align}
\Im m \chi^{\perp}_{\sf nl\sigma m}({\bf q},\omega) &= 
\Im m \chi^{\sf xx}_{\sf nl\sigma m}({\bf q},\omega) +\Im m \chi^{\sf yy}_{\sf nl\sigma m}({\bf q},\omega) = \nonumber \\
& \pi \chi^{\sf xy}_0 \omega^{\sf xy}_{{\bf q},0} \delta(\omega -\omega^{\sf xy}_{{\bf q},0} ) \nonumber \\
\Im m \chi^{\sf zz}_{\sf nl\sigma m}({\bf q},\omega)  &= 
\frac{\pi}{2} \chi^{\sf Q,z}_0 \omega^{\sf Q,z}_{{\bf q},0}  \delta(\omega -\omega^{\sf Q,z}_{{\bf q},0} ),
\end{align}
and,
\begin{align}
&\Im m \chi^{\perp}_{\sf nl\sigma m}({\bf q}_{\sf M}+{\bf q},\omega) = \nonumber \\
& \quad \Im m \chi^{\sf xx}_{\sf nl\sigma m}({\bf q}_{\sf M}+{\bf q},\omega) +\Im m \chi^{\sf yy}_{\sf nl\sigma m}({\bf q}_{\sf M}+{\bf q},\omega) = \nonumber \\
& \qquad \qquad \pi \chi^{\sf xy}_0 \omega^{\sf xy}_{{\bf q},0} \delta(\omega -\omega^{\sf xy}_{{\bf q},0} ) \nonumber \\
&\Im m \chi^{\sf zz}_{\sf nl\sigma m}({\bf q}_{\sf M}+{\bf q},\omega)  = 
 \frac{2\pi}{\chi^{\sf S,z}_0  \omega^{\sf S,z}_{{\bf q},0}} \delta(\omega - \omega^{\sf S,z}_{{\bf q},0}),
\label{eq:chinlsm}
\end{align}
where ${\bf q}\approx 0$.
This is shown in Fig.~\ref{fig:appSpin1chi}, and can be compared to Fig.~\ref{fig:FTspin1-chiqomh}f, where the physics can be seen to be qualitatively the same, despite the different values chosen for the parameters $J_{11}$, $J_{12}$ and $J_{22}$.

In order to study the bond-nematic phase found in $\mathcal{H}^{\sf S=1/2}_{\sf J_1-J_2}$ [Eq.~\ref{eq:HJ1J2}], the mapping described in Section~\ref{sec:spinmapping} can be used.
The field theory thus describes the dynamic spin susceptibility close to ${\bf q}=0$, and this is shown in Fig.~\ref{fig:appSpinhalfchi}.


\subsection{Comparison with lattice gauge theory}


Finally it is interesting to compare the continuum model developed in this Appendix with previous work studying the $h=0$ spin-nematic region of $\mathcal{H}^{\sf S=1/2}_{\sf J_1-J_2}$ [Eq.~\ref{eq:HJ1J2}] \cite{shindou09,shindou11,shindou13}.

In Ref.~[\onlinecite{shindou09}] a matrix-formed action for the gapless modes is written down [Eq.~46 of this reference]. 
Comparing this to Eq.~\ref{eq:S-of-U-square}, one can see that these actions have the same low-energy form if the generator $\lambda_4$ [Eq.~\ref{eq:generators}] is ignored.
The relationship between the hydrodynamic parameters of Eq.~\ref{eq:S-of-U-square} and Ref.~[\onlinecite{shindou09}] is given by,
\begin{align}
\chi^{\sf Q,z}_0=2c_2 &, \quad \chi^{\sf xy}_0=c_2, \quad \rho^{\sf Q,z}_0 =2c_3, \nonumber \\
\rho^{\sf xy}_0 &=c_1+c_3, \quad c_3=c_4.
\end{align}

Comparison can also be made to Ref.~[\onlinecite{shindou13}] in which the imaginary part of the dynamical spin susceptibility is calculated via a $1/N$ expansion scheme. 
At small ${\bf q}$ the field theory developed in this Appendix agrees well with this $1/N$ expansion scheme, which can be see by comparing Fig.~\ref{fig:appSpinhalfchi} with Fig.~9 and Fig.~10 of Ref.~[\onlinecite{shindou13}].
In both cases there are three Goldstone modes at ${\bf q}=0$, a degenerate pair (at low energy) that appear in $\Im m \chi^{\perp}$ and a third in the longitudinal channel $\Im m \chi^{\sf zz}$.
In Ref.~[\onlinecite{shindou13}], there also appear a number of gapped modes close to ${\bf q}=0$ without significant spectral weight, but it is not completely clear if one of these is equivalent to the dynamical spin density wave that appears in $\mathcal{S}_{\sf nl\sigma m}[{\bf U}]$ [Eq.~\ref{eq:S-of-U-square}].
Finally Ref.~[\onlinecite{shindou13}] finds a spinon continuum at high energies, and this cannot be captured by the low-energy continuum theory presented in this Appendix.


\bibliographystyle{apsrev4-1}
\bibliography{bibfile}

\begin{thebibliography}{57}%
\makeatletter
\providecommand \@ifxundefined [1]{%
 \@ifx{#1\undefined}
}%
\providecommand \@ifnum [1]{%
 \ifnum #1\expandafter \@firstoftwo
 \else \expandafter \@secondoftwo
 \fi
}%
\providecommand \@ifx [1]{%
 \ifx #1\expandafter \@firstoftwo
 \else \expandafter \@secondoftwo
 \fi
}%
\providecommand \natexlab [1]{#1}%
\providecommand \enquote  [1]{``#1''}%
\providecommand \bibnamefont  [1]{#1}%
\providecommand \bibfnamefont [1]{#1}%
\providecommand \citenamefont [1]{#1}%
\providecommand \href@noop [0]{\@secondoftwo}%
\providecommand \href [0]{\begingroup \@sanitize@url \@href}%
\providecommand \@href[1]{\@@startlink{#1}\@@href}%
\providecommand \@@href[1]{\endgroup#1\@@endlink}%
\providecommand \@sanitize@url [0]{\catcode `\\12\catcode `\$12\catcode
  `\&12\catcode `\#12\catcode `\^12\catcode `\_12\catcode `\%12\relax}%
\providecommand \@@startlink[1]{}%
\providecommand \@@endlink[0]{}%
\providecommand \url  [0]{\begingroup\@sanitize@url \@url }%
\providecommand \@url [1]{\endgroup\@href {#1}{\urlprefix }}%
\providecommand \urlprefix  [0]{URL }%
\providecommand \Eprint [0]{\href }%
\providecommand \doibase [0]{http://dx.doi.org/}%
\providecommand \selectlanguage [0]{\@gobble}%
\providecommand \bibinfo  [0]{\@secondoftwo}%
\providecommand \bibfield  [0]{\@secondoftwo}%
\providecommand \translation [1]{[#1]}%
\providecommand \BibitemOpen [0]{}%
\providecommand \bibitemStop [0]{}%
\providecommand \bibitemNoStop [0]{.\EOS\space}%
\providecommand \EOS [0]{\spacefactor3000\relax}%
\providecommand \BibitemShut  [1]{\csname bibitem#1\endcsname}%
\let\auto@bib@innerbib\@empty
\bibitem [{\citenamefont {Blume}\ and\ \citenamefont {Hsieh}(1969)}]{blume69}%
  \BibitemOpen
  \bibfield  {author} {\bibinfo {author} {\bibfnamefont {M.}~\bibnamefont
  {Blume}}\ and\ \bibinfo {author} {\bibfnamefont {Y.}~\bibnamefont {Hsieh}},\
  }\href@noop {} {\bibfield  {journal} {\bibinfo  {journal} {Journal of Applied
  Physics}\ }\textbf {\bibinfo {volume} {40}},\ \bibinfo {pages} {1249}
  (\bibinfo {year} {1969})}\BibitemShut {NoStop}%
\bibitem [{\citenamefont {Chen}\ and\ \citenamefont {Levy}(1971)}]{chen71}%
  \BibitemOpen
  \bibfield  {author} {\bibinfo {author} {\bibfnamefont {H.~H.}\ \bibnamefont
  {Chen}}\ and\ \bibinfo {author} {\bibfnamefont {P.~M.}\ \bibnamefont
  {Levy}},\ }\href {\doibase 10.1103/PhysRevLett.27.1383} {\bibfield  {journal}
  {\bibinfo  {journal} {Phys. Rev. Lett.}\ }\textbf {\bibinfo {volume} {27}},\
  \bibinfo {pages} {1383} (\bibinfo {year} {1971})}\BibitemShut {NoStop}%
\bibitem [{\citenamefont {Andreev}\ and\ \citenamefont
  {Grishchuk}(1984)}]{andreev84}%
  \BibitemOpen
  \bibfield  {author} {\bibinfo {author} {\bibfnamefont {A.}~\bibnamefont
  {Andreev}}\ and\ \bibinfo {author} {\bibfnamefont {I.}~\bibnamefont
  {Grishchuk}},\ }\href@noop {} {\bibfield  {journal} {\bibinfo  {journal}
  {JETP}\ }\textbf {\bibinfo {volume} {87}},\ \bibinfo {pages} {467} (\bibinfo
  {year} {1984})}\BibitemShut {NoStop}%
\bibitem [{\citenamefont {Chubukov}(1991)}]{chubukov91-PRB}%
  \BibitemOpen
  \bibfield  {author} {\bibinfo {author} {\bibfnamefont {A.~V.}\ \bibnamefont
  {Chubukov}},\ }\href {\doibase 10.1103/PhysRevB.44.4693} {\bibfield
  {journal} {\bibinfo  {journal} {Phys. Rev. B}\ }\textbf {\bibinfo {volume}
  {44}},\ \bibinfo {pages} {4693} (\bibinfo {year} {1991})}\BibitemShut
  {NoStop}%
\bibitem [{\citenamefont {Shannon}\ \emph {et~al.}(2006)\citenamefont
  {Shannon}, \citenamefont {Momoi},\ and\ \citenamefont
  {Sindzingre}}]{shannon06}%
  \BibitemOpen
  \bibfield  {author} {\bibinfo {author} {\bibfnamefont {N.}~\bibnamefont
  {Shannon}}, \bibinfo {author} {\bibfnamefont {T.}~\bibnamefont {Momoi}}, \
  and\ \bibinfo {author} {\bibfnamefont {P.}~\bibnamefont {Sindzingre}},\
  }\href {\doibase 10.1103/PhysRevLett.96.027213} {\bibfield  {journal}
  {\bibinfo  {journal} {Phys. Rev. Lett.}\ }\textbf {\bibinfo {volume} {96}},\
  \bibinfo {pages} {027213} (\bibinfo {year} {2006})}\BibitemShut {NoStop}%
\bibitem [{\citenamefont {Momoi}\ \emph {et~al.}(2006)\citenamefont {Momoi},
  \citenamefont {Sindzingre},\ and\ \citenamefont {Shannon}}]{momoi06}%
  \BibitemOpen
  \bibfield  {author} {\bibinfo {author} {\bibfnamefont {T.}~\bibnamefont
  {Momoi}}, \bibinfo {author} {\bibfnamefont {P.}~\bibnamefont {Sindzingre}}, \
  and\ \bibinfo {author} {\bibfnamefont {N.}~\bibnamefont {Shannon}},\ }\href
  {\doibase 10.1103/PhysRevLett.97.257204} {\bibfield  {journal} {\bibinfo
  {journal} {Phys. Rev. Lett.}\ }\textbf {\bibinfo {volume} {97}},\ \bibinfo
  {pages} {257204} (\bibinfo {year} {2006})}\BibitemShut {NoStop}%
\bibitem [{\citenamefont {Heidrich-Meisner}\ \emph {et~al.}(2006)\citenamefont
  {Heidrich-Meisner}, \citenamefont {Honecker},\ and\ \citenamefont
  {Vekua}}]{heidrich-meisner06}%
  \BibitemOpen
  \bibfield  {author} {\bibinfo {author} {\bibfnamefont {F.}~\bibnamefont
  {Heidrich-Meisner}}, \bibinfo {author} {\bibfnamefont {A.}~\bibnamefont
  {Honecker}}, \ and\ \bibinfo {author} {\bibfnamefont {T.}~\bibnamefont
  {Vekua}},\ }\href {\doibase 10.1103/PhysRevB.74.020403} {\bibfield  {journal}
  {\bibinfo  {journal} {Phys. Rev. B}\ }\textbf {\bibinfo {volume} {74}},\
  \bibinfo {pages} {020403} (\bibinfo {year} {2006})}\BibitemShut {NoStop}%
\bibitem [{\citenamefont {Lu}\ \emph {et~al.}(2006)\citenamefont {Lu},
  \citenamefont {Wang}, \citenamefont {Qin},\ and\ \citenamefont
  {Xiang}}]{lu06}%
  \BibitemOpen
  \bibfield  {author} {\bibinfo {author} {\bibfnamefont {H.~T.}\ \bibnamefont
  {Lu}}, \bibinfo {author} {\bibfnamefont {Y.~J.}\ \bibnamefont {Wang}},
  \bibinfo {author} {\bibfnamefont {S.}~\bibnamefont {Qin}}, \ and\ \bibinfo
  {author} {\bibfnamefont {T.}~\bibnamefont {Xiang}},\ }\href {\doibase
  10.1103/PhysRevB.74.134425} {\bibfield  {journal} {\bibinfo  {journal} {Phys.
  Rev. B}\ }\textbf {\bibinfo {volume} {74}},\ \bibinfo {pages} {134425}
  (\bibinfo {year} {2006})}\BibitemShut {NoStop}%
\bibitem [{\citenamefont {Ueda}\ and\ \citenamefont {Totsuka}(2007)}]{ueda07}%
  \BibitemOpen
  \bibfield  {author} {\bibinfo {author} {\bibfnamefont {H.}~\bibnamefont
  {Ueda}}\ and\ \bibinfo {author} {\bibfnamefont {K.}~\bibnamefont {Totsuka}},\
  }\href {\doibase 10.1103/PhysRevB.76.214428} {\bibfield  {journal} {\bibinfo
  {journal} {Phys. Rev. B}\ }\textbf {\bibinfo {volume} {76}},\ \bibinfo
  {pages} {214428} (\bibinfo {year} {2007})}\BibitemShut {NoStop}%
\bibitem [{\citenamefont {Vekua}\ \emph {et~al.}(2007)\citenamefont {Vekua},
  \citenamefont {Honecker}, \citenamefont {Mikeska},\ and\ \citenamefont
  {Heidrich-Meisner}}]{vekua07}%
  \BibitemOpen
  \bibfield  {author} {\bibinfo {author} {\bibfnamefont {T.}~\bibnamefont
  {Vekua}}, \bibinfo {author} {\bibfnamefont {A.}~\bibnamefont {Honecker}},
  \bibinfo {author} {\bibfnamefont {H.-J.}\ \bibnamefont {Mikeska}}, \ and\
  \bibinfo {author} {\bibfnamefont {F.}~\bibnamefont {Heidrich-Meisner}},\
  }\href {\doibase 10.1103/PhysRevB.76.174420} {\bibfield  {journal} {\bibinfo
  {journal} {Phys. Rev. B}\ }\textbf {\bibinfo {volume} {76}},\ \bibinfo
  {pages} {174420} (\bibinfo {year} {2007})}\BibitemShut {NoStop}%
\bibitem [{\citenamefont {Kecke}\ \emph {et~al.}(2007)\citenamefont {Kecke},
  \citenamefont {Momoi},\ and\ \citenamefont {Furusaki}}]{kecke07}%
  \BibitemOpen
  \bibfield  {author} {\bibinfo {author} {\bibfnamefont {L.}~\bibnamefont
  {Kecke}}, \bibinfo {author} {\bibfnamefont {T.}~\bibnamefont {Momoi}}, \ and\
  \bibinfo {author} {\bibfnamefont {A.}~\bibnamefont {Furusaki}},\ }\href
  {\doibase 10.1103/PhysRevB.76.060407} {\bibfield  {journal} {\bibinfo
  {journal} {Phys. Rev. B}\ }\textbf {\bibinfo {volume} {76}},\ \bibinfo
  {pages} {060407} (\bibinfo {year} {2007})}\BibitemShut {NoStop}%
\bibitem [{\citenamefont {Kuzian}\ and\ \citenamefont
  {Drechsler}(2007)}]{kuzian07}%
  \BibitemOpen
  \bibfield  {author} {\bibinfo {author} {\bibfnamefont {R.~O.}\ \bibnamefont
  {Kuzian}}\ and\ \bibinfo {author} {\bibfnamefont {S.-L.}\ \bibnamefont
  {Drechsler}},\ }\href {\doibase 10.1103/PhysRevB.75.024401} {\bibfield
  {journal} {\bibinfo  {journal} {Phys. Rev. B}\ }\textbf {\bibinfo {volume}
  {75}},\ \bibinfo {pages} {024401} (\bibinfo {year} {2007})}\BibitemShut
  {NoStop}%
\bibitem [{\citenamefont {Hikihara}\ \emph {et~al.}(2008)\citenamefont
  {Hikihara}, \citenamefont {Kecke}, \citenamefont {Momoi},\ and\ \citenamefont
  {Furusaki}}]{hikihara08}%
  \BibitemOpen
  \bibfield  {author} {\bibinfo {author} {\bibfnamefont {T.}~\bibnamefont
  {Hikihara}}, \bibinfo {author} {\bibfnamefont {L.}~\bibnamefont {Kecke}},
  \bibinfo {author} {\bibfnamefont {T.}~\bibnamefont {Momoi}}, \ and\ \bibinfo
  {author} {\bibfnamefont {A.}~\bibnamefont {Furusaki}},\ }\href {\doibase
  10.1103/PhysRevB.78.144404} {\bibfield  {journal} {\bibinfo  {journal} {Phys.
  Rev. B}\ }\textbf {\bibinfo {volume} {78}},\ \bibinfo {pages} {144404}
  (\bibinfo {year} {2008})}\BibitemShut {NoStop}%
\bibitem [{\citenamefont {Sudan}\ \emph {et~al.}(2009)\citenamefont {Sudan},
  \citenamefont {L\"uscher},\ and\ \citenamefont {L\"auchli}}]{sudan09}%
  \BibitemOpen
  \bibfield  {author} {\bibinfo {author} {\bibfnamefont {J.}~\bibnamefont
  {Sudan}}, \bibinfo {author} {\bibfnamefont {A.}~\bibnamefont {L\"uscher}}, \
  and\ \bibinfo {author} {\bibfnamefont {A.~M.}\ \bibnamefont {L\"auchli}},\
  }\href {\doibase 10.1103/PhysRevB.80.140402} {\bibfield  {journal} {\bibinfo
  {journal} {Phys. Rev. B}\ }\textbf {\bibinfo {volume} {80}},\ \bibinfo
  {pages} {140402} (\bibinfo {year} {2009})}\BibitemShut {NoStop}%
\bibitem [{\citenamefont {Ueda}\ and\ \citenamefont {Totsuka}(2009)}]{ueda09}%
  \BibitemOpen
  \bibfield  {author} {\bibinfo {author} {\bibfnamefont {H.~T.}\ \bibnamefont
  {Ueda}}\ and\ \bibinfo {author} {\bibfnamefont {K.}~\bibnamefont {Totsuka}},\
  }\href {\doibase 10.1103/PhysRevB.80.014417} {\bibfield  {journal} {\bibinfo
  {journal} {Phys. Rev. B}\ }\textbf {\bibinfo {volume} {80}},\ \bibinfo
  {pages} {014417} (\bibinfo {year} {2009})}\BibitemShut {NoStop}%
\bibitem [{\citenamefont {Zhitomirsky}\ and\ \citenamefont
  {Tsunetsugu}(2010)}]{zhitomirsky10}%
  \BibitemOpen
  \bibfield  {author} {\bibinfo {author} {\bibfnamefont {M.~E.}\ \bibnamefont
  {Zhitomirsky}}\ and\ \bibinfo {author} {\bibfnamefont {H.}~\bibnamefont
  {Tsunetsugu}},\ }\href {http://stacks.iop.org/0295-5075/92/i=3/a=37001}
  {\bibfield  {journal} {\bibinfo  {journal} {EPL (Europhysics Letters)}\
  }\textbf {\bibinfo {volume} {92}},\ \bibinfo {pages} {37001} (\bibinfo {year}
  {2010})}\BibitemShut {NoStop}%
\bibitem [{\citenamefont {Syromyatnikov}(2012)}]{syromyatnikov12}%
  \BibitemOpen
  \bibfield  {author} {\bibinfo {author} {\bibfnamefont {A.~V.}\ \bibnamefont
  {Syromyatnikov}},\ }\href {\doibase 10.1103/PhysRevB.86.014423} {\bibfield
  {journal} {\bibinfo  {journal} {Phys. Rev. B}\ }\textbf {\bibinfo {volume}
  {86}},\ \bibinfo {pages} {014423} (\bibinfo {year} {2012})}\BibitemShut
  {NoStop}%
\bibitem [{\citenamefont {Sato}\ \emph {et~al.}(2013)\citenamefont {Sato},
  \citenamefont {Hikihara},\ and\ \citenamefont {Momoi}}]{sato13}%
  \BibitemOpen
  \bibfield  {author} {\bibinfo {author} {\bibfnamefont {M.}~\bibnamefont
  {Sato}}, \bibinfo {author} {\bibfnamefont {T.}~\bibnamefont {Hikihara}}, \
  and\ \bibinfo {author} {\bibfnamefont {T.}~\bibnamefont {Momoi}},\ }\href
  {\doibase 10.1103/PhysRevLett.110.077206} {\bibfield  {journal} {\bibinfo
  {journal} {Phys. Rev. Lett.}\ }\textbf {\bibinfo {volume} {110}},\ \bibinfo
  {pages} {077206} (\bibinfo {year} {2013})}\BibitemShut {NoStop}%
\bibitem [{\citenamefont {Ueda}\ and\ \citenamefont {Momoi}(2013)}]{ueda13}%
  \BibitemOpen
  \bibfield  {author} {\bibinfo {author} {\bibfnamefont {H.~T.}\ \bibnamefont
  {Ueda}}\ and\ \bibinfo {author} {\bibfnamefont {T.}~\bibnamefont {Momoi}},\
  }\href {\doibase 10.1103/PhysRevB.87.144417} {\bibfield  {journal} {\bibinfo
  {journal} {Phys. Rev. B}\ }\textbf {\bibinfo {volume} {87}},\ \bibinfo
  {pages} {144417} (\bibinfo {year} {2013})}\BibitemShut {NoStop}%
\bibitem [{\citenamefont {Starykh}\ and\ \citenamefont
  {Balents}(2014)}]{starykh14}%
  \BibitemOpen
  \bibfield  {author} {\bibinfo {author} {\bibfnamefont {O.~A.}\ \bibnamefont
  {Starykh}}\ and\ \bibinfo {author} {\bibfnamefont {L.}~\bibnamefont
  {Balents}},\ }\href {\doibase 10.1103/PhysRevB.89.104407} {\bibfield
  {journal} {\bibinfo  {journal} {Phys. Rev. B}\ }\textbf {\bibinfo {volume}
  {89}},\ \bibinfo {pages} {104407} (\bibinfo {year} {2014})}\BibitemShut
  {NoStop}%
\bibitem [{\citenamefont {Smerald}\ and\ \citenamefont
  {Shannon}(2013)}]{smerald13}%
  \BibitemOpen
  \bibfield  {author} {\bibinfo {author} {\bibfnamefont {A.}~\bibnamefont
  {Smerald}}\ and\ \bibinfo {author} {\bibfnamefont {N.}~\bibnamefont
  {Shannon}},\ }\href {\doibase 10.1103/PhysRevB.88.184430} {\bibfield
  {journal} {\bibinfo  {journal} {Phys. Rev. B}\ }\textbf {\bibinfo {volume}
  {88}},\ \bibinfo {pages} {184430} (\bibinfo {year} {2013})}\BibitemShut
  {NoStop}%
\bibitem [{\citenamefont {{Smerald}}\ and\ \citenamefont
  {{Shannon}}(2013)}]{smerald-arXiv}%
  \BibitemOpen
  \bibfield  {author} {\bibinfo {author} {\bibfnamefont {A.}~\bibnamefont
  {{Smerald}}}\ and\ \bibinfo {author} {\bibfnamefont {N.}~\bibnamefont
  {{Shannon}}},\ }\href@noop {} {\bibfield  {journal} {\bibinfo  {journal}
  {ArXiv e-prints}\ } (\bibinfo {year} {2013})},\ \Eprint
  {http://arxiv.org/abs/1303.4465} {arXiv:1303.4465 [cond-mat.str-el]}
  \BibitemShut {NoStop}%
\bibitem [{\citenamefont {Smerald}(2013)}]{smerald-thesis}%
  \BibitemOpen
  \bibfield  {author} {\bibinfo {author} {\bibfnamefont {A.}~\bibnamefont
  {Smerald}},\ }\href@noop {} {\emph {\bibinfo {title} {Theory of the nuclear
  magnetic $1/T_1$ relaxation rate in conventional and unconventional
  magnets}}}\ (\bibinfo  {publisher} {Springer Theses},\ \bibinfo {year}
  {2013})\BibitemShut {NoStop}%
\bibitem [{\citenamefont {Ueda}(2015)}]{ueda-arXiv}%
  \BibitemOpen
  \bibfield  {author} {\bibinfo {author} {\bibfnamefont {H.~T.}\ \bibnamefont
  {Ueda}},\ }\href {\doibase 10.7566/JPSJ.84.023601} {\bibfield  {journal}
  {\bibinfo  {journal} {Journal of the Physical Society of Japan}\ }\textbf
  {\bibinfo {volume} {84}},\ \bibinfo {pages} {023601} (\bibinfo {year}
  {2015})},\ \Eprint
  {http://arxiv.org/abs/http://dx.doi.org/10.7566/JPSJ.84.023601}
  {http://dx.doi.org/10.7566/JPSJ.84.023601} \BibitemShut {NoStop}%
\bibitem [{\citenamefont {Nath}\ \emph {et~al.}(2008)\citenamefont {Nath},
  \citenamefont {Tsirlin}, \citenamefont {Rosner},\ and\ \citenamefont
  {Geibel}}]{nath08}%
  \BibitemOpen
  \bibfield  {author} {\bibinfo {author} {\bibfnamefont {R.}~\bibnamefont
  {Nath}}, \bibinfo {author} {\bibfnamefont {A.~A.}\ \bibnamefont {Tsirlin}},
  \bibinfo {author} {\bibfnamefont {H.}~\bibnamefont {Rosner}}, \ and\ \bibinfo
  {author} {\bibfnamefont {C.}~\bibnamefont {Geibel}},\ }\href {\doibase
  10.1103/PhysRevB.78.064422} {\bibfield  {journal} {\bibinfo  {journal} {Phys.
  Rev. B}\ }\textbf {\bibinfo {volume} {78}},\ \bibinfo {pages} {064422}
  (\bibinfo {year} {2008})}\BibitemShut {NoStop}%
\bibitem [{\citenamefont {Sindzingre}\ \emph {et~al.}(2009)\citenamefont
  {Sindzingre}, \citenamefont {Seabra}, \citenamefont {Shannon},\ and\
  \citenamefont {Momoi}}]{sindzingre09}%
  \BibitemOpen
  \bibfield  {author} {\bibinfo {author} {\bibfnamefont {P.}~\bibnamefont
  {Sindzingre}}, \bibinfo {author} {\bibfnamefont {L.}~\bibnamefont {Seabra}},
  \bibinfo {author} {\bibfnamefont {N.}~\bibnamefont {Shannon}}, \ and\
  \bibinfo {author} {\bibfnamefont {T.}~\bibnamefont {Momoi}},\ }\href
  {http://stacks.iop.org/1742-6596/145/i=1/a=012048} {\bibfield  {journal}
  {\bibinfo  {journal} {Journal of Physics: Conference Series}\ }\textbf
  {\bibinfo {volume} {145}},\ \bibinfo {pages} {012048} (\bibinfo {year}
  {2009})}\BibitemShut {NoStop}%
\bibitem [{\citenamefont {Kaul}\ \emph {et~al.}(2004)\citenamefont {Kaul},
  \citenamefont {Rosner}, \citenamefont {Shannon}, \citenamefont
  {Shpanchenko},\ and\ \citenamefont {Geibel}}]{kaul04}%
  \BibitemOpen
  \bibfield  {author} {\bibinfo {author} {\bibfnamefont {E.}~\bibnamefont
  {Kaul}}, \bibinfo {author} {\bibfnamefont {H.}~\bibnamefont {Rosner}},
  \bibinfo {author} {\bibfnamefont {N.}~\bibnamefont {Shannon}}, \bibinfo
  {author} {\bibfnamefont {R.}~\bibnamefont {Shpanchenko}}, \ and\ \bibinfo
  {author} {\bibfnamefont {C.}~\bibnamefont {Geibel}},\ }\href {\doibase
  http://dx.doi.org/10.1016/j.jmmm.2003.12.002} {\bibfield  {journal} {\bibinfo
   {journal} {Journal of Magnetism and Magnetic Materials}\ }\textbf {\bibinfo
  {volume} {272--276, Part 2}},\ \bibinfo {pages} {922 } (\bibinfo {year}
  {2004})},\ \bibinfo {note} {proceedings of the International Conference on
  Magnetism (ICM 2003)}\BibitemShut {NoStop}%
\bibitem [{\citenamefont {Kaul}(2005)}]{kaul05}%
  \BibitemOpen
  \bibfield  {author} {\bibinfo {author} {\bibfnamefont {E.~E.}\ \bibnamefont
  {Kaul}},\ }\href@noop {} {Ph.D. thesis},\ \bibinfo  {school} {Technical
  University Dresden} (\bibinfo {year} {2005})\BibitemShut {NoStop}%
\bibitem [{\citenamefont {Tsirlin}\ \emph {et~al.}(2009)\citenamefont
  {Tsirlin}, \citenamefont {Schmidt}, \citenamefont {Skourski}, \citenamefont
  {Nath}, \citenamefont {Geibel},\ and\ \citenamefont {Rosner}}]{tsirlin09}%
  \BibitemOpen
  \bibfield  {author} {\bibinfo {author} {\bibfnamefont {A.~A.}\ \bibnamefont
  {Tsirlin}}, \bibinfo {author} {\bibfnamefont {B.}~\bibnamefont {Schmidt}},
  \bibinfo {author} {\bibfnamefont {Y.}~\bibnamefont {Skourski}}, \bibinfo
  {author} {\bibfnamefont {R.}~\bibnamefont {Nath}}, \bibinfo {author}
  {\bibfnamefont {C.}~\bibnamefont {Geibel}}, \ and\ \bibinfo {author}
  {\bibfnamefont {H.}~\bibnamefont {Rosner}},\ }\href {\doibase
  10.1103/PhysRevB.80.132407} {\bibfield  {journal} {\bibinfo  {journal} {Phys.
  Rev. B}\ }\textbf {\bibinfo {volume} {80}},\ \bibinfo {pages} {132407}
  (\bibinfo {year} {2009})}\BibitemShut {NoStop}%
\bibitem [{\citenamefont {Skoulatos}\ \emph {et~al.}(2009)\citenamefont
  {Skoulatos}, \citenamefont {Goff}, \citenamefont {Geibel}, \citenamefont
  {Kaul}, \citenamefont {Nath}, \citenamefont {Shannon}, \citenamefont
  {Schmidt}, \citenamefont {Murani}, \citenamefont {Deen}, \citenamefont
  {Enderle},\ and\ \citenamefont {Wildes}}]{skoulatos09}%
  \BibitemOpen
  \bibfield  {author} {\bibinfo {author} {\bibfnamefont {M.}~\bibnamefont
  {Skoulatos}}, \bibinfo {author} {\bibfnamefont {J.~P.}\ \bibnamefont {Goff}},
  \bibinfo {author} {\bibfnamefont {C.}~\bibnamefont {Geibel}}, \bibinfo
  {author} {\bibfnamefont {E.~E.}\ \bibnamefont {Kaul}}, \bibinfo {author}
  {\bibfnamefont {R.}~\bibnamefont {Nath}}, \bibinfo {author} {\bibfnamefont
  {N.}~\bibnamefont {Shannon}}, \bibinfo {author} {\bibfnamefont
  {B.}~\bibnamefont {Schmidt}}, \bibinfo {author} {\bibfnamefont {A.~P.}\
  \bibnamefont {Murani}}, \bibinfo {author} {\bibfnamefont {P.~P.}\
  \bibnamefont {Deen}}, \bibinfo {author} {\bibfnamefont {M.}~\bibnamefont
  {Enderle}}, \ and\ \bibinfo {author} {\bibfnamefont {A.~R.}\ \bibnamefont
  {Wildes}},\ }\href {http://stacks.iop.org/0295-5075/88/i=5/a=57005}
  {\bibfield  {journal} {\bibinfo  {journal} {EPL (Europhysics Letters)}\
  }\textbf {\bibinfo {volume} {88}},\ \bibinfo {pages} {57005} (\bibinfo {year}
  {2009})}\BibitemShut {NoStop}%
\bibitem [{\citenamefont {Nath}\ \emph {et~al.}(2009)\citenamefont {Nath},
  \citenamefont {Furukawa}, \citenamefont {Borsa}, \citenamefont {Kaul},
  \citenamefont {Baenitz}, \citenamefont {Geibel},\ and\ \citenamefont
  {Johnston}}]{nath09}%
  \BibitemOpen
  \bibfield  {author} {\bibinfo {author} {\bibfnamefont {R.}~\bibnamefont
  {Nath}}, \bibinfo {author} {\bibfnamefont {Y.}~\bibnamefont {Furukawa}},
  \bibinfo {author} {\bibfnamefont {F.}~\bibnamefont {Borsa}}, \bibinfo
  {author} {\bibfnamefont {E.~E.}\ \bibnamefont {Kaul}}, \bibinfo {author}
  {\bibfnamefont {M.}~\bibnamefont {Baenitz}}, \bibinfo {author} {\bibfnamefont
  {C.}~\bibnamefont {Geibel}}, \ and\ \bibinfo {author} {\bibfnamefont {D.~C.}\
  \bibnamefont {Johnston}},\ }\href {\doibase 10.1103/PhysRevB.80.214430}
  {\bibfield  {journal} {\bibinfo  {journal} {Phys. Rev. B}\ }\textbf {\bibinfo
  {volume} {80}},\ \bibinfo {pages} {214430} (\bibinfo {year}
  {2009})}\BibitemShut {NoStop}%
\bibitem [{\citenamefont {Bossoni}\ \emph {et~al.}(2011)\citenamefont
  {Bossoni}, \citenamefont {Carretta}, \citenamefont {Nath}, \citenamefont
  {Moscardini}, \citenamefont {Baenitz},\ and\ \citenamefont
  {Geibel}}]{bossoni11}%
  \BibitemOpen
  \bibfield  {author} {\bibinfo {author} {\bibfnamefont {L.}~\bibnamefont
  {Bossoni}}, \bibinfo {author} {\bibfnamefont {P.}~\bibnamefont {Carretta}},
  \bibinfo {author} {\bibfnamefont {R.}~\bibnamefont {Nath}}, \bibinfo {author}
  {\bibfnamefont {M.}~\bibnamefont {Moscardini}}, \bibinfo {author}
  {\bibfnamefont {M.}~\bibnamefont {Baenitz}}, \ and\ \bibinfo {author}
  {\bibfnamefont {C.}~\bibnamefont {Geibel}},\ }\href {\doibase
  10.1103/PhysRevB.83.014412} {\bibfield  {journal} {\bibinfo  {journal} {Phys.
  Rev. B}\ }\textbf {\bibinfo {volume} {83}},\ \bibinfo {pages} {014412}
  (\bibinfo {year} {2011})}\BibitemShut {NoStop}%
\bibitem [{\citenamefont {Shindou}\ and\ \citenamefont
  {Momoi}(2009)}]{shindou09}%
  \BibitemOpen
  \bibfield  {author} {\bibinfo {author} {\bibfnamefont {R.}~\bibnamefont
  {Shindou}}\ and\ \bibinfo {author} {\bibfnamefont {T.}~\bibnamefont
  {Momoi}},\ }\href {\doibase 10.1103/PhysRevB.80.064410} {\bibfield  {journal}
  {\bibinfo  {journal} {Phys. Rev. B}\ }\textbf {\bibinfo {volume} {80}},\
  \bibinfo {pages} {064410} (\bibinfo {year} {2009})}\BibitemShut {NoStop}%
\bibitem [{\citenamefont {Shindou}\ \emph {et~al.}(2011)\citenamefont
  {Shindou}, \citenamefont {Yunoki},\ and\ \citenamefont {Momoi}}]{shindou11}%
  \BibitemOpen
  \bibfield  {author} {\bibinfo {author} {\bibfnamefont {R.}~\bibnamefont
  {Shindou}}, \bibinfo {author} {\bibfnamefont {S.}~\bibnamefont {Yunoki}}, \
  and\ \bibinfo {author} {\bibfnamefont {T.}~\bibnamefont {Momoi}},\ }\href
  {\doibase 10.1103/PhysRevB.84.134414} {\bibfield  {journal} {\bibinfo
  {journal} {Phys. Rev. B}\ }\textbf {\bibinfo {volume} {84}},\ \bibinfo
  {pages} {134414} (\bibinfo {year} {2011})}\BibitemShut {NoStop}%
\bibitem [{\citenamefont {Shindou}\ \emph {et~al.}(2013)\citenamefont
  {Shindou}, \citenamefont {Yunoki},\ and\ \citenamefont {Momoi}}]{shindou13}%
  \BibitemOpen
  \bibfield  {author} {\bibinfo {author} {\bibfnamefont {R.}~\bibnamefont
  {Shindou}}, \bibinfo {author} {\bibfnamefont {S.}~\bibnamefont {Yunoki}}, \
  and\ \bibinfo {author} {\bibfnamefont {T.}~\bibnamefont {Momoi}},\ }\href
  {\doibase 10.1103/PhysRevB.87.054429} {\bibfield  {journal} {\bibinfo
  {journal} {Phys. Rev. B}\ }\textbf {\bibinfo {volume} {87}},\ \bibinfo
  {pages} {054429} (\bibinfo {year} {2013})}\BibitemShut {NoStop}%
\bibitem [{\citenamefont {Chubukov}\ and\ \citenamefont
  {Khveschenko}(1987)}]{chubukov87}%
  \BibitemOpen
  \bibfield  {author} {\bibinfo {author} {\bibfnamefont {A.~V.}\ \bibnamefont
  {Chubukov}}\ and\ \bibinfo {author} {\bibfnamefont {D.~V.}\ \bibnamefont
  {Khveschenko}},\ }\href {http://stacks.iop.org/0022-3719/20/i=22/a=004}
  {\bibfield  {journal} {\bibinfo  {journal} {Journal of Physics C: Solid State
  Physics}\ }\textbf {\bibinfo {volume} {20}},\ \bibinfo {pages} {L505}
  (\bibinfo {year} {1987})}\BibitemShut {NoStop}%
\bibitem [{\citenamefont {Chubukov}(1990)}]{chubukov90}%
  \BibitemOpen
  \bibfield  {author} {\bibinfo {author} {\bibfnamefont {A.~V.}\ \bibnamefont
  {Chubukov}},\ }\href {\doibase {10.1088/0953-8984/2/19/012}} {\bibfield
  {journal} {\bibinfo  {journal} {J. Phys.-Condes. Matter}\ }\textbf {\bibinfo
  {volume} {2}},\ \bibinfo {pages} {4455} (\bibinfo {year} {1990})}\BibitemShut
  {NoStop}%
\bibitem [{\citenamefont {Chubukov}\ and\ \citenamefont
  {Golosov}(1991)}]{chubukov91-JPCM}%
  \BibitemOpen
  \bibfield  {author} {\bibinfo {author} {\bibfnamefont {A.~V.}\ \bibnamefont
  {Chubukov}}\ and\ \bibinfo {author} {\bibfnamefont {D.~I.}\ \bibnamefont
  {Golosov}},\ }\href {\doibase {10.1088/0953-8984/3/1/005}} {\bibfield
  {journal} {\bibinfo  {journal} {J. Phys.-Condes. Matter}\ }\textbf {\bibinfo
  {volume} {3}},\ \bibinfo {pages} {69} (\bibinfo {year} {1991})}\BibitemShut
  {NoStop}%
\bibitem [{\citenamefont {Sato}\ \emph {et~al.}(2009)\citenamefont {Sato},
  \citenamefont {Momoi},\ and\ \citenamefont {Furusaki}}]{sato09}%
  \BibitemOpen
  \bibfield  {author} {\bibinfo {author} {\bibfnamefont {M.}~\bibnamefont
  {Sato}}, \bibinfo {author} {\bibfnamefont {T.}~\bibnamefont {Momoi}}, \ and\
  \bibinfo {author} {\bibfnamefont {A.}~\bibnamefont {Furusaki}},\ }\href
  {\doibase 10.1103/PhysRevB.79.060406} {\bibfield  {journal} {\bibinfo
  {journal} {Phys. Rev. B}\ }\textbf {\bibinfo {volume} {79}},\ \bibinfo
  {pages} {060406} (\bibinfo {year} {2009})}\BibitemShut {NoStop}%
\bibitem [{\citenamefont {Sato}\ \emph {et~al.}(2011)\citenamefont {Sato},
  \citenamefont {Hikihara},\ and\ \citenamefont {Momoi}}]{sato11}%
  \BibitemOpen
  \bibfield  {author} {\bibinfo {author} {\bibfnamefont {M.}~\bibnamefont
  {Sato}}, \bibinfo {author} {\bibfnamefont {T.}~\bibnamefont {Hikihara}}, \
  and\ \bibinfo {author} {\bibfnamefont {T.}~\bibnamefont {Momoi}},\ }\href
  {\doibase 10.1103/PhysRevB.83.064405} {\bibfield  {journal} {\bibinfo
  {journal} {Phys. Rev. B}\ }\textbf {\bibinfo {volume} {83}},\ \bibinfo
  {pages} {064405} (\bibinfo {year} {2011})}\BibitemShut {NoStop}%
\bibitem [{\citenamefont {Svistov}\ \emph {et~al.}(2011)\citenamefont
  {Svistov}, \citenamefont {Fujita}, \citenamefont {Yamaguchi}, \citenamefont
  {Kimura}, \citenamefont {Omura}, \citenamefont {Prokofiev}, \citenamefont
  {Smirnov}, \citenamefont {Honda},\ and\ \citenamefont
  {Hagiwara}}]{svistov10}%
  \BibitemOpen
  \bibfield  {author} {\bibinfo {author} {\bibfnamefont {L.}~\bibnamefont
  {Svistov}}, \bibinfo {author} {\bibfnamefont {T.}~\bibnamefont {Fujita}},
  \bibinfo {author} {\bibfnamefont {H.}~\bibnamefont {Yamaguchi}}, \bibinfo
  {author} {\bibfnamefont {S.}~\bibnamefont {Kimura}}, \bibinfo {author}
  {\bibfnamefont {K.}~\bibnamefont {Omura}}, \bibinfo {author} {\bibfnamefont
  {A.}~\bibnamefont {Prokofiev}}, \bibinfo {author} {\bibfnamefont
  {A.}~\bibnamefont {Smirnov}}, \bibinfo {author} {\bibfnamefont
  {Z.}~\bibnamefont {Honda}}, \ and\ \bibinfo {author} {\bibfnamefont
  {M.}~\bibnamefont {Hagiwara}},\ }\href
  {http://dx.doi.org/10.1134/S0021364011010073} {\bibfield  {journal} {\bibinfo
   {journal} {JETP Letters}\ }\textbf {\bibinfo {volume} {93}},\ \bibinfo
  {pages} {21} (\bibinfo {year} {2011})}\BibitemShut {NoStop}%
\bibitem [{\citenamefont {B\"uttgen}\ \emph {et~al.}(2014)\citenamefont
  {B\"uttgen}, \citenamefont {Nawa}, \citenamefont {Fujita}, \citenamefont
  {Hagiwara}, \citenamefont {Kuhns}, \citenamefont {Prokofiev}, \citenamefont
  {Reyes}, \citenamefont {Svistov}, \citenamefont {Yoshimura},\ and\
  \citenamefont {Takigawa}}]{buttgen14}%
  \BibitemOpen
  \bibfield  {author} {\bibinfo {author} {\bibfnamefont {N.}~\bibnamefont
  {B\"uttgen}}, \bibinfo {author} {\bibfnamefont {K.}~\bibnamefont {Nawa}},
  \bibinfo {author} {\bibfnamefont {T.}~\bibnamefont {Fujita}}, \bibinfo
  {author} {\bibfnamefont {M.}~\bibnamefont {Hagiwara}}, \bibinfo {author}
  {\bibfnamefont {P.}~\bibnamefont {Kuhns}}, \bibinfo {author} {\bibfnamefont
  {A.}~\bibnamefont {Prokofiev}}, \bibinfo {author} {\bibfnamefont {A.~P.}\
  \bibnamefont {Reyes}}, \bibinfo {author} {\bibfnamefont {L.~E.}\ \bibnamefont
  {Svistov}}, \bibinfo {author} {\bibfnamefont {K.}~\bibnamefont {Yoshimura}},
  \ and\ \bibinfo {author} {\bibfnamefont {M.}~\bibnamefont {Takigawa}},\
  }\href {\doibase 10.1103/PhysRevB.90.134401} {\bibfield  {journal} {\bibinfo
  {journal} {Phys. Rev. B}\ }\textbf {\bibinfo {volume} {90}},\ \bibinfo
  {pages} {134401} (\bibinfo {year} {2014})}\BibitemShut {NoStop}%
\bibitem [{\citenamefont {Papanicolaou}(1984)}]{papanicolau84}%
  \BibitemOpen
  \bibfield  {author} {\bibinfo {author} {\bibfnamefont {N.}~\bibnamefont
  {Papanicolaou}},\ }\href {\doibase 10.1016/0550-3213(84)90268-2} {\bibfield
  {journal} {\bibinfo  {journal} {Nuclear Physics B}\ }\textbf {\bibinfo
  {volume} {240}},\ \bibinfo {pages} {281 } (\bibinfo {year}
  {1984})}\BibitemShut {NoStop}%
\bibitem [{\citenamefont {Papanicolaou}(1988)}]{papanicolau88}%
  \BibitemOpen
  \bibfield  {author} {\bibinfo {author} {\bibfnamefont {N.}~\bibnamefont
  {Papanicolaou}},\ }\href@noop {} {\bibfield  {journal} {\bibinfo  {journal}
  {Nuclear Physics B}\ }\textbf {\bibinfo {volume} {305}},\ \bibinfo {pages}
  {367} (\bibinfo {year} {1988})}\BibitemShut {NoStop}%
\bibitem [{\citenamefont {Onufrieva}(1985)}]{onufrieva85}%
  \BibitemOpen
  \bibfield  {author} {\bibinfo {author} {\bibfnamefont {F.~P.}\ \bibnamefont
  {Onufrieva}},\ }\href@noop {} {\bibfield  {journal} {\bibinfo  {journal} {Zh.
  Eksp. Teor. Fiz.}\ }\textbf {\bibinfo {volume} {89}},\ \bibinfo {pages}
  {2270} (\bibinfo {year} {1985})}\BibitemShut {NoStop}%
\bibitem [{\citenamefont {L\"auchli}\ \emph {et~al.}(2006)\citenamefont
  {L\"auchli}, \citenamefont {Mila},\ and\ \citenamefont {Penc}}]{lauchli06}%
  \BibitemOpen
  \bibfield  {author} {\bibinfo {author} {\bibfnamefont {A.}~\bibnamefont
  {L\"auchli}}, \bibinfo {author} {\bibfnamefont {F.}~\bibnamefont {Mila}}, \
  and\ \bibinfo {author} {\bibfnamefont {K.}~\bibnamefont {Penc}},\ }\href
  {\doibase 10.1103/PhysRevLett.97.087205} {\bibfield  {journal} {\bibinfo
  {journal} {Phys. Rev. Lett.}\ }\textbf {\bibinfo {volume} {97}},\ \bibinfo
  {pages} {087205} (\bibinfo {year} {2006})}\BibitemShut {NoStop}%
\bibitem [{\citenamefont {Tsunetsugu}\ and\ \citenamefont
  {Arikawa}(2006)}]{tsunetsugu06}%
  \BibitemOpen
  \bibfield  {author} {\bibinfo {author} {\bibfnamefont {H.}~\bibnamefont
  {Tsunetsugu}}\ and\ \bibinfo {author} {\bibfnamefont {M.}~\bibnamefont
  {Arikawa}},\ }\href {\doibase 10.1143/JPSJ.75.083701} {\bibfield  {journal}
  {\bibinfo  {journal} {Journal of the Physical Society of Japan}\ }\textbf
  {\bibinfo {volume} {75}},\ \bibinfo {pages} {083701} (\bibinfo {year}
  {2006})}\BibitemShut {NoStop}%
\bibitem [{\citenamefont {Smerald}\ and\ \citenamefont
  {Shannon}()}]{smerald-unpub}%
  \BibitemOpen
  \bibfield  {author} {\bibinfo {author} {\bibfnamefont {A.}~\bibnamefont
  {Smerald}}\ and\ \bibinfo {author} {\bibfnamefont {N.}~\bibnamefont
  {Shannon}},\ }\href@noop {} {}\bibinfo {note} {Unpublished}\BibitemShut
  {NoStop}%
\bibitem [{\citenamefont {Ivanov}\ and\ \citenamefont
  {Kolezhuk}(2003)}]{ivanov03}%
  \BibitemOpen
  \bibfield  {author} {\bibinfo {author} {\bibfnamefont {B.~A.}\ \bibnamefont
  {Ivanov}}\ and\ \bibinfo {author} {\bibfnamefont {A.~K.}\ \bibnamefont
  {Kolezhuk}},\ }\href {\doibase 10.1103/PhysRevB.68.052401} {\bibfield
  {journal} {\bibinfo  {journal} {Phys. Rev. B}\ }\textbf {\bibinfo {volume}
  {68}},\ \bibinfo {pages} {052401} (\bibinfo {year} {2003})}\BibitemShut
  {NoStop}%
\bibitem [{\citenamefont {Ivanov}\ and\ \citenamefont
  {Khymyn}(2007)}]{ivanov07}%
  \BibitemOpen
  \bibfield  {author} {\bibinfo {author} {\bibfnamefont {B.~A.}\ \bibnamefont
  {Ivanov}}\ and\ \bibinfo {author} {\bibfnamefont {R.~S.}\ \bibnamefont
  {Khymyn}},\ }\href {\doibase {10.1134/S106377610702015X}} {\bibfield
  {journal} {\bibinfo  {journal} {J. Exp. Theor. Phys.}\ }\textbf {\bibinfo
  {volume} {104}},\ \bibinfo {pages} {307} (\bibinfo {year}
  {2007})}\BibitemShut {NoStop}%
\bibitem [{\citenamefont {Ivanov}\ \emph {et~al.}(2008)\citenamefont {Ivanov},
  \citenamefont {Khymyn},\ and\ \citenamefont {Kolezhuk}}]{ivanov08}%
  \BibitemOpen
  \bibfield  {author} {\bibinfo {author} {\bibfnamefont {B.~A.}\ \bibnamefont
  {Ivanov}}, \bibinfo {author} {\bibfnamefont {R.~S.}\ \bibnamefont {Khymyn}},
  \ and\ \bibinfo {author} {\bibfnamefont {A.~K.}\ \bibnamefont {Kolezhuk}},\
  }\href {\doibase 10.1103/PhysRevLett.100.047203} {\bibfield  {journal}
  {\bibinfo  {journal} {Phys. Rev. Lett.}\ }\textbf {\bibinfo {volume} {100}},\
  \bibinfo {pages} {047203} (\bibinfo {year} {2008})}\BibitemShut {NoStop}%
\bibitem [{SM-()}]{SM-spin1chiqomh}%
  \BibitemOpen
  \href@noop {} {}\bibinfo {howpublished} {Animated version of
  Fig.~\ref{fig:bbq-chiqomh}, showing flavour-wave predictions for the
  imaginary part of the dynamic spin susceptibility of a spin-1,
  partially-polarised, 2-sublattice, spin-nematic state in applied magnetic
  field. The transverse, $\Im m \chi^{\perp}({\bf q},\omega)$
  [Eq.~\ref{eq:bbq-chiqomh}] and longitudinal, $\Im m \chi^{\sf zz}({\bf
  q},\omega) $ [Eq.~\ref{eq:bbq-chiqomh}] components are shown together. The
  same parameter set, $J_{11}=1$, $J_{12}=0.1$ and $J_{22}=2$ is used as in
  Fig.~\ref{fig:bbq-chiqomh}. Magnetic field is varied from an initial value of
  $h=1.5h_{\sf sat}$ to a final value of $h=0$. Dashed red lines show
  $\omega_{{\bf q}+{\bf q}_{\sf M},h}^{\sf a}$ and $\omega_{{\bf q},h}^{\sf b}$
  at all $h$ and $\omega_{{\bf q}+{\bf q}_{\sf M},h}^{\sf b}$ for $h\leq h_{\sf
  sat}$ [see Eq.~\ref{eq:omega-satPM}, Eq.~\ref{eq:omegakha},
  Eq.~\ref{eq:omegakhb}]. All predictions have been convoluted with a gaussian
  to mimic experimental resolution. The circuit $\Gamma$-{\sf X}-{\sf
  M}-$\Gamma$ in the bond-centred Brillouin zone is shown in
  Fig.~\ref{fig:bz}.}\BibitemShut {Stop}%
\bibitem [{\citenamefont {Batyev}\ and\ \citenamefont
  {Braginskii}(1984)}]{batyev84}%
  \BibitemOpen
  \bibfield  {author} {\bibinfo {author} {\bibfnamefont {E.~G.}\ \bibnamefont
  {Batyev}}\ and\ \bibinfo {author} {\bibfnamefont {L.~S.}\ \bibnamefont
  {Braginskii}},\ }\href@noop {} {\bibfield  {journal} {\bibinfo  {journal}
  {Zh. Eksp. Teor. Fiz.}\ }\textbf {\bibinfo {volume} {87}},\ \bibinfo {pages}
  {1361} (\bibinfo {year} {1984})}\BibitemShut {NoStop}%
\bibitem [{\citenamefont {Batyev}(1985)}]{batyev85}%
  \BibitemOpen
  \bibfield  {author} {\bibinfo {author} {\bibfnamefont {E.~G.}\ \bibnamefont
  {Batyev}},\ }\href@noop {} {\bibfield  {journal} {\bibinfo  {journal} {Zh.
  Eksp. Teor. Fiz.}\ }\textbf {\bibinfo {volume} {89}},\ \bibinfo {pages} {308}
  (\bibinfo {year} {1985})}\BibitemShut {NoStop}%
\bibitem [{\citenamefont {Nikuni}\ and\ \citenamefont
  {Shiba}(1995)}]{nikuni95}%
  \BibitemOpen
  \bibfield  {author} {\bibinfo {author} {\bibfnamefont {T.}~\bibnamefont
  {Nikuni}}\ and\ \bibinfo {author} {\bibfnamefont {H.}~\bibnamefont {Shiba}},\
  }\href {\doibase 10.1143/JPSJ.64.3471} {\bibfield  {journal} {\bibinfo
  {journal} {Journal of the Physical Society of Japan}\ }\textbf {\bibinfo
  {volume} {64}},\ \bibinfo {pages} {3471} (\bibinfo {year}
  {1995})}\BibitemShut {NoStop}%
\bibitem [{\citenamefont {Nakanishi}(1969)}]{nakanishi69}%
  \BibitemOpen
  \bibfield  {author} {\bibinfo {author} {\bibfnamefont {N.}~\bibnamefont
  {Nakanishi}},\ }\href {\doibase 10.1143/PTPS.43.1} {\bibfield  {journal}
  {\bibinfo  {journal} {Progress of Theoretical Physics Supplement}\ }\textbf
  {\bibinfo {volume} {43}},\ \bibinfo {pages} {1} (\bibinfo {year} {1969})},\
  \Eprint
  {http://arxiv.org/abs/http://ptps.oxfordjournals.org/content/43/1.full.pdf+html}
  {http://ptps.oxfordjournals.org/content/43/1.full.pdf+html} \BibitemShut
  {NoStop}%
\bibitem [{\citenamefont {{Ueda}}\ and\ \citenamefont
  {{Totsuka}}(2014)}]{ueda-arXiv2}%
  \BibitemOpen
  \bibfield  {author} {\bibinfo {author} {\bibfnamefont {H.~T.}\ \bibnamefont
  {{Ueda}}}\ and\ \bibinfo {author} {\bibfnamefont {K.}~\bibnamefont
  {{Totsuka}}},\ }\href@noop {} {\bibfield  {journal} {\bibinfo  {journal}
  {ArXiv e-prints}\ } (\bibinfo {year} {2014})},\ \Eprint
  {http://arxiv.org/abs/1406.1960} {arXiv:1406.1960 [cond-mat.str-el]}
  \BibitemShut {NoStop}%
\end{thebibliography}%



\end{document}